# Solar Elemental Abundances

Katharina Lodders
Department of Earth & Planetary Sciences and McDonnell Center for the Space Sciences
Washington University, Campus Box 1169, 1 Brookings Drive, St. Louis MO 63130

## Table of Contents

## Summary

Solar elemental abundances, or solar system elemental abundances refer to the complement of chemical elements in the entire solar system. The sun contains more than 99-percent of the mass in the solar system and therefore the composition of the sun is a good proxy for the composition of the overall solar system. The solar system composition can be taken as the overall composition of the molecular cloud within the interstellar medium from which the solar system formed 4.567 billion years ago. Active research areas in astronomy and cosmochemistry model collapse of a molecular cloud of solar composition into a star with a planetary system, and the physical and chemical fractionation of the elements during planetary formation and differentiation. The solar system composition is the initial composition from which all solar system objects (the sun, terrestrial planets, gas giant planets, planetary satellites and moons, asteroids, Kuiper-belt objects, and comets) were derived.

Other dwarf stars (with hydrostatic Hydrogen-burning in their cores) like our Sun (type G2V dwarf star) within the solar neighborhood have compositions similar to our Sun and the solar system composition. In general, differential comparisons of stellar compositions provide insights about stellar evolution as functions of stellar mass and age, and ongoing nucleosynthesis; but also about galactic chemical evolution when elemental compositions of stellar populations across our Milky Way Galaxy is considered. Comparisons to solar composition can reveal element destruction (e.g., Li) in the sun and in other dwarf stars. The comparisons also show element production of e.g., C, N, O, and the heavy elements made by the s-process in low- to intermediate mass stars (3-7 solar masses) after these evolved from their dwarf-star stage into red giant stars (where hydrogen and helium burning can occur in shells around their cores). The solar system abundances are and have been a critical test composition for nucleosynthesis models and models of Galactic chemical evolution, which aim ultimately to track the production of the elements heavier than hydrogen and helium in the generation of stars that came forth after the Big Bang 13.4 billion years ago.

## Definitions

**Element concentrations** usually refer to the elemental content of a given system by mass-fraction or mass-percent of total system mass present. Common units are gram per gram (g/g) units, where 0.01g/g = 1 weight% = 1 mass%, milli-gram per gram (mg/g = $10^{-3}$ g/g = 0.1 mass%), micro-gram per gram (µg/g) or parts per million, ppm (1 µg/g = $10^{-6}$ g/g = 1 ppm = 0.0001 mass%), nano-gram per gram (ng/g) or parts per billion, (ppb), where 1 ng/g = $10^{-9}$ g/g = 1 ppb = 0.001 ppm, and parts per trillion (ppt = $10^{-12}$ g/g = 0.001 ppb).Sometimes atomic or molar concentrations are used, these refer to the number of atoms or molecules as part of total atoms or molecules present, common units are atom-percent (at%) or mole fractions.

**Elemental abundances** refer to the *relative* elemental contents of a given system (e.g., the sun, a planet, meteorite or comet). Abundances are typically reported on an atomic scale by number relative to the number of hydrogen atoms (astronomical abundance scale) or silicon atoms (cosmochemical abundance scale). Relative abundances by mass are rarely used. Sometimes the terms abundances and concentrations are used inter-exchangeably. Concentrations refer to content as a part of the total system, and abundances refer to content in relation to contents of a reference species.

**Solar mass-fractions X, Y, and Z of the elements**. The mass-fraction of hydrogen and helium in the Sun are called “X” and “Y” respectively. The symbol “Z” is the mass fraction of all other elements heavier than helium (Li to U). Astronomers lump together all heavier elements because these only make about Z = 0.0148 of the sun whereas hydrogen has X = 0.7389 and helium Y= 0.2463. The solar mass fractions are important quantities for the standard solar model that describes how the sun works and evolved over time. The mass fractions quoted here are for the outer convection zone of the Sun and are based on helioseismic measurements (Basu & Antia 2004) and the Z/X ratio (0.020) is derived independently from the abundances in the photosphere (see below) which is the observable top layer of the Sun's convection zone.

**Solar system abundances** refer to the elemental inventory of the solar system at the time of solar system formation 4.567 billion years ago. This includes the sun, the planets and their moons, asteroids, comets, meteorites, and interplanetary dust. Historically, the abundances are mainly derived from meteorites and spectroscopy of the solar photosphere. Sometimes solar system abundances are referred to as "solar abundances", but technically there are important differences in how solar and solar system abundances are derived.

**Solar or solar photospheric abundances** refer to the elemental inventory of the spectroscopically accessible parts of the outer convective envelope of the present-day sun, where "present day" is in respect to the young sun 4.6 billion years ago. In addition to the photosphere, solar abundances are obtained from sun-spots, the solar corona, and the solar wind. In the photosphere the elements are mainly present as neutral or singly-ionized atoms, a few absorption bands from $C_2$, CH, CN, NH, CO, and OH are known. These molecules occur in larger quantities in sun-spots in the cooler sun-spots where MgH, CaH, TiO, BO, HS, HF, and HCl are also present. Photospheric abundances reflect the composition of the sun's outer convection zone (about 2% of the sun's total mass) and are not representative of the sun's composition 4.6 Ga ago because of diffusion and gravitational settling into the sun's interior over the sun's lifetime. Roughly 10-20% (depending on models adopted, see below) of each element heavier than hydrogen was lost from the photosphere by while hydrogen diffused up. The sun is losing mass (about $10^{-14}$ solar masses per year) through the sun's outer layers in the form of solar wind. Assuming about similar mass-loss in the past, this mass-loss from the convection zone (about 0.2% total of the current convection zone mass) is negligible for the sun. The elemental abundances in the solar wind are fractionated relative to photospheric values, and mass-loss from the sun must contribute to relative element abundance fractionations within the photosphere over time, although on a much smaller scale than fractionations stemming from element settling.

**Meteoritic abundances** are derived from the composition of meteorites. In the context of solar system abundances, meteoritic abundances typically refer to the abundances of condensable elements derived from CI-chondrites (see below and the article on meteorites in this encyclopedia). In the older literature, meteoritic abundances were often derived from mass-weighted average compositions of the silicate, metal and sulfide phases present in chondritic meteorites (for chondrite classification see Table 1 below). This approach suffered from the necessary assumption of representative phase compositions, and the values for representative mass-fractions of silicate, metal, and sulfide corresponding to the initial solar system composition.

## Abundance Scales

The astronomical abundance scale uses hydrogen, the most abundant element in the sun, as reference element. Relative abundances in the photosphere or other stars are frequently given in the epsilon notation:

$$\log_{10} \varepsilon E = \log_{10} (E/H) = \log_{10}( n(E)/n(H)) \qquad (1)$$

for the atomic ratio of an element "E" relative to hydrogen (H). Sometimes the element symbol is taken as shorthand to express the number of atoms, n(E). Another frequently used notation is

$$A(E) = 12 + \log_{10} \varepsilon E = 12 + \log_{10} (n(E)/n(H)) \quad (2)$$

which expresses abundances on a logarithmic scale relative to $n(H)=10^{12}$ atoms. The predominance of hydrogen in the sun and other stars was recognized in the 1920s, and Cecilia Payne (1925a,b) found A(H) > 11 for the sun and A(H) = 12.5 for her set of stars analyzed. In her thesis, Payne (1925a) defined the fractional $\log_{10}$ abundance to A(H) = 11. The normalization with hydrogen to A(H)=12 was apparently first done by Claas (1951) in order to avoid negative logarithmic values for elements of low abundance. Frequently the term "dex" is used in the astronomical literature to designate "decadic logarithmic units" where 1 dex stands for a factor of 10 ($=10^1$).

Atomic abundances on the cosmochemical scale are frequently normalized to the number of silicon atoms set to $N(Si) = 10^6$. Abundances on this scale are sometimes designated as N(M). These abundances can be calculated from the absolute concentrations by mass.

Goldschmidt (1937) introduced the normalization to silicon, the most abundant cation in the crust (oxygen is typically the most abundant element by number in the crust and rocks, but is difficult to measure routinely). Cations are a better choice to normalize the rock-based abundance scale. Originally, this scale was set to N(Si) = 100 (Goldschmidt 1938), then N(Si) = 10,000 (Brown 1949). With more refined analyses, smaller quantities of many elements could be measured, and it became more practical to increase the normalization. In their seminal paper, Suess & Urey (1956) changed the scale to $N(Si) = 10^6$ explaining, "We use $10^6$ in order to get values for the rarer elements which can be written without negative exponentials or awkward decimal fractions."

For comparing the astronomical and cosmochemical abundance scales, re-normalizing meteoritic values to H is inappropriate because H forms volatile compounds and is highly depleted in meteorites. It is more practical to normalize photospheric abundances (which are always relative and not absolute) to $Si=10^6$ atoms. Any other well-determined refractory elements can serve to link the meteoritic and photospheric scales, and other elements such as Ca Holweger 2001), Mg (Larimer & Anders 1967, Goles 1969), and Ti (Greenland & Lovering 1965) have been suggested over time.

The abundances on the astronomical log scale, A(E) are related to the cosmochemical scale, n(E) using the relationships:

$$A(E) = AF + \log n(E) \quad (3)$$

Or

$$n(E) = 10^{A(E)}/10^{AF} = 10^{(A(E)-AF)} \quad (4)$$

The term "AF" is often called "abundance scaling factor" and determined by comparing elements that are well determined in the sun and in meteorites.

Sometimes only one element (traditionally silicon) is used to determine the scaling factor, (e.g., Lodders 2003). On the other hand, this requires re-computation of all elemental abundances if the abundance for silicon for meteorites and/or in the photosphere is changed. For example, Lodders et al. (2009) recommended a photospheric Si abundances for which Asplund et al. 2009 found a value different by 0.02 dex, and Asplund et al. (2009) then changed all meteoritic abundances from Lodders et al. (2009) by 0.02 dex to obtain their meteoritic values on the astronomical log scale. On the other hand, a mean scaling factor derived from several elements typically makes the coupling of the scales more robust and less sensitive to changes in individual elemental abundances. This approach was introduced by Urey (1967) and Cameron (1968), who used fifteen and ten non-volatile elements, respectively. This practice has been widely followed (e.g., Anders & Grevesse 1989, Palme & Beer 1993, Lodders et al. 2009, Palme et al. 2014). Typically, the scaling factor derived from several elements has an uncertainty of up to ±0.04 dex (an absolute factor of 1.1), which should be considered when gauging how well meteoritic and solar abundances agree once placed onto the same relative abundance scale. Using a scaling factor from several elements may also lead to abundances scales that are nominally normalized to $Si=10^6$ but where the actual silicon abundance does not equal $10^6$ (e.g., in Anders & Grevesse 1989). Here the recommended values below use the scaling factor from silicon to obtain clean normalized scales. The photospheric and meteoritic abundances on the two atomic scales are summarized in Tables 4 and 6, and recommended solar system abundances on both scales are given in Table 8.

The uncertainties of the logarithmic abundance scale and linear abundance scale are compared using the relationship $U\ (\%) = \pm 100\ (10^{\pm a} - 1)$ where "a" is the uncertainty in decadic logarithmic "dex"-units quoted for photospheric abundances on the logarithmic scale, and "U" is the uncertainty of the meteoritic abundance in percent. The uncertainty in dex is an uncertainty factor, hence the percent uncertainty is smaller for –a than for +a. Vice versa, a given percent uncertainty on a linear scale yields two different uncertainty factors on the logarithmic scale. For a conservative approach, the larger percentage value U (from a given uncertainty "a") or the larger uncertainty "a" (from a given U) is taken for comparison here. Uncertainties of meteoritic as well as photospheric and stellar abundances on the logarithmic scale are frequently quoted as 1- or 2-sigma confidence intervals, however, the original papers do not always specify which confidence intervals were used for the reported uncertainties. However, both statistical and systematic errors must be considered. The statistical deviation in solar abundances from the line-to-line scatter (when several lines are used to determine elemental abundances) is typically much smaller than the uncertainties introduced by parameters of the solar atmospheric model used for the analyses; e.g., Lawler et al. 2014, 2018, Scott et al. 2015a,b; Sneden et al. 2009 and the review by Allende-Prieto (2016).

## Historical Background

About a century ago, the era of modern quantitative studies about the abundances and distribution of the chemical elements started as almost all naturally occurring elements of the periodic table had been discovered. Chemical analyses were no longer limited to major elements occurring in rocks, and quantum mechanics had opened the door to quantitative spectral analyses of the sun and stars. There was and remains a strong desire to know which elements exist and their quantities on the surface of the Earth, within the Earth and Planets, the Sun and the Cosmos. The nucleosynthetic, geochemical, and cosmochemical principles governing formation and distribution of the elements remain active research areas in astrophysics and cosmochemistry. Traditionally, the knowledge about elemental abundances comes from two disciplines – astronomy with studies about the composition of the sun and other stars, and cosmochemistry with hands-on studies of meteorites, interplanetary, and interstellar dust. Both disciplines complement each other to validate, supplement, and optimize solar system abundance data which are widely used as a reference composition in both disciplines.

Like others (Chladni 1819, Farrington 1915, Harkins 1917, Merrill 1909, 1915,1916) before him Victor Moritz Goldschmidt (1923, 1937, 1938) realized that stony meteorites seem to be the least differentiated rocky objects compared to the Earth. The Earth has differentiated into a crust, a silicate mantle, and a metallic core and has ongoing geological activity (volcanism, plate tectonics). Goldschmidt and others thought meteorites presumably have been idle on their parent asteroids since shortly after their formation 4.567 Ga ago, yet even the so-called "primitive" meteorites were exposed to space weathering and often bear witness to impact processes they experienced on their parent asteroids (see e.g., article in this encyclopedia on meteorites by K. Wang & R. Korotev).

Goldschmidt's work was quite influential as he developed the geochemical classification of the elements in a series of papers on "Geochemische Verteilungsgesetze der Elemente" (Geochemical Distribution Laws of the Elements) which describes the affinities of the elements for certain phases: Rock-loving (lithophile) elements such as Ca, Al, Mg, Si, and, to some extent, Fe usually occur in oxide-form in rocks. Other elements such as noble metals, Ni, Co, and Fe appear in metal alloys (siderophile or metal-loving elements), and other elements are preferentially associated with sulfide phases (chalcophile elements such as Zn, S, Se, and Fe). The group of atmophile elements contains the noble gases and elements in compounds present in the atmosphere (C, N, O, H) and easily volatilized under typical conditions near the surface of the Earth.

Many analyses of meteorites by Goldschmidt and coworkers, his direct competitors I. and W. Noddack (1930), and the solar abundances of H,C,N, and O from Russell (1929) provided Goldschmidt (1937) with the data needed to assemble a comprehensive table of the elemental abundances in cosmic matter. Although Goldschmidt's (1937) table was instrumental for further work on nucleosynthesis, others predated his work and laid the foundations.

In 1914, Russell made four lists of the elements in order of line intensities in the solar photosperic and coronal spectra and of elemental abundances in the Earth's crust and stony meteorites. He found qualitative resemblances in the ordering for fifteen out of the sixteen abundant metallic elements between the sun, crust, and meteorites. In 1914, the large abundances of hydrogen and helium in the sun and stars was not recognized yet, and Russell (1914) noted, that with some exceptions "the agreement of the solar and terrestrial lists is such as to confirm very strongly Rowland's opinion that, if the earth's crust should be raised to the temperature of the sun's atmosphere, it would give a very similar absorption spectrum. A moderate admixture of meteoritic material would make the similarity even closer".

In this context, it is worth adding that Beaumont in 1847 (before spectral analyses was invented) already found 16 abundant elements common to crustal rocks, ore deposits, volcanic emanations, mineral & ocean waters, meteorites, and organic matter. Thus, by 1914, Rowland and Russell extended the qualitative comparison to the sun. Plaskett (1922) and in particular Payne (1925a) made this comparison more quantitatively. In her thesis, Payne (1925a) also had recognized the large abundances of H and He (but that result was discouraged by her advisor Russell). Thus, she limited the comparison to abundant elements in common, and describes "The relative abundances, in the stellar atmosphere and the earth, of the elements that are known to occur in both, display a striking numerical parallelism." She further noted the better agreement when stellar compositions are compared to the total Earth composition instead to the Earth's crust.

By 1929, Russell had accepted the overwhelmingly large abundances of hydrogen and He and presented the first extended and more-or -less quantitative abundance determinations of 56 elements in the solar photosphere. Russell compared his solar abundance table to meteorite abundances by Merrill (1915) and to abundances in the Earth's crust (Clarke & Washington 1922); these abundances also had improved since 1914. Russell noted stronger similarities in the relative abundances of certain elements in the solar photosphere and meteorites than between photospheric abundances and those in the Earth's crust.

Subsequent abundance tables of the elements became more complete and refined through better analytical techniques for meteorites and solar spectroscopy, but also through better theoretical understandings of the underlying principles of atomic and nuclear physics and nucleosynthesis.

## Meteoritic Abundances

Many meteorites contain silicates, metal and sulfides in proportions that one might expect for the Earth and other terrestrial planets, where silicate rocks and metal have separated to form a silicate mantle and crust, and a core. As early as Chladni (1794) this observation triggered the idea that meteorites may represent "left-over" material from the birth of the solar system, and thus provide information about the elemental composition of the matter from which the solar system objects formed. Here a brief summary about chondritic meteorites follows, and the reader can find more details about meteorites, including definitions and explanations about their classifications in the Meteorite article of this encyclopedia, in Scott & Krot (2014), Lodders & Fegley (1998, 2011), in the Meteoritical Bulletin from The Meteoritical Society, on the excellent webpage about meteorites by Randy Korotev, and in the article by Wang & Korotev (this encyclopedia).

**Table 1. Major Chondrite Meteorite Classes**

| Designation | Classification Notes |
|---|---|
| *Carbonaceous Chondrites* | |
| CI | I for "Ivuna" type, old: Carbonaceous I (Wiik 1956) |
| CM | M for "Murray" type, old: Carbonaceous II (Wiik 1956)) |
| CV | V for "Vigarano" type |
| CO | O for "Ornans" type, old: Carbonaceous III, Ornanites (Wiik 1956) |
| CK | K for "Karoonda" type |
| CR | R for "Renazzo" type |
| CB | B for "Bencubbin" type |
| CH | H for "High metal content" |
| *Ordinary Chondrites* | |
| H | H for "High iron content" (Urey & Craig 1953), old, bronzite-olivine, or Cronstad-type (Prior 1916) |
| L | L for "Low total iron content" (Urey & Craig 1953), old, hyperstene-olivine or Soko-Banja type (Prior 1916) |
| LL | LL for "Low total iron & low total metal content" (Urey & Craig 1953, Keil & Fredriksson 1964), old: hyperstene-olivine or Baroti-type (Prior 1916), amphoterites (Mason & Wiik 1964) |
| K | K for Kakangari |
| R | R for Rumuruti |
| *Enstatite Chondrites* | |
| EH | EH for "Enstatite Chondrite w. High-Fe metal content", old, HH (Wiik 1956) and old type I (Anders 1964) |
| EL | EL for "Enstatite chondrite w. Low-Fe metal content", old, type II (Anders 1964) |

Rose (1864) recognized the major meteorite chondrite groups and introduced the name "chondrites" for stony meteorites containing the silicate spheres. Chondrites are chemically

"primitive" or "undifferentiated", i.e., they did not melt and part into silicates, metal and sulfide phases by density separation. Prior (1916,1920) established the basics of the classification scheme (Table 1) which was expanded upon by Urey and Craig (1953), Wiik (1956), and Mason (1962) as chemical analyses revealed characteristic differences (see Mason 1971). The broad classes, based on composition and silicate mineralogy are carbonaceous chondrites, ordinary or common chondrites, and enstatite chondrites. The carbonaceous meteorite class is subdivided depending on overall carbon and water contents and mineralogy. Wiik (1956) used types "I, II, and III" but these are now named after a type-specimen meteorite (e.g., the CI-chondrites are named after the meteorite fall near Ivuna, Tanzania). A similar name-based designation existed once for other chondrite classes but was abandoned in favor of a designation from their silicate mineralogy as enstatite chondrites, and for ordinary chondrites as bronzite and hyperstene chondrities (and later additionally amphoterites). The ordinary chondrites, based on their relative metal contents and iron concentrations in the silicates, are H, L, and LL (Urey and Craig 1953, Keil and Fredrrikson 1964), and EH and EL for enstatite chondrites. This practice has been used to name a more recently designated group of metal-rich carbonaceous chondrites as "CH", but another recent class "CB" for "Bencubbinites" was named after the Bencubbin meteorite.
The numerical sub-classification for the petrologic type by van Schmus & Wood (1967) describes successive mineral alteration that meteorites experienced on their parent asteroid. The most primitive types were designated "type 3" (for which subdivisions from 3.0 to 3.9 were introduced later). Increasing numbers from 4 to 7 describe mineralogical alteration effects from mild to more severe heating close to melting. Numbers below 3 designate mineralogical alteration caused by aqueous fluids, with type 2 having partial mineral transformation, and type 1 more or less complete mineral transformation from "dry" to hydrous minerals, and essentially oxidation of metal to magnetite. The review by Scott & Krot (2014) describes chondrites in more detail. None of the meteorite groups has escaped some processing on their parent asteroid so the chemical and mineralogical constitution of the original matter in the solar system has been altered or lost. However, the elemental composition was retained in cases where isochemical reactions took place, such as generally assumed for CI-chondrites, which were established as the preferred solar system abundance standards for non-volatile elements in the 1970s. It took some time to investigate and characterize the meteorite classes and it is no coincidence that the selection of the CI-chondrites culminated at about same time when analytical advances provided the information to solidify the chondrite meteorite classification.

Early studies derived meteoritic elemental abundances from weighted averages of elemental concentrations in meteoritic silicates, metal, and sulfide, irrespective of meteorite type, which meant occasionally counting stony meteorites as "silicate" and iron meteorites as "metal" phase. In part this choice was dictated by the way wet-chemical analyses were done, and the analyses were mainly restricted to more abundant elements. Minor elements could be analyzed only when large quantities of a meteoritic metal. Silicates, or sulfide were available. Given the scarcity of meteorites, small amounts of samples were subjected to destructive analyses, and only developments in micro-analytical techniques brought improvement in quantitative analyses to allow the use of whole-rock (bulk) analyses for deriving solar-system abundances from certain meteorite groups. Microscopy for petrological studies, and spectral absorption and emission spectroscopy, and colorimetric analyses had revolutionized chemical analyses since the 1860s, and similarly, analytical developments now can provide measurements of low atomic concentrations on micro-to nano-scale samples and in some instances even beyond. Over time, analytical tools for elemental and isotopic measurements applied to abundance studies in meteorites include X-ray diffraction, X-ray spectrography, and X-ray fluorescence (XRF), mass-spectroscopy (MS) since the early 1900s, radiochemical methods with neutron-activation analysis (NAA), isotope dilution (ID) and electron microprobe (EMP) analyses since the 1950s, secondary-ion-probe mass-spectroscopy (SIMS) since the 1970s, inductively coupled plasma mass spectroscopy (ICP-MS) since the 1990s, and multi-collector (MC-)ICPMS since the 2000s. Currently much more sensitive analytical techniques (e.g., ICP-MS) boosted detection limits towards part-per-trillion (ppt) concentrations and provide elemental and isotopic abundance determinations of meteorites and their components.

Having the analytical data leads to the question of how to derive representative abundances from them. Many early estimates of representative meteorite compositions listed in Table 2 used measurements that provided concentrations in the silicate, metal, and sulfide portions. Then for a representative weighted average, the proportions of silicate, metal, and sulfide had to be assumed. The leading thought was of an original "planet" from which all meteorites had come and this planet must have had a density much like the Earth and the other terrestrial planets. The approach was to select meteoritic silicate, metal and sulfide in proportions to match the densities of the Earth or the average terrestrial planet densities. Then the overall meteoritic composition could be calculated if the elemental compositions of each phase was known. Farrington (1915), Noddack & Noddack (1930,1935), Saslawsky (1933), Goldschmidt (1937, 1938), Brown (1949), Levin et al. (1956), Urey (1952), and Mason (1962) and others followed this procedure. More details about these approaches are in the footnote to Table 2.

The alternative was to assume chondrites with their silicates, metal, and sulfides are already a representative mixture, because their chemical similarities, especially when compared on a volatile-compound free (i.e., no carbon, water, or nitrogen bearing compounds) basis, were noted as early as Wahl (1910). According to Mason (1960) chondrites are not left-over debris from a disrupted planet, but "have always been independent and individual objects". The approach of taking grand-averages of available data for chondrites as representative was used by Merrill (1909, 1915,1916), Harkins (1917) and Saslawsky (1933). The Noddacks in the 1930s used a more pragmatic approach to obtain an average chondrite composition by analyzing a physical mixture of many different pulverized chondrites (but they still adhered to using weighted phase average to derive representative meteoritic compositions). Urey and Craig (1953) adopted chondrites to search out what needed to be done with more modern analysis tools at the time. As more meteorites were analyzed chemically and mineralogically the question arose which meteorite group is closest to the average composition of condensable elements of the solar system.

**Table 2. Early Estimates of Representative Meteoritic Compositions (ppm by mass)**

| Z | E | Average Phases [Mas62] | Average Phases [Lev56] | Average Phases [U52] | Average Phases [Bro49] | Stony Mets. [Budd46] | Average Phases [Gold37] | Stony Mets. [NN34] | Average Phases [Sas33] | Average Phases [NN30] | Stony Mets. [Har17] | Stony Mets. [Mer16] | Average Phases [Far15] | Stony Mets. [Mer09] |
|---|---|---|---|---|---|---|---|---|---|---|---|---|---|---|
| 1 | H | … | … | … | 380 | … | … | … | … | … | … | 830 | … | 220 |
| 2 | He | … | … | … | … | … | … | … | … | … | … | … | … | … |
| 3 | Li | 2 | 3.2 | 5 | 2 | 5 | 4 | 5 | … | 3 | … | … | … | … |
| 4 | Be | [1] | 0.09 | 1 | 0.6 | 10 | 1 | 10 | … | 5 | … | … | … | … |
| 5 | B | [2] | 2.6 | 1.5 | 2 | … | 1.5 | | … | … | … | … | … | … |
| 6 | C | … | … | … | 700 | 340 | 300 | 340 | 300 | 330 | 600 | 1500 | 400 | 600 |
| 7 | N | … | 1 | … | 0.5 | … | … | … | … | … | … | … | … | … |
| 8 | O | 330000 | 346000 | … | 246100 | 420400 | 323000 | 420400 | 318100 | 230400 | 360200 | 367000 | 101000 | 388400 |
| 9 | F | [30] | 40 | 40 | 20 | … | 28 | … | … | … | … | … | … | … |
| 10 | Ne | … | … | … | … | … | … | … | … | … | … | … | … | … |
| 11 | Na | 6800 | 7000 | 7500 | 4700 | 7180 | 5950 | 7180 | 5400 | 3950 | 5900 | 1560 | 1700 | 6000 |
| 12 | Mg | 138600 | 139000 | 135500 | 95000 | 159000 | 123000 | 159000 | 120400 | 87800 | 134300 | 136700 | 38000 | 135200 |
| 13 | Al | 11000 | 14000 | 14300 | 10000 | 16100 | 13800 | 16100 | 12200 | 8800 | 13900 | 15200 | 3900 | 15870 |
| 14 | Si | 169500 | 178000 | 179600 | 123000 | 214300 | 163000 | 214300 | 163700 | 12070 | 184100 | 180800 | 52000 | 182900 |
| 15 | P | 1300 | 16000 | 1500 | 1800 | 506 | 1050 | 506 | 800 | 988 | 600 | 1140 | 1400 | 520 |
| 16 | S | 20700 | 20000 | 20100 | 10800 | 20100 | 21200 | 20100 | 33400 | 18650 | 19800 | 18000 | 4900 | 19800 |
| 17 | Cl | [100] | 800 | 470 | 500 | 904 | 1000-1500 | 904 | … | 500 | … | 800 | … | … |
| 18 | Ar | … | … | … | … | … | … | … | … | … | … | … | … | … |
| 19 | K | 1000 | 900 | 900 | 1200 | 2630 | 1540 | 2630 | 2000 | 1430 | 1700 | 1740 | 400 | 1660 |
| 20 | Ca | 13900 | 16000 | 14300 | 12000 | 19200 | 13300 | 19200 | 14600 | 11000 | 16500 | 17300 | 4600 | 16500 |
| 21 | Sc | 9 | 5 | 5 | 3.5 | 110 | 4 | 5.8 | … | 60 | … | … | … | … |
| 22 | Ti | 800 | 700 | 580 | 600 | 2100 | 1320 | 2100 | … | 1150 | 100 | 1080 | … | 120 |
| 23 | V | 65 | 80 | 50 | 56 | 300 | 39 | 300 | … | 165 | … | … | … | … |
| 24 | Cr | 3000 | 2500 | 2700 | 2200 | 5000 | 3430 | 5000 | 3900 | 2900 | 2800 | 3220 | 900 | 2800 |
| 25 | Mn | 2000 | 2000 | 2400 | 1900 | 2050 | 2080 | 2050 | 1600 | 1260 | 1400 | 2250 | 300 | 1400 |
| 26 | Fe | 286000 | 256000 | 241000 | 457000 | 127600 | 288000 | 127600 | 299500 | 471830 | 243200 | 233100 | 720600 | 213900 |
| 27 | Co | 1000 | 900 | 1100 | 2600 | 181 | 1200 | 181 | 1300 | 2250 | 500 | 1300 | 4400 | 500 |
| 28 | Ni | 16800 | 14000 | 14500 | 35100 | 2010 | 15680 | 20100 | 19000 | 35600 | 13100 | 16300 | 65000 | 13600 |
| 29 | Cu | 100 | [40] | 170 | 130 | 1.55 | 170 | 1.55 | … | 346 | 100 | 140 | … | … |
| 30 | Zn | 50 | 20 | 76 | 480 | 3.4 | 138 | 3.4 | … | 126 | … | … | … | … |
| 31 | Ga | 5 | 8 | 4.6 | 20 | 5 | 4.2 | … | … | 5 | … | … | … | … |
| 32 | Ge | 10 | [40] | 53 | 82 | 33 | 79 | 33 | … | 170 | … | … | … | … |
| 33 | As | 3 | 70 | 18 | 160 | 20 | … | 20 | … | 200 | … | … | … | … |
| 34 | Se | 9 | 9 | 6.7 | 9 | … | 7 | … | … | 46 | … | … | … | … |
| 35 | Br | [10] | 22 | 23 | 1.5 | 1 | 20 | 1 | … | 0.55 | … | … | … | … |
| 36 | Kr | … | … | … | … | … | … | … | … | … | … | … | … | … |
| 37 | Rb | 3 | 8 | 8 | 2.7 | 4.5 | 3.5 | 4.5 | … | 2.5 | … | … | … | … |
| 38 | Sr | 11 | 22 | 23 | 16 | 72 | 20 | 72 | … | 40 | … | … | … | … |
| 39 | Y | 4 | 5 | 5.5 | 4 | 34 | 4.72 | 6.13 | … | 8.2 | … | … | … | … |
| 40 | Zr | 33 | 90 | 80 | 60 | 100 | 73 | 100 | … | 55 | … | … | … | … |
| 41 | Nb | 0.5 | 0.5 | 0.41 | 0.38 | 2 | … | 2 | … | 1.5 | … | … | … | … |
| 42 | Mo | 1.6 | 5 | 3.6 | 8.1 | 2.5 | 5.3 | 2.5 | … | 33 | … | … | … | … |
| 43 | Tc | … | … | … | … | … | … | … | … | | … | … | … | … |
| 44 | Ru | … | … | … | 4.2 | … | … | … | … | 9 | … | … | … | … |
| 45 | Rh | 0.2 | 0.6 | 0.47 | 1.6 | … | 0.8 | … | … | 1.9 | … | … | … | … |
| 46 | Pd | [1] | 0.5 | 0.92 | 1.5 | … | 1.53 | … | … | 7.1 | … | … | … | … |
| 47 | Ag | 0.1 | 0.5 | 1.35 | 1.3 | … | 2 | … | … | 2 | … | … | … | … |
| 48 | Cd | [0.5] | 2 | 1.6 | 1.3 | … | … | … | … | 4.6 | … | … | … | … |
| 49 | In | 0.001 | 0.2 | 0.2 | 0.5 | … | 0.15 | … | … | 0.044 | … | … | … | … |
| 50 | Sn | 1 | 20 | 14 | 33 | 4 | 20 | 4 | … | 130 | … | … | … | … |
| 51 | Sb | 0.1 | 0.4 | 0.64 | 0.9 | 0.1 | … | 0.1 | … | 1.3 | … | … | … | … |
| 52 | Te | 1 | 0.14 | 0.13 | … | … | [0.1] | … | … | 0.93 | … | … | … | … |
| 53 | I | 0.04 | 1 | 1.25 | 1 | … | 1 | … | … | … | … | … | … | … |
| 54 | Xe | … | … | … | … | … | … | … | … | … | … | … | … | … |
| 55 | Cs | 0.1 | 0.08 | 1.1 | 0.06 | 0.1 | 0.08 | 0.1 | … | 0.055 | … | … | … | … |
| 56 | Ba | 3.4 | 7 | 3.9 | 5.4 | 20 | 6.9 | 20 | … | 11 | … | … | … | … |
| 57 | La | 0.33 | 0.2 | 1.9 | 1.3 | … | 1.58 | 2.05 | … | … | … | … | … | … |
| 58 | Ce | 0.51 | 2 | 2.1 | 1.5 | 4 | 1.77? | 2.3 | … | 2.2 | … | … | … | … |
| 59 | Pr | 0.12 | 0.8 | 0.88 | 0.6 | … | 0.75 | 0.97 | … | … | … | … | … | … |
| 60 | Nd | 0.63 | 3 | 3 | 2.2 | 3 | 2.59 | 3.37 | … | 1.65 | … | … | … | … |
| 61 | Pm | … | … | … | … | … | … | … | … | … | … | … | … | … |
| 62 | Sm | 0.22 | 1 | 1.1 | 0.78 | 3 | 0.95 | 1.23 | … | 1.65 | … | … | … | … |
| 63 | Eu | 0.083 | 0.3 | 0.27 | 0.2 | … | 0.25 | 0.3 | … | … | … | … | … | … |
| 64 | Gd | 0.34 | 1.6 | 1.7 | 1.2 | … | 1.42 | 1.84 | … | … | … | … | … | … |
| 65 | Tb | 0.37 | 2 | 0.52 | 0.38 | … | 0.45 | 0.58 | … | … | … | … | … | … |
| 66 | Dy | 0.37 | 2 | 2.1 | 1.5 | … | 1.8 | 2.34 | … | … | … | … | … | … |
| 67 | Ho | 0.075 | 0.6 | 0.6 | 0.43 | … | 0.51 | 0.66 | … | … | … | … | … | … |
| 68 | Er | 0.19 | 1.6 | 1.7 | 1.3 | … | 1.48 | 1.93 | … | … | … | … | … | … |
| 69 | Tm | 0.038 | 0.3 | 0.31 | 0.22 | … | 0.26 | 0.34 | … | … | … | … | … | … |
| 70 | Yb | 0.19 | 1.6 | 1.7 | 1.2 | … | 1.42 | 1.85 | … | … | … | … | … | … |
| 71 | Lu | 0.036 | 0.5 | 0.54 | 0.39 | … | 0.46 | 0.6 | … | … | … | … | … | … |
| 72 | Hf | 1.4 | 0.8 | 1.6 | 0.6 | 1 | 1.6 | 1 | … | 0.55 | … | … | … | … |
| 73 | Ta | 0.02 | 0.3 | 0.28 | 0.25 | 0.7 | … | 0.7 | … | 0.0077 | … | … | … | … |
| 74 | W | 0.14 | 17 | 16 | 14 | 18 | 15 | 18 | … | 12.6 | … | … | … | … |
| 75 | Re | 0.08 | 0.0018 | 0.08 | 0.34 | 0.0008 | 0.002 | 0.0008 | … | 0.0036 | … | … | … | … |
| 76 | Os | 0.5 | 1.1 | 1.2 | 3 | … | 1.92 | … | … | 3.84 | … | … | … | … |
| 77 | Ir | 0.5 | 0.6 | 0.38 | 1.2 | … | 0.65 | … | … | 0.88 | … | … | … | … |
| 78 | Pt | [5] | 3 | 1.9 | 7.6 | 0.083 | 3.25 | 0.083 | … | 6.6 | … | … | … | … |
| 79 | Au | 0.3 | 0.26 | 0.25 | 0.72 | … | 0.65 | … | … | 0.55 | … | … | … | … |
| 80 | Hg | [0.1] | [0.009] | <0.01 | … | … | … | … | … | … | … | … | … | … |
| 81 | Tl | [0.0004] | [0.14] | 0.15 | … | … | … | … | … | 0.016 | … | … | … | … |
| 82 | Pb | 0.15 | [2] | … | 25 | 5 | 11 | … | … | 61 | … | … | … | … |
| 83 | Bi | [0.003] | [0.16] | 0.02 | 0.2 | … | … | … | … | 0.3 | … | … | … | … |
| 90 | Th | 0.04 | 0.2 | … | 1.2 | 2 | 0.8 | … | … | 1.1 | … | … | … | … |
| 92 | U | 0.014 | 0.05 | … | 2.8 | 1 | 0.36 | … | … | … | … | … | … | … |
| | Sum | 100.27 | 101.43 | [63.7] | 100.68 | 100.04 | 97.85 | 100.00 | 100.00 | 100.00 | 99.88 | 100.04 | 99.95 | 100.00 |

**Footnote for Table 2**: [Mas62] Mason 1962, mass-ratios assumed were silicate,metal,sulfide = 74.7,19.6,5.7, volatile-free. [Lev56] Levin et al. 1956, silicate,metal = 6,1. volatile-free. [U52] Urey 1952, volatile-free chondrite average, for a few elements weighted average of meteoritic silicate,metal,sulfide 85,9.5,5.5, following Goldschmidt, who did not use these observed proportions from Prior 1916. A previous, incorrect version of Urey 1952 appeared in the first edition of Urey's book the Planets" where he used silicate,metal,sulfide 85,9.5,5.5, and Brown's 1949 "silicates" (= stony meteorites) and "metal (=iron meteorites). Urey had taken the sulfide composition listed by Brown 1949 in addition to the stony and iron meteorites but did not realized that the sulfide portion was already included with the metal portion there. Thus the sulfide component was about twice too high. He corrected his mistake in Urey (1952) and corrected the table in the 2nd edition of his book. [Bro49] Brown 1949, weighted average of stony- and iron meteorites to match mantle and core mass of Earth, silicate,metal, sulfide = 100,67,0 (67,33,0), sulfide automatically determined from sulfides native to stony and iron meteorites. [Budd46] Buddhue 1946, aerolites (excluding iron meteorites). [Gold37] Goldschmidt 1937, weighted average of meteoritic silicate,metal,sulfide = 100,20,10 (equivalent to 77,153,1.1). [NN34] Noddack & Noddack 1934,1935, average stony meteorites. [Sas33] Saslawsky, 1933, weighted average of meteoritic silicate,metal,sulfide 100,25,7 to match meteorite mean density of 3.34 g/cm$^3$. [NN30] Noddack & Noddack 1930, weighted average of meteoritic silicate,metal,sulfide = 100,68,9.8 = 82.5,12,5.5 to match average density of terrestrial planets taken as 5.10 g/cm$^3$, but see Saslawasky (1933) who noted errors in their mass-balance computation and density assumptions. [Har17] Harkins 1917, average of 125 chondrite analyses. [Mer16] Merrill 1916, average of 53 chondrite analyses. [Far15] Farrington 1915, weighted mixture of stony and iron meteorites 100,257 to match the Earth's density taken as 5.57 g/cm$^3$. [Mer09] Merrill 1909, average of stony meteorites.

The CI-chondrites are the most volatile element-rich among the known chondritic meteorite groups (Mason, 1960, 1971). The arguments for using CI-chondrites as proxies for the composition of the primordial solar-system matter are their good correspondence with elemental abundances, including volatile elements (but not highly volatile elements such as H, C,N,O, noble gases), measured in the solar photosphere, and their continuity in elemental and isotopic abundance trends with mass number. Figure 1 displays the close correspondence of elemental abundances in the sun and in CI-chondrites.

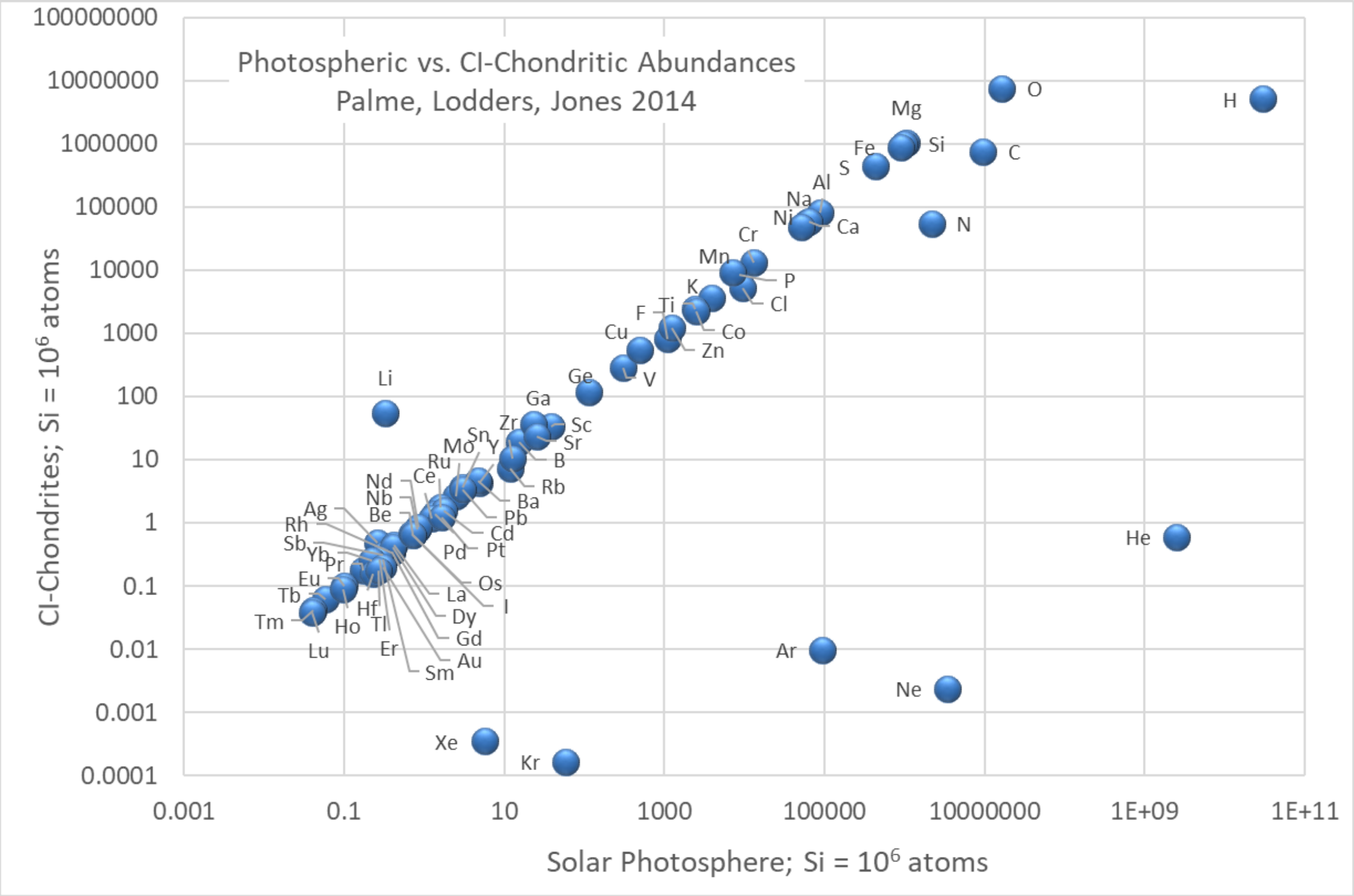

**Figure 1.** Atomic abundances of the elements in CI-chondrites versus abundances in the Sun (mainly photosphere). Both abundance sets are normalized to 10$^6$ silicon atoms. Perfect agreement would be along a 1:1 line and most elements plot along such a line within 10-20%. Despite some small scatter around the perfect correlation line, the correspondence over 13-orders of magnitude is impressive. Notable exceptions are Lithium, which is normal in meteorites but lost from the photosphere by diffusion into the sun and destroyed there; the other exceptions are elements (the noble gases, H, C, N, and O) that form highly volatile gases not retained in meteorites. Data from Palme et al. (2014).

Other chondrite groups have elemental fractionations relative to CI-chondrites and are not as good a match for the solar photospheric abundances. Carbonaceous chondrites other than CI-chondrites show clear volatility-related fractionations for some elements (e.g., alkali elements, halogens, In, Tl, Bi) that correlate with condensation temperatures computed for a gas of overall solar composition (see, e.g., Palme et al. 2014). The ordinary and enstatite chondrites also show volatility related elemental fractionations and these chondrites further accreted metal and silicate phases in non-solar proportions as seen from their Si/Fe or Mg/Fe concentration ratios (Si and Mg are typically in silicates, and larger amounts of Fe are typically in the form of metal and sulfide in chondrites). Therefore, these chondrites may not provide good matches to the sun. However, more recent photospheric abundances measurements using more detailed 3D-atmospheric models (see Asplund et al. 2009) gave different results for some diagnostic element ratios such as Si/Fe or Mg/Si and then the correlation in Figure 1 would show more scatter. However, the basic feature remains – a correlation over 13-orders of magnitude for all elements except the most volatile ones and Lithium. In addition, there are uncertainties in the photospheric abundances (typically 7% (1-sigma) or higher) and the CI-chondrite abundances (typically 3-5% (1-sigma) or higher) which must be considered for the comparison. As described below, measurements of the solar wind from the Genesis mission can provide high-level accuracy measurements for some major elements and initial results seem to indicate better agreement with CI-chondrites than with the spectroscopic 3D photospheric results (e. g. Burnett et al. 2017). The use of CI-chondrites as abundance standard started during the 1950s but was questioned in the interim and ample literature about this issue exists. The earlier analysts of CI-chondrites – Berzelius in 1835 investigating the Alais CI-chondrite (fall in 1806) and Daubree (1864) the CI-chondrite Orgueil (fall in 1864) - already suspected these rare carbonaceous chondrites could be close to "primeval matter" mainly because of their relative high contents of organic compounds. Urey (1952) concluded the carbonaceous chondrites should be the best representatives of non-volatile solar system material because they contain abundant water, S, C, alkalies and other volatiles, but he (Urey & Edwards 1956) abandoned this idea because the carbonaceous chondrites showed more variation in volatile element contents from his modern analyses of alkali elements than Urey expected from assumed uniform primordial condensations. Then Urey favored ordinary chondrites for a brief time. However, as Ringwood (1961) remarks, Urey used compositional averages of all different types of carbonaceous chondrites instead of only CI-chondrites to conclude this. Ordinary chondrites can have higher alkali element contents than some types of carbonaceous chondrites, but this is not true for larger samples of CI-chondrites.

The use of CI-chondrites became firmly established after measurements by Reed et al. (1960) for the volatile elements Bi, Pb, and Tl in CI-chondrites provided a much better match to the theoretical solar system abundances estimated by Suess and Urey (1956) and Cameron (1959) from nuclear systematics. In his subsequent abundance tables Cameron (1968, 1973) made extensive use of CI-chondrite data, and set the use of CI-chondrites into motion.

It took into the 1970s to completely settle this issue because there was a severe disagreement between the CI-chondritic (and other chondrites) and the photospheric iron abundance (e.g., Withbroe 1971, Anders 1971). Abundant elements such as Mg and Si agreed reasonably well, but the solar iron value appeared to be a factor of ten times lower than in meteorites. Further, the photospheric and coronal Fe abundances were discordant, and the discovery of the forbidden transition lines of singly ionized Fe (written as [Fe II]) in the photosphere hinted at an iron abundance an order of magnitude larger than previously thought (e.g., Grevesse & Swings 1969, Nussbaumer & Swings 1970). Eventually the atomic parameters (the gf-values) for iron measured in the laboratory were found to be off by a factor of ten because the temperature calibration of the furnace used in the measurements was incorrect. New laboratory oscillator strengths for permitted neutral (Fe I) and ionized (Fe II) iron lines were measured and the Fe abundance was redetermined (Garz & Koch 1969, Garz et al. 1969ab and Baschek et al. 1970). The abundances of Fe from the Fe I and Fe II lines then agreed far better with values from the forbidden [Fe II] lines, the corona, and meteorites. This solidified the use of CI-chondritic compositions as the solar system standard of condensable elemental abundances.

There are five known CI-chondrite falls (Alais, Ivuna, Orgueil, Tonk, Revelstoke) and all are of petrologic type 1.(For a while the Tagish Lake meteorite fall was thought to be a CI2 chondrite, but this was incorrect). As noted above, the petrological type 1 designates completely aqueously altered meteorites, and the mineralogy of CI-chondrites is not that of pristine condensates from the solar nebula. Their mineralogy was changed on their parent asteroid after condensates accreted, but, importantly, their overall *elemental* composition was not changed. Thus, the mineral alteration occurred in a closed system without element additions or losses (isochemical alteration).

## Abundances of the Elements in CI-Chondrites

The elemental composition by mass of CI-chondrites from various sources with somewhat different approaches is given in Table 3. The Orgueil CI-chondrite fell 1864 in France, and about 12 kg are left of Orgueil in meteorite collections. This is more than 90% of all mass preserved from all five known CI-chondrite falls. The number of analyses of the Orgueil dominates the abundance statistics, and several authors mainly use only Orgueil data (indicated by "Orgueil" in Table 3). Others use straight or weighted compositions of all five CI-chondrites to calculate the group meteorite composition (e.g., Wasson & Kallemeyn 1988, Lodders 2003, this review). For some elements, some authors prefer to derive concentrations of one element from element concentration ratios measured in other meteorite groups. This approach exploits the chemical similarities (leading to often constant concentration ratios) of elements during fractionation processes in meteorites. Then abundances can be computed from well-determined ratios in other chondrites if an element in the ratio is hard to measure in CI-chondrites (see Anders & Ebihara 1982, Anders & Grevesse 1989, Lodders et al. 2009, Palme et al. 2014). The "ratio-approach" was often simply a necessity because good analytical data for CI-chondrites were sparse (only five meteorites of which only two to three are available in sufficient quantity), and for several elements, the number of reliable analyses remains precariously sparse (see Lodders 2003).

An often-unappreciated complication for analyses of CI-chondrites is their high water content (around 15-20%) in the form of crystal water (in epsomite) and in hydrated silicates. Sometimes analytical results are reported for dried samples so reported absolute element concentrations will be larger than in the native meteorite. In other instances, analytical results are reported for de-volatilized samples, which were water-free and carbon and nitrogen free (carbonates, N-bearing organics, free carbon, and ammonia were driven off by heating, and in some instances, all or some portions of sulfur were also removed). Therefore, individual analyses must be checked for sample processing before they can be included into evaluations of mean CI-chondrite compositions. Older average meteorite compositions listed in Table 2 sometimes refer to "volatile-free" compositions, which is indicated in the table footnote.

**Table 3. Elemental Compositions by Mass (ppm) of CI-Chondrites**.

| Z | E | This Review | 2σ | Orgueil [PLJ14] | Orgueil [LPG09] | Orgueil [PJ03] | Orgueil [L03] | all CI [L03] | Orgueil [MS95] | Orgueil [PB93] | Orgueil [AG89] | all CI [WK88] | Orgueil [P88] | all CI [W85] | All CI [AE82] | Orgueil [PSZ81] | all CI [Yav68] |
|---|---|---|---|---|---|---|---|---|---|---|---|---|---|---|---|---|---|
| 1 | H | 18600 | 3440 | 19700 | 19700 | 20200 | 19700 | 21015 | … | 20200 | 20200 | 20000 | … | 20000 | 20200 | … | … |
| 2 | He | 0.00917 | … | 0.00917 | 0.00917 | 0.00917 | 0.00917E | 0.00917 | … | … | 0.00917 | … | … | 0.018 | 0.00917 | … | … |
| 3 | Li | 1.51 | 0.12 | 1.45 | 1.47 | 1.49 | 1.45 | 1.46 | 1.5 | 1.49 | 1.5 | 1.57 | 1.45 | 1.57 | 1.59 | 1.45 | 1.3 |
| 4 | Be | 0.0220 | 0.0016 | 0.0219 | 0.0210 | 0.0249 | 0.019 | 0.0252 | 0.025 | 0.0249 | 0.0249 | 0.027 | 0.025 | 0.027 | 0.027 | 0.025 | … |
| 5 | B | 0.744 | 0.172 | 0.775 | 0.775 | 0.69 | 0.775 | 0.710 | 0.9 | 0.87 | 0.87 | 1.2 | 0.27 | 1 | 1.25 | 0.27 | … |
| 6 | C | 41300 | 8400 | 34800 | 34800 | 32200 | 34800 | 35180 | 35000 | 32200 | 34500 | 32000 | 35000 | 32000 | 34500 | 35000 | … |
| 7 | N | 2500 | 660 | 2950 | 2950 | 3180 | 2950 | 2940 | 3180 | 3180 | 3180 | 1500 | 3180 | 1500 | 3180 |  | … |
| 8 | O | 453840 | 20000 | 459000 | 458500 | 465000 | 462300 | 458190 | … | 465000 | 464000 | 460000 | 470000 | 460000 | 464000 | 470000 | … |
| 9 | F | 92 | 40 | 58.2 | 58.2 | 58.2 | 58.2 | 60.6 | 60 | 58.2 | 60.7 | 64 | 54 | 64 | 58.2 | 54 | … |
| 10 | Ne | 1.80E-4 | … | 1.80E-4 | 1.80E-4 | 1.68E-04 | 1.80E-4 | 1.80E-4 | … | … | 1.68E-4 | … | … | 3.0E-04 | 1.68E-4 | … | … |
| 11 | Na | 5100 | 500 | 4962 | 4990 | 4982 | 4990 | 5010 | 5100 | 4982 | 5000 | 4900 | 5020 | 4800 | 4830 | 5020 | 5500 |
| 12 | Mg | 95170 | 4000 | 95400 | 95800 | 96100 | 96000 | 95870 | 96500 | 96100 | 98900 | 97000 | 94000 | 97000 | 95500 | 93600 | 96000 |
| 13 | Al | 8370 | 600 | 8400 | 8500 | 8490 | 8580 | 8500 | 8600 | 8650 | 8680 | 8600 | 8200 | 8600 | 8620 | 8200 | 8600 |
| 14 | Si | 107740 | 7200 | 107000 | 107000 | 106800 | 105700 | 106500 | 106500 | 106800 | 106400 | 105000 | 107000 | 105000 | 106700 | 106800 | 106000 |
| 15 | P | 978 | 120 | 985 | 967 | 926 | 924 | 920 | 1080 | 1105 | 1220 | 1020 | 1010 | 1020 | 1180 | 1010 | … |
| 16 | S | 53600 | 4400 | 53500 | 54100 | 54100 | 53500 | 54100 | 54000 | 52500 | 62500 | 59000 | 58000 | 59000 | 52500 | 58000 | 61300 |
| 17 | Cl | 717 | 220 | 698 | 698 | 698 | 700 | 704 | 680 | 698 | 704 | 680 | 678 | 680 | 698 | 678 | … |
| 18 | Ar | 0.00133 | … | 0.00133 | 0.00133 | 0.00123 | 0.00133 | 0.00133 | … | … | 0.00123 | … | … | 0.0013 | 0.00123 | … | … |
| 19 | K | 539 | 48 | 546 | 544 | 544 | 543 | 531 | 550 | 544 | 558 | 560 | 517 | 560 | 569 | 517 | 560 |
| 20 | Ca | 8840 | 700 | 9110 | 9220 | 9320 | 9080 | 9070 | 9250 | 9510 | 9280 | 9200 | 9000 | 9200 | 9020 | 9000 | 11100 |
| 21 | Sc | 5.83 | 0.40 | 5.81 | 5.90 | 5.9 | 5.84 | 5.83 | 5.92 | 5.9 | 5.82 | 5.8 | 5.9 | 5.8 | 5.76 | 5.9 | 5.6 |
| 22 | Ti | 450 | 30 | 447 | 451 | 458 | 439 | 440 | 440 | 441 | 436 | 420 | 440 | 420 | 436 | 440 | 420 |
| 23 | V | 53.6 | 4.0 | 54.6 | 54.3 | 54.3 | 56.0 | 55.7 | 56 | 54.3 | 56.5 | 55 | 55.6 | 56 | 56.7 | 55.6 | … |
| 24 | Cr | 2610 | 200 | 2623 | 2650 | 2646 | 2630 | 2590 | 2650 | 2646 | 2660 | 2650 | 2670 | 2650 | 2650 | 2670 | 2400 |
| 25 | Mn | 1896 | 160 | 1916 | 1930 | 1933 | 1920 | 1910 | 1920 | 1933 | 1990 | 1900 | 1820 | 1900 | 1960 | 1820 | 1600 |
| 26 | Fe | 185620 | 13000 | 186600 | 185000 | 184300 | 183500 | 182800 | 181000 | 182300 | 190400 | 182000 | 183000 | 182000 | 185100 | 183000 | … |
| 27 | Co | 508 | 30 | 513 | 506 | 506 | 507 | 502 | 500 | 506 | 502 | 508 | 501 | 508 | 509 | 501 | … |
| 28 | Ni | 10950 | 700 | 10910 | 10800 | 10770 | 10670 | 10640 | 10500 | 10770 | 11000 | 10700 | 10800 | 10700 | 11000 | 10800 | … |
| 29 | Cu | 130 | 20 | 133 | 131 | 131 | 131 | 127 | 120 | 131 | 126 | 121 | 108 | 121 | 112 | 108 | … |
| 30 | Zn | 311 | 20 | 309 | 323 | 323 | 318 | 310 | 310 | 323 | 312 | 312 | 347 | 312 | 308 | 347 | … |
| 31 | Ga | 9.45 | 0.70 | 9.62 | 9.71 | 9.71 | 9.57 | 9.51 | 9.2 | 9.71 | 10 | 9.8 | 9.1 | 9.8 | 10.1 | 9.1 | … |
| 32 | Ge | 33.4 | 3.0 | 32.6 | 32.6 | 32.6 | 33.1 | 33.2 | 31 | 32.6 | 32.7 | 33 | 31.3 | 33 | 32.2 | 31.3 | … |
| 33 | As | 1.77 | 0.16 | 1.74 | 1.74 | 1.81 | 1.70 | 1.73 | 1.85 | 1.81 | 1.86 | 1.84 | 1.85 | 1.84 | 1.91 | 1.85 | … |
| 34 | Se | 20.4 | 1.6 | 20.3 | 20.7 | 21.4 | 19.5 | 19.7 | 21 | 21.3 | 18.6 | 19.6 | 18.9 | 19.6 | 18.2 | 19.9 | … |
| 35 | Br | 3.77 | 1.80 | 3.26 | 3.26 | 3.5 | 3.18 | 3.43 | 3.57 | 3.5 | 3.57 | 3.6 | 3.56 | 3.6 | 3.56 | 2.53 | … |
| 36 | Kr | 5.22E-5 | … | 5.22E-5 | 5.22E-5 | 2.98E-5 | 5.22E-5 | 5.22E-5 | … | … | 2.98E-5 | … | … | 3.3E-5 | 2.98E-5 | … | … |
| 37 | Rb | 2.22 | 0.18 | 2.32 | 2.31 | 2.32 | 2.12 | 2.13 | 2.3 | 2.32 | 2.3 | 2.22 | 2.06 | 2.22 | 2.30 | 2.06 | 2.3 |
| 38 | Sr | 7.79 | 0.50 | 7.79 | 7.81 | 7.26 | 7.67 | 7.74 | 7.25 | 7.26 | 7.8 | 7.9 | 8.6 | 7.9 | 7.91 | 8.6 | <1 |
| 39 | Y | 1.50 | 0.10 | 1.46 | 1.53 | 1.56 | 1.6 | 1.53 | 1.57 | 1.57 | 1.56 | 1.44 | 1.57 | 1.44 | 1.50 | 1.44 | 1.6 |
| 40 | Zr | 3.79 | 0.28 | 3.63 | 3.62 | 3.86 | 4.02 | 3.96 | 3.82 | 3.87 | 3.94 | 3.8 | 3.87 | 3.8 | 3.69 | 3.82 | 11 |
| 41 | Nb | 0.279 | 0.015 | 0.283 | 0.279 | 0.247 | 0.259 | 0.265 | 0.24 | 0.246 | 0.246 | 0.27 | 0.246 | 0.27 | 0.250 | 0.3 | … |
| 42 | Mo | 0.976 | 0.050 | 0.961 | 0.973 | 0.928 | 0.929 | 1.02 | 0.9 | 0.928 | 0.928 | 0.92 | 0.92 | 0.92 | 0.92 | 0.92 | … |
| 43 | Tc | … | … | … | … | … | … | … | … | … | … | … | … | … | | … | … |
| 44 | Ru | 0.666 | 0.04 | 0.690 | 0.686 | 0.683 | 0.725 | 0.692 | 0.71 | 0.714 | 0.712 | 0.71 | 0.69 | 0.71 | 0.714 | 0.69 | … |
| 45 | Rh | 0.133 | 0.008 | 0.132 | 0.139 | 0.14 | 0.139 | 0.141 | 0.13 | 0.134 | 0.134 | 0.134 | 0.13 | 0.134 | 0.134 | 0.13 | … |
| 46 | Pd | 0.558 | 0.030 | 0.560 | 0.558 | 0.556 | 0.600 | 0.588 | 0.55 | 0.556 | 0.56 | 0.56 | 0.53 | 0.56 | 0.557 | 0.53 | … |
| 47 | Ag | 0.204 | 0.008 | 0.201 | 0.201 | 0.197 | 0.203 | 0.201 | 0.2 | 0.197 | 0.199 | 0.208 | 0.21 | 0.208 | 0.220 | 0.21 | … |
| 48 | Cd | 0.679 | 0.024 | 0.674 | 0.674 | 0.68 | 0.675 | 0.675 | 0.71 | 0.68 | 0.686 | 0.65 | 0.77 | 0.65 | 0.673 | 0.77 | … |
| 49 | In | 0.0786 | 0.0040 | 0.0778 | 0.0778 | 0.078 | 0.0779 | 0.0788 | 0.08 | 0.0778 | 0.08 | 0.08 | 0.08 | 0.08 | 0.0778 | 0.08 | … |
| 50 | Sn | 1.63 | 0.16 | 1.63 | 1.63 | 1.68 | 1.66 | 1.68 | 1.65 | 1.68 | 1.72 | 1.72 | 1.75 | 1.72 | 1.68 | 1.75 | … |
| 51 | Sb | 0.169 | 0.018 | 0.145 | 0.149 | 0.133 | 0.148 | 0.152 | 0.14 | 0.133 | 0.142 | 0.153 | 0.13 | 0.153 | 0.155 | 0.13 | … |
| 52 | Te | 2.31 | 0.18 | 2.28 | 2.28 | 2.27 | 2.26 | 2.33 | 2.33 | 2.27 | 2.32 | 2.4 | 2.34 | 2.4 | 2.28 | 2.34 | … |
| 53 | I | 0.77 | 0.62 | 0.53 | 0.53 | 0.43 | 0.48 | 0.480 | 0.45 | 0.433 | 0.433 | 0.5 | 0.56 | 0.5 | 0.43 | 0.56 | … |
| 54 | Xe | 1.74E-4 | … | 1.74E-4 | 1.74E-4 | 4.62E-05 | 1.74E-4 | 1.74E-4 | … | … | 4.62E-5 | … | … | 3.2E-5 | 4.62E-5 | … | … |
| 55 | Cs | 0.188 | 0.012 | 0.188 | 0.188 | 0.188 | 0.184 | 0.185 | 0.19 | 0.188 | 0.187 | 0.183 | 0.19 | 0.183 | 0.186 | 0.19 | 0.18 |
| 56 | Ba | 2.39 | 0.16 | 2.42 | 2.41 | 2.41 | 2.30 | 2.31 | 2.41 | 2.41 | 2.34 | 2.3 | 2.6 | 2.3 | 2.27 | 2.2 | 2.4 |
| 57 | La | 0.244 | 0.016 | 0.241 | 0.242 | 0.245 | 0.238 | 0.232 | 0.237 | 0.245 | 0.2347 | 0.236 | 0.245 | 0.236 | 0.236 | 0.245 | 0.19 |
| 58 | Ce | 0.627 | 0.052 | 0.619 | 0.622 | 0.638 | 0.637 | 0.621 | 0.613 | 0.638 | 0.6032 | 0.616 | 0.638 | 0.616 | 0.619 | 0.638 | 0.62 |
| 59 | Pr | 0.0951 | 0.0066 | 0.0939 | 0.0946 | 0.0964 | 0.0925 | 0.0928 | 0.0928 | 0.0964 | 0.0891 | 0.0929 | 0.096 | 0.0929 | 0.090 | 0.096 | 0.09 |
| 60 | Nd | 0.472 | 0.036 | 0.474 | 0.471 | 0.474 | 0.459 | 0.457 | 0.457 | 0.474 | 0.4524 | 0.457 | 0.474 | 0.457 | 0.462 | 0.474 | 0.42 |
| 61 | Pm | … | … | … | … | … | … | … | … | … | … | … | … | … | ... | … | … |
| 62 | Sm | 0.153 | 0.012 | 0.154 | 0.152 | 0.154 | 0.145 | 0.145 | 0.148 | 0.154 | 0.1471 | 0.149 | 0.154 | 0.149 | 0.142 | 0.154 | 0.13 |
| 63 | Eu | 0.0577 | 0.0050 | 0.0588 | 0.0578 | 0.058 | 0.0548 | 0.0546 | 0.0563 | 0.058 | 0.056 | 0.056 | 0.058 | 0.056 | 0.0543 | 0.058 | 0.053 |
| 64 | Gd | 0.208 | 0.018 | 0.207 | 0.205 | 0.204 | 0.201 | 0.198 | 0.199 | 0.204 | 0.1966 | 0.197 | 0.204 | 0.197 | 0.196 | 0.204 | 0.24 |
| 65 | Tb | 0.0380 | 0.0030 | 0.0380 | 0.0384 | 0.0375 | 0.0372 | 0.0356 | 0.0361 | 0.0375 | 0.0363 | 0.0355 | 0.037 | 0.0355 | 0.0353 | 0.037 | 0.044 |
| 66 | Dy | 0.252 | 0.020 | 0.256 | 0.255 | 0.254 | 0.239 | 0.238 | 0.246 | 0.254 | 0.2427 | 0.245 | 0.254 | 0.245 | 0.242 | 0.254 | 0.22 |
| 67 | Ho | 0.0563 | 0.0044 | 0.0564 | 0.0572 | 0.0567 | 0.0555 | 0.0562 | 0.0546 | 0.0567 | 0.0556 | 0.0547 | 0.057 | 0.0547 | 0.054 | 0.057 | 0.056 |
| 68 | Er | 0.164 | 0.012 | 0.166 | 0.163 | 0.166 | 0.167 | 0.162 | 0.16 | 0.166 | 0.1589 | 0.16 | 0.166 | 0.159 | 0.160 | 0.166 | 0.14 |
| 69 | Tm | 0.0259 | 0.0024 | 0.0261 | 0.0261 | 0.0256 | 0.0245 | 0.0237 | 0.0247 | 0.0256 | 0.0242 | 0.0247 | 0.026 | 0.0247 | 0.022 | 0.026 | 0.022 |
| 70 | Yb | 0.167 | 0.014 | 0.169 | 0.169 | 0.165 | 0.165 | 0.163 | 0.161 | 0.165 | 0.1625 | 0.159 | 0.165 | 0.159 | 0.166 | 0.165 | 0.13 |
| 71 | Lu | 0.0249 | 0.0020 | 0.0250 | 0.0253 | 0.0254 | 0.0235 | 0.0237 | 0.0246 | 0.0254 | 0.0243 | 0.0245 | 0.025 | 0.0245 | 0.0243 | 0.025 | 0.023 |
| 72 | Hf | 0.106 | 0.008 | 0.107 | 0.106 | 0.107 | 0.116 | 0.115 | 0.103 | 0.107 | 0.104 | 0.12 | 0.106 | 0.12 | 0.119 | 0.12 | … |
| 73 | Ta | 0.0148 | 0.0014 | 0.015 | 0.0145 | 0.0142 | 0.0145 | 0.0144 | 0.0136 | 0.014 | 0.0142 | 0.016 | 0.014 | 0.016 | 0.017 | 0.014 | … |
| 74 | W | 0.102 | 0.014 | 0.096 | 0.0960 | 0.0903 | 0.089 | 0.089 | 0.093 | 0.095 | 0.0926 | 0.1 | 0.093 | 0.1 | 0.089 | 0.089 | … |
| 75 | Re | 0.0369 | 0.0028 | 0.040 | 0.0393 | 0.0395 | 0.0379 | 0.0371 | 0.04 | 0.0383 | 0.0365 | 0.037 | 0.037 | 0.037 | 0.037 | 0.037 | … |
| 76 | Os | 0.475 | 0.020 | 0.495 | 0.493 | 0.506 | 0.487 | 0.486 | 0.49 | 0.486 | 0.486 | 0.49 | 0.49 | 0.49 | 0.699 | 0.49 | … |
| 77 | Ir | 0.474 | 0.020 | 0.469 | 0.469 | 0.48 | 0.470 | 0.470 | 0.455 | 0.459 | 0.481 | 0.46 | 0.48 | 0.46 | 0.473 | 0.48 | … |
| 78 | Pt | 0.931 | 0.072 | 0.925 | 0.947 | 0.982 | 0.962 | 1.00 | 1.01 | 0.994 | 0.999 | 0.99 | 1.05 | 0.99 | 0.953 | 1.05 | … |
| 79 | Au | 0.147 | 0.024 | 0.148 | 0.146 | 0.148 | 0.145 | 0.146 | 0.14 | 0.152 | 0.14 | 0.144 | 0.14 | 0.144 | 0.145 | 0.14 | … |
| 80 | Hg | 0.288 | 0.140 | 0.35 | 0.350 | 0.31 | 0.345 | 0.314 | 0.3 | 0.31 | 0.258 | 0.39 | (5.3) | 0.39 | 0.39 | (5.3) | … |
| 81 | Tl | 0.141 | 0.014 | 0.140 | 0.142 | 0.143 | 0.144 | 0.143 | 0.14 | 0.143 | 0.142 | 0.142 | 0.14 | 0.142 | 0.143 | 0.14 | … |
| 82 | Pb | 2.64 | 0.16 | 2.62 | 2.63 | 2.53 | 2.58 | 2.56 | 2.47 | 2.53 | 2.47 | 2.4 | 2.43 | 2.4 | 2.43 | 2.43 | … |

| | | | | | | | | | | | | | | | | |
|---|---|---|---|---|---|---|---|---|---|---|---|---|---|---|---|---|
| 83 | Bi | 0.113 | 0.016 | 0.110 | 0.110 | 0.111 | 0.108 | 0.110 | 0.11 | 0.111 | 0.114 | 0.11 | 0.11 | 0.11 | 0.111 | 0.11 | … |
| 90 | Th | 0.0298 | 0.0030 | 0.0300 | 0.0310 | 0.0298 | 0.0312 | 0.0309 | 0.029 | 0.0298 | 0.0294 | 0.029 | 0.029 | 0.029 | 0.0286 | 0.029 | 0.065 |
| 92 | U | 0.00816 | 0.00106 | 0.0081 | 0.0081 | 0.0078 | 0.0081 | 0.0084 | 0.0074 | 0.0078 | 0.0081 | 0.0082 | 0.0082 | 0.0082 | 0.0081 | 0.0082 | 0.024 |
| Sum | % | 100.00 | … | 100.07 | 99.98 | 100.38 | 100.00 | 99.81 | [51.81] | 100.07 | 102.28 | 99.83 | 98.77 | 99.82 | 100.38 | 98.77 | [29.4] |

**Footnote for Table 3:** This review: Weighted average of all CI-chondrites as done in Lodders 2003. Uncertainties are either the deviation of the weighted group mean or the standard deviation found for individual meteorites, whichever is larger. The quoted uncertainties of CI-chondritic abundances thus should reflect upper limits to the characteristic variations in elemental abundances obtained on different samples and by different analytical methods.

[AE82] Anders & Ebihara 1982. [AG89] Anders & Grevesse 1989. [L03] Lodders 2003. [LPG09] Lodders, Palme Gail 2009. [MS95] McDonough & Sun 1995. [PJ03] Palme & Jones 2003. [PLJ14] Palme, Lodders, Jones 2014. [PSZ81] Palme, Suess, Zeh 1981. [W85] Wasson 1985. [WK88] Wasson & Kallemeyn 1988. [Yav68] Yavnel 1968.

The CI-chondrite compositions by mass are easily converted into atomic abundances, which traditionally is done by setting the number of silicon to $10^6$ atoms. All data are converted using the atomic weights calculated for the adopted isotopic composition of each element using the individual isotope masses from Pfeiffer et al. (2014). Elements can have different solar isotopic compositions than on the Earth so the atomic weights will be different than those listed for terrestrial materials (e.g., C,N,O, noble gases). For example, the atomic weight of Ar in air on Earth is 39.948 amu (atomic mass units where $^{12}C$=12.0) and is dominated by the mass of abundant $^{40}Ar$. For solar, the relative masses of $^{36}Ar$ and $^{38}Ar$ dominate to give 36.275 amu because $^{40}Ar$ is extremely rare in the solar wind).

Atomic abundance ratios have been used since the 1930s, and the use of atom-percentages is not deemed practical for modelling solar system abundances. Using a relative atomic scale circumvents the problems with compositions reported on a "volatile-free" basis since the atomic ratios of refractory elements relative to silicon are unaffected. Data from several studies listed in Tables 2 and 3 converted to the atomic scale relative to silicon are used in Table 6 below.

## Solar Abundances of the Elements, Photosphere, Sunspots, Corona, and Solar Wind

As mentioned earlier, the sun contains more than 99 mass-percent of solar system materials and therefore the sun's composition should represent a good average composition of the elements within the entire solar system. There is one notable exception, after the sun formed from the proto-solar nebula and before hydrogen fusion began, the young sun was fully convective and was burning deuterium during its T-Tauri stage, which is the reason for the absence of deuterium in the solar photosphere. The current sun is not uniform in composition and has a radial abundance gradient. The sun is a main-sequence star and it operates by nuclear fusion of hydrogen in the in its center. The hydrogen-burning reactions through the so-called pp-fusion chain (ca 97%) and CNO-burning cycle (ca. 3%) steadily decrease the amount of hydrogen and increase the amounts of helium in the center. The interior of the sun is not fully convective, and energy is transported from the core outward through radiation. Only the upper envelope of the sun is now convective (see review by Hanasoge et al. 2016) and believed to be relatively homogeneous in composition. The convective envelope (a shell about 1/3 of the sun's total radius), and because the convective layer only has minimal exchange of matter with the central sun, the photosphere near the top of the convective zone should give a good representative composition for the sun (and therefore the solar system) at the time of its formation.

Over the Sun's lifetime, some helium and all other heavy elements gravitationally settled from the convective zone into the solar interior. Thus the photosphere (near the top of the convection zone) no longer represents the original proto-solar composition. Gravitational settling is in part counteracted by radiation levitation, which acts differently on different ions (Eddington 1926). These effects were investigated by e.g. Turcotte et al. (1998), Vauclair (1998), Turcotte & Wimmer-Schweingruber, (2002) and Yang (2016, 2019). Yang (2019) took the effects of rotation on mixing into account (see below). In the present-day convective envelope, the amounts of and the heavier elements are about 10-20% smaller than at the time the sun formed (depending on models, see below). However, except for Li, the relative abundances of the elements heavier than helium are assumed to be more or less unchanged and any potential differences to meteoritic values as a function of mass are not resolvable with current abundances data. Therefore heavy element abundance ratios are taken as approximately constant (e.g., Piersanti et al., 2007, Asplund et al. 2009, Lodders et al., 2009). The notable exception is lithium. The Lithium relative abundance in the solar photosphere is 170-times lower than that in meteorites. This is due to pre-mainsequence lithium destruction and ongoing settling plus the nuclear destruction of the fragile lithium nuclei at the hot bottom of the solar convection zone.

The most reliable abundances for the present-day sun are obtained from photospheric absorption spectra. Even today not all elements can be detected or determined quantitatively in the Sun because the solar spectrum is crowded with thousands of lines, e.g., neutral Fe and ionized Fe lines are prolific. Crowded spectra complicate the spectral analyses because of line blending, which makes difficult the detection of lines from elements with low excitation potentials because their absorptions lines are very weak. Some elements with low abundances may only have few useful spectral lines at wavelengths difficult to access (e.g., in the ultraviolet (UV) part of spectrum that is not accessible through the Earth's atmosphere). Other elements (e.g., F, Cl, Tl) can only be measured from the spectra of the cooler sun-spots where many more atomic and molecular lines complicate the spectra. For some elements (As, Se, Br, Te, I, Cs, Ta, Re, Hg, Bi, Th, noble gases) no or only unreliable photospheric determinations exist because there are no observable or only heavily blended lines in the solar spectrum. The noble gases He, Ne, Ar, Kr, and Xe have no lines in the photospheric spectrum and their abundances could only be derived from observations of coronal sources such the solar wind (SW), solar flares, or solar energetic particles (SEP).

The He abundance is derived indirectly by solar models in conjunction with observations from helioseismology. The solar surface shows up- and downward oscillations caused by some sound waves created in the turbulent convection zone and reflected at the photospheric region ("the solar surface"). Helioseismology can probe the entire structure of the sun through observations of the different sound modes, and because sound speed is dependent on the composition of the medium that sound passes through, constraints on composition (especially the mass fractions of X, Y, and Z noted above) are obtained. However, matching the seismic observations (and to derive the helium abundance, which is a free parameter in the solar models) depends on the details of the standard solar models. The standard models must explain how the sun works over time (solar radius, density structure, neutrino flux, depth of convection zone, see, e.g., Basu & Antia, 2004, 2008, Serenelli & Basu 2010, Serenelli et al. 2011, Basu et al. 2015). The overall solar composition derived from the elements in the photosphere is extremely important for the standard solar model because the abundant elements C, N, O, Ne, Fe, and lesser abundant heavy elements determine the opacity of the solar interior which in turn determines how energy is transported outward (radiation, convection). The C, N, and O abundances were revised substantially downward since the early 2000s. The elements C, N and O are about 2/3 of all heavy elements by mass. The lower abundances gave lower opacities that conflicted with those required from helioseismology. A surge of new measurements and theoretical work has not yet resolved all disagreements (see, e.g., Basu et al. 2015, Serenelli & Basu 2010, Yang 2016, 2019, Vinyoles et al. 2017).

The helium abundance in Table 4 for the recommended 3D- and 1D- photospheric compositions is obtained as follows. The photospheric abundances relative to hydrogen for all other elements can be independently converted from atomic abundances to mass-abundances which gives the mass fraction ratio for all elements heavier than helium relative to the hydrogen mass as $Z/X = 0.0149$ for the recommended 3D photospheric values in Table 4. Basu & Antia (2004) derived he hydrogen mass fraction X in the solar convection zone from the observed solar oscillation frequencies by the Global Oscillations Network Group (GONG; Hill et al. 1996) and by the Michelson Doppler Imager (MDI; Schou et al. 1998). They found a hydrogen mass fraction X=0.7389, which was insensitive to the mass fraction ratio Z/X assumed in their two calibration models (with Z/X = 0.0171 and 0.0218). The Y/X values for the recommended abundances here are between the values used in the calibration models by Basu & Antia (2004). Using their determined mass fraction X, the mass fraction of helium is then easily obtained as Y = 0.2462 (since Y/X = 1/X -1 -Z/X and X+Y+Z =1). Once the Y/X ratio is converted back to an atomic ratio and properly scaled, the photospheric helium abundance is A(He) = 10.924 (the 1D abundances with X/Z =0.0153 give Y = 0.2458 and A(He) = 10.923). More related to the solar mass fractions is in the section on Solar System Abundances below.

## Photospheric Abundances

The quantitative determination of photospheric abundances involves construction of a numerical model atmosphere, calculation of the emitted spectrum based on the model atmosphere, and comparison of this spectrum with the observed spectrum. A comprehensive review about techniques, problems, and results from solar and stellar spectroscopy is given by Allende-Prieto (2016) and the reader is encouraged to consult this review for reference.

The photosphere is the top-layer above the solar convection zone in which the solar gas is well mixed. Suitable atmospheric models describing the temperature, densities, total pressure, composition, and convective mixing as a function of depth were developed to describe the spectral line formation regions in the photosphere and sun-spots. Strömgren (1940) constructed the first photospheric model atmosphere with temperature, density, and electron pressure as function of optical depth after Wildt (1939) identified negatively charged hydrogen ($H^-$) as the cause of the continuous opacity in the solar atmosphere. This model allowed for the first time the determination of elements relative to hydrogen because the equivalent widths of the observed absorption lines in the spectra depend on the continuum absorption from hydride ions. Unsöld (1948) and Claas (1951) did the first detailed abundance determinations using model atmospheres. Model atmospheres evolved from 1D hydrostatic models (e.g., Holweger & Müller 1974, Holweger 2001) to 3D models and can take time-dependent hydrodynamics and granulation into account (Asplund et. al., 2009, Caffau et al., 2011). It is still debated if 3D models produce more accurate solar abundances than 1D models. Differences among various 3D-models are apparently small if the same line selections and local thermodynamic equilibrium (LTE) corrections are applied (below), so line selections, number of lines included in the analyses, and atomic properties remain major issues when 3D results are compared.

Emission from the deeper regions of the photosphere is absorbed in overlying layers and leads to the absorption lines in photospheric spectra. The models of spectral line formation require atomic physics parameters such as accurate wavelengths, ionization potentials, atomic term structures, partition functions, oscillator strengths, transition probabilities and life-times, and hyperfine splitting, which must be determined experimentally, especially for the heavy elements because quantum mechanical computations become extremely complex. Goldberg et al. (1960) first applied experimental oscillator strengths to the absorption lines and could obtain refined abundances for 42 elements. Much progress had been made, and laboratory measurements of transition probabilities and lifetimes are ongoing and remain an active research area in astronomy (e.g., see papers by Lawler (2014, 2015, 2017, 2019, Sneden et al. 2009, 2016 and references therein) and solar abundance determinations will continue to improve.

An important consideration in solar abundance determinations are departures from local thermodynamic equilibrium (LTE), i.e., whether “the quantum-mechanical states of atoms, ions, and molecules are populated according to the relations of Boltzmann and Saha, valid strictly in thermodynamic equilibrium” (see Holweger, 2001, Allende-Prieto 2016). The assumption of LTE may not be justified for a highly inhomogeneous and dynamic plasma permeated by an intense, anisotropic radiation. Consequences of deviations from LTE are changes in line intensities and in the line shapes of strong lines, and these effects are apparent when observations from the centre of the solar disc are compared to those from the limb. Newer spectrum synthesis calculations take effects of NLTE (non-local thermodynamic equilibrium) more regularly into account which, however, are computationally intensive.

Table 4 summarizes photospheric abundance compilations from Russell (1929) onward. The original tables should be consulted for the measurement details and/or sources of compiled data. The photospheric abundances changed considerably since the abundance reviews by Anders & Grevesse (1989) and Grevesse & Sauval (1998) which is largely due to the application of 3D model atmospheres and NLTE corrections. Downward revisions of elemental abundances started with the photospheric oxygen abundance (Allende Pietro et al. 2000), and other elements followed as 3D model atmospheres were applied in solar spectrum analyses (see, e.g., Asplund et al. 2005, 2009, Caffau et al. 2011). Differences between different 3D models are ascribed to different selections of lines and atomic parameters as evaluated by different groups. Another change is the more frequent application of NLTE corrections to 1D and 3D analyses to account for collisional line excitations. The problems arising from the lower new solar “3D abundances” of oxygen and other abundant highly volatile elements have become known as the “oxygen crisis” and the well-working standard solar models have an “opacity problem” (e.g., Basu & Antia 2005, 2008; Vinyoles et al. 2017, and below).

Asplund et al. (2009) presented self-consistently photospheric abundances based on their 3D model atmospheres with much improved, realistic physics, and what they regarded as the best-available atomic parameters and lines selections. Scott et al. (2015a,b) and Grevesse et al. (2015), subsequently gave the complete descriptions of the analyses by Asplund et al. (2009); these are also shown in Table 4.

Palme et al. (2014) did not adopt the photospheric abundances based on 3D models by Asplund et al. (2009) because at the time the detailed descriptions for the Asplund et al. (2009) solar abundances were pending. Palme et al. (2014) compiled photospheric abundances from different papers (as in Lodders 2004 and Lodders et al. 2009). This approach relies on the recommendations given in various papers, and sometimes averages of several studies were used. This may not have a self-consistent base for atmospheric models, since 3D results from different groups, as well as 1D and 3D model results were included. The recommended photospheric abundances here are listed for “1D” and “3D” separately in Table 4 which also illustrates the generally lower photospheric abundances from 3D models.

The photospheric 3D-abundances from Asplund et al. (2009) of several rock-forming elements compare somewhat worse to CI-chondrites (Lodders 2019) and to results from the solar wind Genesis measurements whereas the latter correspond better to CI-chondrites (e.g., Heber et al. 2014, Burnett et al. 2017). Figure 2 illustrates this situation for Mg, Si and Fe, which are the three most abundant rock-forming elements (after oxygen).

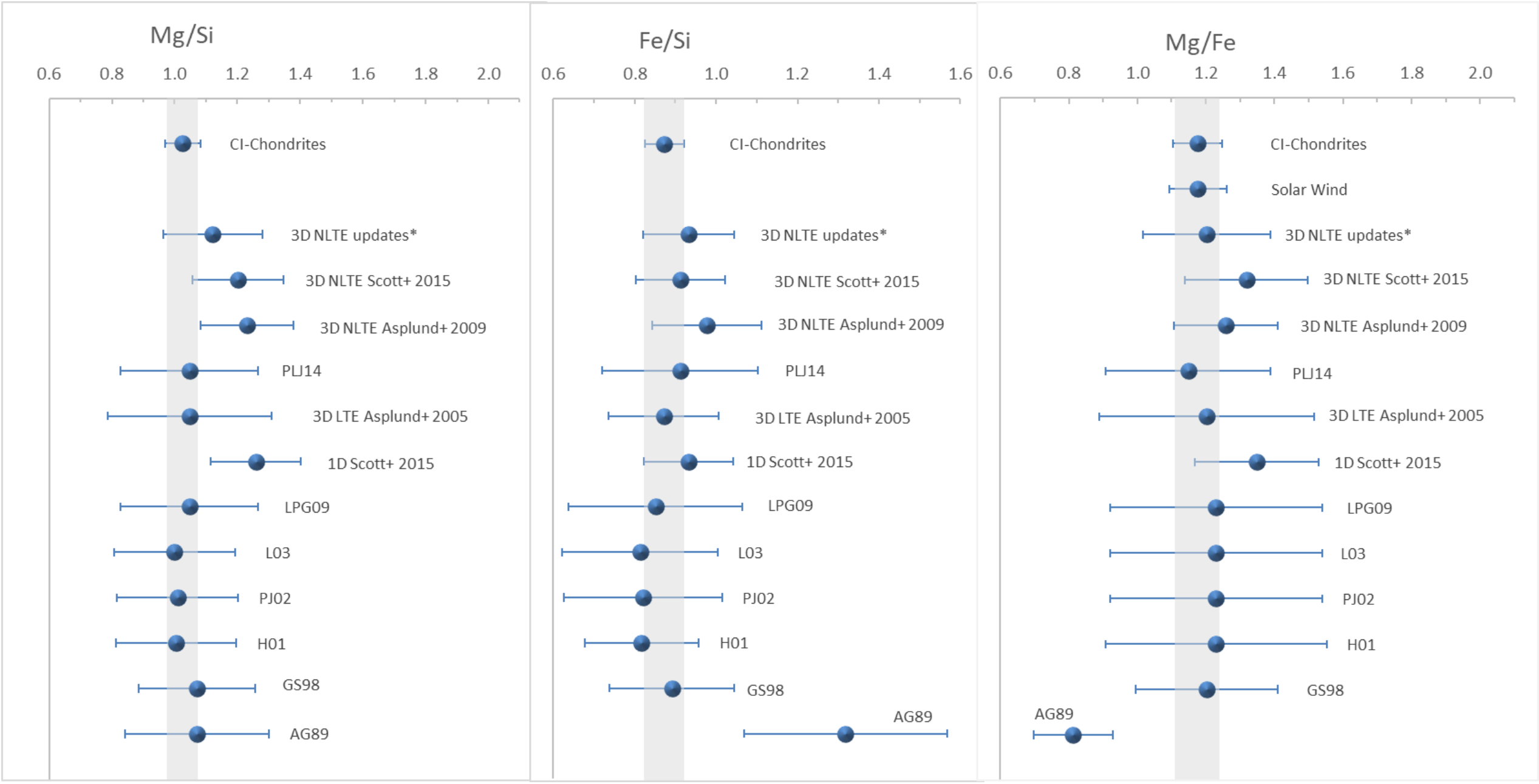

**Figure 2**. Major element ratios in CI-chondrites (top row), and from analyses of the solar photosphere. The solar wind Mg/Fe datum is from Genesis data, see Table 5 for references. The block of photospheric abundances is in reverse chronological order. Near the top is for 3D LTE (Asplund et al. 2005), and for 3D NLTE results (Asplund et al. 2009), revised/updated values (Scott et al. 2015ab), and further updates from Armasi & Asplund (2017), Bergemann et al. (2017), and Lind et al. (2017). Scott et al. (215ab) also reported 1D results for comparison. The lower block shows the ratios from earlier compilations which includes many analyses with 1D model atmosphere and limited NLTE corrections (see footnote to Table 4 for abbreviations of references used). The error bars of the ratios are calculated by quadratic combination of the elemental uncertainties reported in the references. The grey bar shows the 1-sigma uncertainties of the CI-chondrite ratios from the recommended values in Table 3

The top row in Figure 2 shows Mg/Si, Fe/Si and Mg/Fe ratios in CI-chondrites, and the Mg/Fe ratio for the solar wind. The lower block of data in Figure 2 are photospheric abundances mainly from 1D and LTE analyses, and shows the Mg/Si, Fe/Si and Mg/Fe ratios seem to agree better with respective ratios in CI-chondrites; this includes the 3D LTE results from Asplund et al. (2005). (The Fe/Si and Fe/Mg from Anders & Grevesse 1989 used an incorrect Fe, see above about the Fe abundance controversy in meteorites and the sun). However, 3D NLTE abundances by Asplund et al. (2009) deviate more from the CI-chondritic ratios and their Mg/Si ratio is higher than in CI-chondrites or any chondrite group. Subsequent updates to the solar 3D NLTE abundances seem to shift the abundance ratios closer to the CI-chondritic values. The agreement of major element abundances between the solar photosphere and CI-chondrites was one justification to use CI-chondrites as an abundance standard, and to use CI-chondrite abundances for elements that cannot be determined for the sun. If real compositional differences exist between the photosphere, the solar wind, and CI-chondrites that are not already understood within the framework of current models, approaches to derive a representative solar system composition will have to be revised. More work on photospheric abundances is needed to understand differences for other elements between meteoritic and 3D NLTE photospheric abundances.

**Table 4. Elemental Abundances in the Solar Photosphere.** Atomic abundances scaled to: A(E) = 12 + $\log_{10}(n(E)/n(H))$

Part 1.

| Z | E | Recommended 3D | | Recommended 1D | | [Z16] 1D NLTE | | [SL+] 1D | | [SSG15] 3D NLTE | | [SSG15] 1D NLTE | | [PLJ14] | | [C11] 3D NLTE | | [LPG09] | | [AGSS09] 3D | | [GAS07] | | [AGS05] | | [L03] | |
|---|---|---|---|---|---|---|---|---|---|---|---|---|---|---|---|---|---|---|---|---|---|---|---|---|---|---|---|
| | | A(E) | | A(E) | | A(E) | | A(E) | | A(E) | | A(E) | | A(E) | | A(E) | | A(E) | | A(E) | | A(E) | | A(E) | | A(E) | |
| 1 | H | 12 | | 12 | | 12 | | 12 | | 12 | | 12 | | 12 | | 12 | | 12 | | 12 | | 12 | | 12 | | 12 | |
| 2 | He | [10.924] in | 0.02 | [10.923] in | 0.02 | … | | … | | … | | … | | 10.925 | 0.02 | | | 10.925 | 0.02 | 10.930 | 0.01 | 10.93 | 0.01 | 10.93 | 0.01 | 10.899 | 0.01 |
| 3 | Li | 1.04 | 0.10 | 1.04 | 0.10 | … | | … | | … | | … | | 1.03 | 0.03 | 1.03 | 0.03 | 1.10 | 0.10 | 1.05 | 0.10 | 1.05 | 0.10 | 1.05 | 0.10 | 1.10 | 0.10 |
| 4 | Be | 1.38 | 0.09 | 1.38 | 0.09 | … | | … | | … | | … | | 1.38 | 0.09 | … | | 1.38 | 0.09 | 1.38 | 0.09 | 1.38 | 0.09 | 1.38 | 0.09 | 1.15 | 0.20 |
| 5 | B | 2.70 | 0.20 | 2.70 | 0.20 | … | | … | | … | | … | | 2.70 | 0.17 | … | | 2.70 | 0.17 | 2.70 | 0.20 | 2.70 | 0.20 | 2.70 | 0.20 | 2.70 | 0.21 |
| 6 | C | 8.47 | 0.06 | 8.58 | 0.11 | 8.43 | 0.02 | … | | … | | … | | 8.50 | 0.06 | 8.50 | 0.06 | 8.39 | 0.04 | 8.43 | 0.05 | 8.39 | 0.05 | 8.39 | 0.05 | 8.39 | 0.04 |
| 7 | N | 7.85 | 0.12 | 8.00 | 0.11 | … | | … | | … | | … | | 7.86 | 0.12 | 7.86 | 0.12 | 7.86 | 0.12 | 7.83 | 0.05 | 7.78 | 0.06 | 7.78 | 0.06 | 7.83 | 0.11 |
| 8 | O | 8.71 | 0.04 | 8.76 | 0.06 | 8.78 | 0.04 | … | | … | | … | | 8.73 | 0.07 | 8.76 | 0.07 | 8.73 | 0.07 | 8.69 | 0.03 | 8.66 | 0.05 | 8.66 | 0.05 | 8.69 | 0.05 |
| 9 | F | [4.40]sp | 0.25 | [4.40]sp | 0.25 | … | | … | | … | | … | | 4.56 | 0.30 | … | | 4.56 | 0.30 | 4.56 | 0.30 | 4.56 | 0.30 | 4.56 | 0.30 | 4.56 | 0.30 |
| 10 | Ne | [8.15]in | 0.10 | [8.10]in | 0.10 | … | | … | | … | | … | | 8.05 | 0.10 | … | | 8.05 | 0.10 | 7.93 | 0.10 | 7.84 | 0.06 | 7.84 | 0.06 | 7.87 | 0.10 |
| 11 | Na | 6.21 | 0.04 | 6.28 | 0.04 | 6.29 | 0.02 | … | | 6.21 | 0.04 | 6.27 | 0.04 | 6.30 | 0.03 | … | | 6.30 | 0.03 | 6.24 | 0.04 | 6.17 | 0.04 | 6.17 | 0.04 | 6.30 | 0.03 |
| 12 | Mg | 7.56 | 0.05 | 7.50 | 0.05 | 7.51 | 0.05 | … | | 7.59 | 0.04 | 7.63 | 0.04 | 7.54 | 0.06 | … | | 7.54 | 0.06 | 7.60 | 0.03 | 7.53 | 0.09 | 7.53 | 0.09 | 7.54 | 0.06 |
| 13 | Al | 6.43 | 0.04 | 6.41 | 0.03 | 6.39 | 0.02 | … | | 6.43 | 0.04 | 6.48 | 0.04 | 6.47 | 0.07 | … | | 6.47 | 0.07 | 6.45 | 0.03 | 6.37 | 0.06 | 6.37 | 0.06 | 6.47 | 0.07 |
| 14 | Si | 7.51 | 0.03 | 7.52 | 0.06 | 7.50 | 0.03 | … | | 7.51 | 0.03 | 7.53 | 0.03 | 7.52 | 0.06 | … | | 7.52 | 0.06 | 7.51 | 0.04 | 7.51 | 0.04 | 7.51 | 0.04 | 7.54 | 0.05 |
| 15 | P | 5.41 | 0.03 | 5.42 | 0.03 | … | | … | | 5.41 | 0.03 | 5.42 | 0.03 | 5.46 | 0.04 | 5.46 | 0.04 | 5.46 | 0.04 | 5.41 | 0.03 | 5.36 | 0.04 | 5.36 | 0.04 | 5.49 | 0.05 |
| 16 | S | 7.15 | 0.05 | 7.12 | 0.05 | … | | … | | 7.12 | 0.03 | 7.12 | 0.03 | 7.16 | 0.05 | 7.16 | 0.05 | 7.14 | 0.01 | 7.12 | 0.03 | 7.14 | 0.05 | 7.14 | 0.05 | 7.21 | 0.04 |
| 17 | Cl | [5.25] sp | 0.12 | [5.25]sp | 0.12 | … | | … | | … | | | | 5.50 | 0.30 | … | | 5.50 | 0.30 | 5.50 | 0.30 | 5.50 | 0.30 | 5.50 | 0.30 | 5.50 | 0.30 |
| 18 | Ar | [6.50]in | 0.10 | [6.50]in | 0.10 | … | | … | | … | | | | 6.50 | 0.10 | … | | 6.50 | 0.10 | 6.40 | 0.13 | 6.18 | 0.08 | 6.18 | 0.08 | 6.55 | 0.08 |
| 19 | K | 5.04 | 0.05 | 5.10 | 0.07 | 5.14 | 0.03 | … | | 5.04 | 0.05 | 5.06 | 0.05 | 5.11 | 0.09 | 5.11 | 0.09 | 5.12 | 0.03 | 5.03 | 0.09 | 5.08 | 0.07 | 5.08 | 0.07 | 5.12 | … |
| 20 | Ca | 6.32 | 0.03 | 6.34 | 0.06 | 6.34 | 0.08 | … | | 6.32 | 0.03 | 6.34 | 0.03 | 6.33 | 0.07 | … | | 6.33 | 0.07 | 6.34 | 0.04 | 6.31 | 0.04 | 6.31 | 0.04 | 6.36 | 0.02 |
| 21 | Sc | 3.16 | 0.04 | 3.15 | 0.06 | 3.07 | 0.04 | 3.15 | 0.06 | 3.16 | 0.04 | 3.22 | 0.04 | 3.10 | 0.10 | … | | 3.10 | 0.10 | 3.14 | 0.04 | 3.17 | 0.10 | 3.05 | 0.08 | 3.17 | 0.10 |
| 22 | Ti | 4.93 | 0.04 | 4.97 | 0.04 | 4.98 | 0.05 | 4.97 | 0.04 | 4.93 | 0.04 | 4.99 | 0.04 | 4.90 | 0.06 | … | | 4.90 | 0.06 | 4.95 | 0.05 | 4.90 | 0.06 | 4.90 | 0.06 | 5.02 | 0.06 |
| 23 | V | 3.89 | 0.08 | 3.96 | 0.04 | … | | 3.96 | 0.04 | 3.89 | 0.08 | 4.07 | 0.08 | 4.00 | 0.02 | … | | 4.00 | 0.02 | 3.95 | 0.10 | 4.00 | 0.02 | 4.00 | 0.02 | 4.00 | 0.02 |
| 24 | Cr | 5.62 | 0.04 | 5.62 | 0.07 | … | | 5.62 | 0.07 | 5.62 | 0.04 | 5.65 | 0.04 | 5.64 | 0.01 | … | | 5.64 | 0.01 | 5.64 | 0.04 | 5.64 | 0.10 | 5.64 | 0.10 | 5.64 | 0.13 |
| 25 | Mn | 5.52 | 0.03 | 5.45 | 0.05 | … | | 5.45 | 0.05 | 5.42 | 0.04 | 5.47 | 0.04 | 5.37 | 0.05 | … | | 5.37 | 0.05 | 5.43 | 0.04 | 5.39 | 0.03 | 5.39 | 0.03 | 5.39 | 0.03 |
| 26 | Fe | 7.48 | 0.04 | 7.52 | 0.05 | … | | 7.52 | 0.05 | 7.47 | 0.04 | 7.5 | 0.04 | 7.48 | 0.06 | 7.52 | 0.06 | 7.45 | 0.08 | 7.50 | 0.04 | 7.45 | 0.05 | 7.45 | 0.05 | 7.45 | 0.08 |
| 27 | Co | 4.93 | 0.05 | 4.96 | 0.06 | … | | 4.96 | 0.06 | 4.93 | 0.05 | 4.99 | 0.05 | 4.92 | 0.08 | … | | 4.92 | 0.08 | 4.99 | 0.07 | 4.92 | 0.08 | 4.92 | 0.08 | 4.92 | 0.08 |
| 28 | Ni | 6.20 | 0.04 | 6.28 | 0.06 | … | | 6.28 | 0.06 | 6.20 | 0.04 | 6.24 | 0.04 | 6.23 | 0.04 | … | | 6.23 | 0.04 | 6.22 | 0.04 | 6.23 | 0.04 | 6.23 | 0.04 | 6.22 | 0.13 |
| 29 | Cu | 4.18 | 0.05 | 4.21 | 0.03 | 4.10 | 0.06 | 4.21 | 0.03 | 4.18 | 0.05 | | 0.05 | 4.21 | 0.04 | … | | 4.21 | 0.04 | 4.19 | 0.04 | 4.21 | 0.04 | 4.21 | 0.04 | 4.21 | 0.04 |
| 30 | Zn | 4.56 | 0.05 | 4.61 | 0.09 | … | | 4.61 | 0.09 | 4.56 | 0.05 | | 0.05 | 4.62 | 0.15 | … | | 4.62 | 0.15 | 4.56 | 0.05 | 4.60 | 0.03 | 4.60 | 0.03 | 4.62 | 0.15 |
| 31 | Ga | 3.02 | 0.05 | 3.09 | 0.05 | … | | … | | 3.02 | 0.05 | | 0.05 | 2.88 | 0.10 | … | | 2.88 | 0.10 | 3.04 | 0.09 | 2.88 | 0.10 | 2.88 | 0.10 | 2.88 | 0.10 |
| 32 | Ge | 3.63 | 0.07 | 3.62 | 0.07 | … | | … | | 3.63 | 0.07 | | 0.07 | 3.58 | 0.05 | … | | 3.58 | 0.05 | 3.65 | 0.10 | 3.58 | 0.05 | 3.58 | 0.05 | 3.58 | 0.05 |
| 33 | As | … | | … | | … | | … | | … | | … | | … | | … | | … | | … | | … | | … | | … | |
| 34 | Se | … | | … | | … | | … | | … | | … | | … | | … | | … | | … | | … | | … | | … | |
| 35 | Br | … | | … | | … | | … | | … | | … | | … | | … | | … | | … | | … | | … | | … | |
| 36 | Kr | [3.25]in | 0.08 | [3.25]in | 0.08 | … | | … | | 3.25 | 0.06 | | 0.06 | 3.28 | 0.08 | … | | 3.28 | 0.08 | 3.25 | 0.06 | 3.25 | 0.08 | 3.28 | 0.08 | 3.28 | 0.08 |
| 37 | Rb | 2.47 | 0.07 | 2.57 | 0.07 | … | | … | | 2.47 | 0.07 | 2.57 | 0.07 | 2.60 | 0.10 | … | | 2.60 | 0.10 | 2.52 | 0.10 | 2.60 | 0.15 | 2.60 | 0.15 | 2.60 | … |
| 38 | Sr | 2.83 | 0.06 | 2.92 | 0.05 | 2.95 | 0.08 | … | | 2.83 | 0.06 | 2.85 | 0.06 | 2.92 | 0.05 | … | | 2.92 | 0.05 | 2.87 | 0.07 | 2.92 | 0.05 | 2.92 | 0.05 | 2.92 | 0.05 |
| 39 | Y | 2.21 | 0.05 | 2.20 | 0.05 | … | | … | | 2.21 | 0.05 | 2.20 | 0.05 | 2.21 | 0.02 | … | | 2.21 | 0.02 | 2.21 | 0.05 | 2.21 | 0.02 | 2.21 | 0.02 | 2.21 | 0.02 |
| 40 | Zr | 2.59 | 0.04 | 2.59 | 0.06 | 2.60 | 0.10 | … | | 2.59 | 0.04 | 2.59 | 0.04 | 2.62 | 0.06 | … | | 2.58 | 0.02 | 2.58 | 0.04 | 2.58 | 0.02 | 2.59 | 0.04 | 2.59 | 0.04 |
| 41 | Nb | 1.47 | 0.06 | 1.49 | 0.06 | … | | … | | 1.47 | 0.06 | 1.49 | 0.06 | 1.44 | 0.06 | … | | 1.42 | 0.06 | 1.46 | 0.04 | 1.42 | 0.06 | 1.42 | 0.06 | 1.42 | 0.06 |
| 42 | Mo | 1.88 | 0.09 | 2.04 | 0.05 | … | | … | | 1.88 | 0.09 | 2.04 | 0.09 | 1.92 | 0.05 | … | | 1.92 | 0.05 | 1.88 | 0.08 | 1.92 | 0.05 | 1.92 | 0.05 | 1.92 | 0.05 |
| 43 | Tc | … | | … | | … | | … | | … | | | | … | | … | | … | | … | | … | | … | | … | |
| 44 | Ru | 1.75 | 0.08 | 1.91 | 0.10 | … | | … | | 1.75 | 0.08 | 1.91 | 0.08 | 1.72 | 0.10 | … | | 1.84 | 0.07 | 1.75 | 0.08 | 1.84 | 0.07 | 1.84 | 0.07 | 1.84 | 0.07 |
| 45 | Rh | 0.89 | 0.08 | 1.07 | 0.08 | … | | … | | 0.89 | 0.08 | 1.07 | 0.08 | 1.12 | 0.12 | … | | 1.12 | 0.12 | 0.91 | 0.10 | 1.12 | 0.12 | 1.12 | 0.12 | 1.12 | 0.12 |
| 46 | Pd | 1.55 | 0.06 | 1.61 | 0.06 | … | | … | | 1.55 | 0.06 | 1.61 | 0.06 | 1.66 | 0.04 | … | | 1.66 | 0.04 | 1.57 | 0.10 | 1.66 | 0.04 | 1.69 | 0.04 | 1.69 | 0.04 |
| 47 | Ag | 0.96 | 0.1 | 1.04 | 0.10 | … | | … | | 0.96 | 0.1 | 1.04 | 0.1 | 0.94 | 0.30 | … | | 0.94 | 0.30 | 0.94 | 0.10 | 0.94 | 0.25 | 0.94 | 0.24 | 0.94 | uncert. |
| 48 | Cd | 1.77 | 0.15 | 1.79 | 0.15 | … | | … | | 1.77 | 0.15 | 1.79 | 0.15 | 1.77 | 0.11 | … | | 1.77 | 0.11 | … | | 1.77 | 0.11 | 1.77 | 0.11 | 1.77 | 0.11 |
| 49 | In | … | | [0.8]sp | 0.20 | … | | … | | … | | 0.8 | 0.2 | ≤1.5 | | … | | ≤1.5 | | 0.80 | 0.20 | 1.60 | 0.20 | 1.60 | 0.20 | 1.56 | 0.20 |
| 50 | Sn | 2.02 | 0.1 | 2.04 | 0.10 | … | | … | | 2.02 | 0.1 | 2.04 | 0.1 | 2.00 | 0.30 | … | | 2.00 | 0.30 | 2.04 | 0.10 | 2.00 | 0.30 | 2.00 | 0.30 | 2.00 | 0.30 |
| 51 | Sb | … | | … | | … | | … | | … | | … | | 1.00 | 0.30 | … | | 1.00 | 0.30 | … | | 1.00 | 0.30 | 1.00 | 0.30 | 1.00 | 0.30 |
| 52 | Te | … | | … | | … | | … | | … | | … | | … | | … | | … | | … | | … | | … | | … | |
| 53 | I | … | | … | | … | | … | | … | | … | | … | | … | | … | | … | | … | | … | | … | |
| 54 | Xe | [2.25]in | 0.08 | [2.25]in | 0.08 | … | | … | | [2.24] | 0.06 | [2.24] | 0.06 | 2.27 | 0.08 | … | | 2.27 | 0.08 | 2.24 | 0.06 | 2.24 | 0.02 | 2.27 | 0.02 | 2.27 | 0.02 |
| 55 | Cs | … | | … | | … | | … | | … | | | | | | … | | … | | … | | … | … | … | … | … | … |
| 56 | Ba | 2.25 | 0.07 | 2.19 | 0.07 | 2.20 | 0.04 | … | | 2.25 | 0.07 | 2.18 | 0.07 | 2.20 | 0.1 | … | | 2.17 | 0.07 | 2.18 | 0.09 | 2.17 | 0.07 | 2.17 | 0.07 | 2.17 | 0.07 |
| 57 | La | 1.11 | 0.04 | 1.14 | 0.03 | … | | 1.14 | 0.03 | 1.11 | 0.04 | 1.14 | 0.04 | 1.14 | 0.03 | … | | 1.14 | 0.03 | 1.10 | 0.03 | 1.13 | 0.05 | 1.13 | 0.05 | 1.13 | 0.03 |
| 58 | Ce | 1.58 | 0.04 | 1.61 | 0.06 | … | | 1.61 | 0.06 | 1.58 | 0.04 | 1.61 | 0.04 | 1.61 | 0.06 | … | | 1.61 | 0.06 | 1.58 | 0.04 | 1.70 | 0.10 | 1.58 | 0.09 | 1.58 | 0.09 |
| 59 | Pr | 0.72 | 0.04 | 0.76 | 0.04 | … | | 0.76 | 0.04 | 0.72 | 0.04 | 0.76 | 0.04 | 0.76 | 0.04 | … | | 0.76 | 0.04 | 0.72 | 0.04 | 0.58 | 0.10 | 0.71 | 0.08 | 0.71 | 0.08 |
| 60 | Nd | 1.42 | 0.04 | 1.45 | 0.05 | … | | 1.45 | 0.05 | 1.42 | 0.04 | 1.45 | 0.04 | 1.45 | 0.05 | … | | 1.45 | 0.05 | 1.42 | 0.04 | 1.45 | 0.05 | 1.45 | 0.05 | 1.50 | 0.12 |
| 61 | Pm | … | … | … | … | … | | … | | … | … | | … | … | … | … | | … | … | … | … | … | … | … | … | … | … |
| 62 | Sm | 0.95 | 0.04 | 1.00 | 0.05 | … | | 1 | 0.05 | 0.95 | 0.04 | 0.99 | 0.04 | 1.00 | 0.05 | … | | 1.00 | 0.05 | 0.96 | 0.04 | 1.00 | 0.03 | 1.01 | 0.06 | 0.99 | … |
| 63 | Eu | 0.52 | 0.04 | 0.52 | 0.04 | 0.53 | 0.02 | 0.52 | 0.04 | 0.52 | 0.04 | 0.55 | 0.04 | 0.52 | 0.04 | 0.52 | 0.03 | 0.52 | 0.04 | 0.52 | 0.04 | 0.52 | 0.06 | 0.52 | 0.06 | 0.52 | 0.04 |
| 64 | Gd | 1.08 | 0.04 | 1.11 | 0.05 | … | | 1.11 | 0.05 | 1.08 | 0.04 | 1.11 | 0.04 | 1.11 | 0.05 | … | | 1.11 | 0.05 | 1.07 | 0.04 | 1.11 | 0.03 | 1.12 | 0.04 | 1.12 | 0.04 |
| 65 | Tb | 0.31 | 0.1 | 0.28 | 0.10 | … | | 0.28 | 0.10 | 0.31 | 0.1 | 0.28 | 0.1 | 0.28 | 0.05 | … | | 0.28 | 0.10 | 0.30 | 0.10 | 0.28 | 0.30 | 0.28 | 0.30 | 0.28 | 0.30 |
| 66 | Dy | 1.1 | 0.04 | 1.13 | 0.06 | … | | 1.13 | 0.06 | 1.1 | 0.04 | 1.13 | 0.04 | 1.13 | 0.06 | … | | 1.13 | 0.06 | 1.10 | 0.04 | 1.14 | 0.08 | 1.14 | 0.08 | 1.14 | 0.08 |
| 67 | Ho | 0.48 | 0.11 | 0.51 | 0.10 | … | | 0.51 | 0.10 | 0.48 | 0.11 | 0.51 | 0.11 | 0.51 | 0.10 | … | | 0.51 | 0.10 | 0.48 | 0.11 | 0.51 | 0.10 | 0.51 | 0.10 | 0.53 | 0.10 |
| 68 | Er | 0.93 | 0.05 | 0.96 | 0.06 | … | | 0.96 | 0.06 | 0.93 | 0.05 | 0.96 | 0.05 | 0.96 | 0.05 | … | | 0.96 | 0.06 | 0.92 | 0.05 | 0.93 | 0.06 | 0.93 | 0.06 | 0.93 | 0.06 |
| 69 | Tm | 0.11 | 0.04 | 0.14 | 0.04 | … | | 0.14 | 0.04 | 0.11 | 0.04 | 0.14 | 0.04 | 0.14 | 0.04 | … | | 0.14 | 0.04 | 0.10 | 0.04 | 0.00 | 0.15 | 0.00 | 0.15 | 0.00 | 0.15 |
| 70 | Yb | 0.85 | 0.11 | 0.86 | 0.10 | … | | 0.86 | 0.10 | 0.85 | 0.11 | 0.86 | 0.11 | 0.86 | 0.10 | … | | 0.86 | 0.10 | 0.84 | 0.11 | 1.08 | 0.15 | 1.08 | 0.15 | 1.08 | 0.15 |
| 71 | Lu | 0.1 | 0.09 | 0.12 | 0.08 | … | | 0.12 | 0.08 | 0.1 | 0.09 | 0.12 | 0.09 | 0.12 | 0.08 | … | | 0.12 | 0.08 | 0.10 | 0.09 | 0.06 | 0.10 | 0.06 | 0.10 | 0.06 | 0.10 |
| 72 | Hf | 0.85 | 0.05 | 0.88 | 0.03 | … | | 0.88 | 0.03 | 0.85 | 0.05 | 0.88 | 0.05 | 0.88 | 0.05 | 0.87 | 0.04 | 0.88 | 0.08 | 0.85 | 0.04 | 0.88 | 0.08 | 0.88 | 0.08 | 0.88 | 0.08 |
| 73 | Ta | … | | … | | … | | … | | … | | | | … | | … | | … | | … | | … | | … | | … | |
| 74 | W | 0.83 | 0.11 | 1.03 | 0.11 | … | | … | | 0.83 | 0.11 | 1.03 | 0.11 | 1.11 | 0.15 | … | | 1.11 | 0.15 | 0.85 | 0.12 | 1.11 | 0.15 | 1.11 | 0.15 | 1.11 | 0.15 |
| 75 | Re | … | | … | | … | | … | | … | | | | … | | … | | … | | … | | … | | … | | … | |
| 76 | Os | 1.4 | 0.05 | 1.5 | 0.05 | … | | … | | 1.4 | 0.05 | 1.5 | 0.05 | 1.36 | 0.19 | 1.36 | 0.19 | 1.45 | 0.11 | 1.40 | 0.08 | 1.25 | 0.11 | 1.45 | 0.10 | 1.45 | 0.10 |
| 77 | Ir | 1.42 | 0.07 | 1.46 | 0.07 | … | | … | | 1.42 | 0.07 | 1.46 | 0.07 | 1.38 | 0.05 | … | | 1.38 | 0.05 | 1.38 | 0.07 | 1.38 | 0.05 | 1.38 | 0.05 | 1.38 | 0.05 |
| 78 | Pt | … | | … | | … | | … | | … | | … | | 1.74 | 0.30 | … | | 1.74 | 0.30 | … | | … | | … | | 1.74 | uncert. |
| 79 | Au | 0.91 | 0.08 | 0.93 | 0.08 | … | | … | | 0.91 | 0.08 | 0.93 | 0.08 | 1.01 | 0.18 | … | | 1.01 | 0.18 | 0.92 | 0.10 | 1.01 | 0.15 | 1.01 | 0.15 | 1.01 | 0.18 |
| 80 | Hg | … | | … | | … | | … | | … | | | | … | | … | | | | … | | … | | … | | … | |
| 81 | Tl | … | | [0.95]sp | [0.2] | … | | … | | | | [0.9] | 0.2 | 0.95 | 0.20 | … | | 0.95 | 0.20 | 0.90 | 0.20 | 0.90 | 0.20 | 0.90 | 0.20 | 0.72-1.1 | uncert. |
| 82 | Pb | 1.92 | 0.08 | 2.05 | 0.08 | … | | … | | 1.92 | 0.08 | 2.05 | 0.08 | 2.00 | 0.06 | … | | 2.00 | 0.06 | 1.75 | 0.10 | 2.00 | 0.06 | 2.00 | 0.06 | 2.00 | 0.06 |
| 83 | Bi | … | | … | | … | | … | | … | | | | … | | … | | … | | … | | … | | … | | … | |
| 90 | Th | ≤0.03 | uncert | ≤ 0.07 | uncert. | … | | … | | 0.03 | 0.1 | 0.07 | 0.1 | < 0.08 | | 0.08 | 0.03 | 0.08 | 0.03 | 0.02 | 0.10 | uncert. | | uncert. | | 0.12 | uncert. |
| 92 | U | … | | … | | … | | … | | … | | … | | < -0.47 | | … | | < -0.47 | | | | < -0.47 | | < -0.47 | | < -0.47 | |

**Table 4. Elemental Abundances in the Solar Photosphere.** Atomic abundances scaled to: $A(E) = 12 + \log_{10}(n(E)/n(H))$
Part 2.

| Z | E | [PJ03] | [GS02] | [H01] | [GS98] | [GNS96] | [GN93] | [AG89] | [G84] | [P79] | [E77] | [RA76] | [W71] | [M68] | [GBB68] | [M65] | [GMA60] | [C51] | [U48] | [U47] | [U45] | [Me43] | [R29] | [P25] |
|---|---|---|---|---|---|---|---|---|---|---|---|---|---|---|---|---|---|---|---|---|---|---|---|---|
| | | A(E) | A(E) | A(E) | A(E) | A(E) | A(E) | A(E) | A(E) | A(E) | A(E) | A(E) | A(E) | A(E) | A(E) | A(E) | A(E) | A(E) | A(E) | A(E) | A(E) | A(E) | A(E) | A(E) |
| 1 | H | 12 | 12 | 12 | 12 | 12 | 12 | 12 | 12 | 12 | 12 | 12 | 12 | 12 | 12 | 12 | 12 | 12 | 12 | 12 | 12 | 12 | 12 | 12 |
| 2 | He | 10.99 | 10.99 | … | 10.99 | 10.99 | 10.93 | 10.99 | 11 | 10.8 | 10.9 | 10.8 | … | … | … | … | … | … | … | 11.2 | 11 | … | 10.5 | 9.3 |
| 3 | Li | 1.10 | 1.10 | … | 1.10 | 1.16 | 1.15 | 1.16 | 1 | 1 | 1 | 1 | 0.6 | <0.9 | 1 | 1.54 | 0.96 | 1.08 | … | … | … | … | 2.5 | 1.0 |
| 4 | Be | 1.40 | 1.40 | … | 1.40 | 1.15 | 1.15 | 1.15 | 1.15 | 1.2 | 1.1 | 1.15 | 1.06 | 2.34 | 2.33/1.24 | 2.34 | 2.36 | … | … | … | … | … | 2.3 | … |
| 5 | B | 2.55 | 2.70 | … | 2.55 | 2.60 | 2.60 | 2.60 | 2.60 | 2.4 | 2.3 | <2.1 | <2.8 | <3.6 | … | … | … | … | … | … | … | … | … | … |
| 6 | C | 8.59 | 8.52 | 8.592 | 8.52 | 8.55 | 8.55 | 8.56 | 8.69 | 8.66 | 8.7 | 8.62 | 8.57 | 8.51 | 8.55 | 8.62 | 8.72 | … | 7.78 | 8.64 | 8.4 | 7.56 | 7.9 | 5.5 |
| 7 | N | 7.93 | 7.95 | 7.931 | 7.92 | 7.97 | 7.97 | 8.05 | 7.99 | 7.93 | 7.9 | 7.94 | 8.06 | 8.06 | 8.93 | 8.86 | 7.98 | … | 8.1 | 8.86 | 8.73 | 8.09 | 8.1 | … |
| 8 | O | 8.69 | 8.73 | 8.736 | 8.83 | 8.87 | 8.87 | 8.93 | 8.91 | 8.88 | 8.8 | 8.84 | 8.83 | 8.83 | 8.93 | 8.86 | 8.96 | 8.65 | 8.22 | 8.98 | 8.99 | 8.56 | 9.5 | 7.1 |
| 9 | F | 4.56 | 4.56 | … | 4.56 | 4.56 | 4.56 | 4.56 | 4.56 | 4.6 | 4.6 | 4.56 | 4.56 | … | … | … | … | … | … | … | … | … | … | … |
| 10 | Ne | 8.00 | 8.06 | 8.001 | 8.08 | 8.08 | 8.08 | 8.09 | 8.00 | 7.8 | 7.7 | 7.57 | 7.45 | … | … | … | … | … | … | 9.06 | 9.06 | … | … | … |
| 11 | Na | 6.33 | 6.33 | … | 6.33 | 6.33 | 6.33 | 6.33 | 6.33 | 6.29 | 6.3 | 6.28 | 6.24 | 6.3 | … | … | 6.30 | 6.33 | 5.77 | 6.53 | 6.36 | 6.56 | 7.7 | 6.2 |
| 12 | Mg | 7.54 | 7.58 | 7.538 | 7.58 | 7.58 | 7.58 | 7.58 | 7.58 | 7.6 | 7.6 | 7.6 | 7.54 | 7.36 | … | … | 7.40 | 7.57 | 7 | 7.76 | 7.79 | 8.39 | 8.3 | 6.6 |
| 13 | Al | 6.47 | 6.47 | … | 6.47 | 6.47 | 6.47 | 6.47 | 6.47 | 6.4 | 6.4 | 6.52 | 6.4 | 6.2 | … | … | 6.20 | 6.17 | 5.82 | 6.58 | 6.58 | 6.39 | 6.9 | 6.0 |
| 14 | Si | 7.54 | 7.55 | 7.536 | 7.55 | 7.55 | 7.55 | 7.55 | 7.55 | 7.6 | 7.6 | 7.65 | 7.55 | 7.70/7.24 | 7.74 | 7.5 | 7.50 | 7.12 | 6.78 | 7.7 | 7.75 | 7.87 | 7.8 | 6.6 |
| 15 | P | 5.45 | 5.43 | … | 5.45 | 5.45 | 5.45 | 5.45 | 5.45 | 5.5 | 5.5 | 5.5 | 5.43 | 5.34 | … | … | 5.34 | … | … | … | … | … | … | … |
| 16 | S | 7.33 | 7.33 | … | 7.33 | 7.33 | 7.21 | 7.21 | 7.21 | 7.2 | 7.2 | 7.2 | 7.21 | 7.3 | … | … | 7.30 | … | 6.41 | 7.17 | 7.17 | 7.57 | 6.2 | … |
| 17 | Cl | 5.5 | 5.5 | … | 5.5 | 5.5 | 5.5 | 5.5 | 5.5 | 5.4 | 5.5 | 5.5 | 5.65 | … | … | … | … | … | … | … | … | … | … | … |
| 18 | Ar | 6.4 | 6.4 | … | 6.4 | 6.52 | 6.52 | 6.56 | 6.58 | 6.4 | 6.0 | 6.00 | 6.65 | … | … | … | … | … | … | … | … | … | … | … |
| 19 | K | 5.12 | 5.12 | … | 5.12 | 5.12 | 5.12 | 5.12 | 5.12 | 5.15 | 5.2 | 5.16 | 5.05 | 4.7 | … | … | 4.70 | 5.01 | 4.69 | 5.45 | … | 5.09 | 7.3 | 4.4 |
| 20 | Ca | 6.36 | 6.36 | … | 6.36 | 6.36 | 6.36 | 6.36 | 6.36 | 6.3 | 6.3 | 6.35 | 6.33 | 6.04 | … | … | 6.15 | 6.46 | 5.72 | 6.48 | 6.48 | 6.57 | 7.2 | 5.8 |
| 21 | Sc | 3.17 | 3.17 | … | 3.17 | 3.17 | 3.17 | 3.1 | 3.1 | 3.0 | 3.1 | 3.04 | 3.07 | 2.85 | 3.24 | 2.8 | 2.82 | … | 2.82 | 3.58 | … | … | 4.1 | … |
| 22 | Ti | 5.02 | 5.02 | … | 5.02 | 5.02 | 5.02 | 4.99 | 5.02 | 4.8 | 5.0 | 5.05 | 4.74 | 4.63/4.81 | 4.85 | 4.58 | 4.68 | 7.56 | 4.45 | 5.21 | … | 4.57 | 5.7 | 5.1 |
| 23 | V | 4.00 | 4.00 | … | 4.00 | 4.00 | 4.00 | 4.00 | 4.00 | 4.0 | 4.1 | 4.02 | 4.1 | 4.17 | 4.28 | 4.12 | 3.70 | … | 3.54 | 4.3 | … | 4.09 | 5.5 | 4.0 |
| 24 | Cr | 5.67 | 5.67 | … | 5.67 | 5.67 | 5.67 | 5.67 | 5.67 | 5.7 | 5.7 | 5.71 | 5.7 | 5.12/5.01 | 5.12 | 5.07 | 5.36 | … | 5.07 | 5.83 | … | 4.87 | 6.2 | 4.9 |
| 25 | Mn | 5.39 | 5.39 | … | 5.39 | 5.39 | 5.39 | 5.39 | 5.45 | 5.4 | 5.4 | 5.42 | 5.2 | 4.85 | 5.01 | 4.80 | 4.90 | … | 4.95 | 5.71 | … | 5.09 | 6.4 | 5.6 |
| 26 | Fe | 7.45 | 7.5 | 7.448 | 7.50 | 7.50 | 7.50 | 7.67 | 7.67 | 7.5 | 7.6 | 7.50 | 7.40 | 6.80 | 6.79 | 6.70 | 6.57 | 7.16 | 7.21 | 7.97 | 7.97 | 6.99 | 7.7 | 5.8 |
| 27 | Co | 4.92 | 4.92 | … | 4.92 | 4.92 | 4.92 | 4.92 | 4.92 | 5.0 | 5.0 | 4.90 | 4.50 | 4.70 | 4.56 | 4.40 | 4.64 | … | 4.52 | 5.28 | … | 4.69 | 6.1 | … |
| 28 | Ni | 6.25 | 6.25 | … | 6.25 | 6.25 | 6.25 | 6.25 | 6.25 | 6.3 | 6.3 | 6.28 | 6.28 | 5.77 | 5.64 | 5.44 | 5.91 | … | 3.64 | 6.2 | … | 6.39 | 6.5 | … |
| 29 | Cu | 4.21 | 4.21 | … | 4.21 | 4.21 | 4.21 | 4.21 | 4.21 | 4.1 | 4.1 | 4.06 | 4.45 | 4.45 | 4.18 | 3.5 | 5.04 | 4.8 | 3.72 | 4.48 | … | 4.39 | 5.5 | … |
| 30 | Zn | 4.6 | 4.6 | … | 4.6 | 4.6 | 4.60 | 4.6 | 4.6 | 4.4 | 4.4 | 4.45 | 4.42 | 3.52 | 3.23 | 3.52 | 4.40 | 4.52 | 4.27 | 5.03 | … | 5.57 | 5.4 | 5.2 |
| 31 | Ga | 2.88 | 2.88 | … | 2.88 | 2.88 | 2.88 | 2.88 | 2.88 | 2.8 | 2.8 | 2.8 | 2.84 | 2.72 | … | … | 2.36 | … | … | … | … | … | 2.5 | … |
| 32 | Ge | 3.41 | 3.41 | … | 3.41 | 3.41 | 3.41 | 3.41 | 3.63 | 3.4 | 3.4 | 3.5 | 3.32 | 2.49 | … | … | 3.29 | … | … | … | … | … | 3.5 | … |
| 33 | As | … | … | … | … | … | … | … | … | … | … | … | … | … | … | … | … | … | … | … | … | … | … | … |
| 34 | Se | … | … | … | … | … | … | … | … | … | … | … | … | … | … | … | … | … | … | … | … | … | … | … |
| 35 | Br | … | … | … | … | … | … | … | … | … | … | … | … | … | … | … | … | … | … | … | … | … | … | … |
| 36 | Kr | 3.3 | 3.31 | … | 3.31 | … | … | … | | … | … | … | … | … | … | … | … | … | … | … | … | … | | … |
| 37 | Rb | 2.6 | 2.6 | … | 2.6 | 2.60 | 2.60 | 2.60 | 2.60 | 2.6 | 2.6 | 2.6 | 2.63 | 2.48 | … | … | 2.48 | … | | … | … | … | 2.2 | … |
| 38 | Sr | 2.97 | 2.92 | … | 2.97 | 2.97 | 2.90 | 2.90 | 2.90 | 2.9 | 2.9 | 2.9 | 2.82 | 3.02 | … | … | 2.60 | 2.88 | 2.84 | 3.7 | … | … | 3.8 | 2.6 |
| 39 | Y | 2.21 | 2.24 | … | 2.24 | 2.24 | 2.24 | 2.24 | 2.24 | 2.1 | 2.1 | 2.1 | 1.62 | 3.2 | 2.33/2.42 | 3.2 | 2.25 | … | 2.7 | 3.56 | … | … | 3.1 | … |
| 40 | Zr | 2.59 | 2.6 | … | 2.6 | 2.60 | 2.60 | 2.60 | 2.56 | 2.8 | 2.8 | 2.75 | … | 2.65 | 3.31/2.28 | 2.65 | 2.23 | … | 1.86 | 2.72 | … | … | 3 | … |
| 41 | Nb | 1.42 | 1.42 | … | 1.42 | 1.42 | 1.42 | 1.42 | 2.1 | 2.0 | 2.0 | 1.9 | 2.3 | 2.3 | 2.73 | 2.3 | 1.95 | … | … | … | … | … | 1.5 | … |
| 42 | Mo | 1.92 | 1.92 | … | 1.92 | 1.92 | 1.92 | 1.92 | 1.92 | 2.2 | 2.2 | 2.16 | 1.9 | 2.3 | 2.52 | 2.3 | 1.90 | … | 1.27 | 2.13 | … | … | 1.9 | … |
| 43 | Tc | … | … | … | … | … | … | … | … | … | … | … | … | … | … | … | … | … | … | … | … | … | … | … |
| 44 | Ru | 1.84 | 1.84 | … | 1.84 | 1.84 | 1.84 | 1.84 | 1.84 | 1.8 | 1.9 | 1.83 | 1.57 | 1.82 | 1.98 | 1.82 | 1.43 | … | … | … | … | … | 2.2 | … |
| 45 | Rh | 1.12 | 1.12 | … | 1.12 | 1.12 | 1.12 | 1.12 | 1.12 | 1.5 | 1.5 | 1.4 | 1.55 | 1.37 | 1.55 | 1.37 | 0.78 | … | … | … | … | … | 1 | … |
| 46 | Pd | 1.69 | 1.69 | … | 1.69 | 1.69 | 1.69 | 1.69 | 1.69 | 1.5 | 1,5 | 1.5 | 1.57 | 1.57 | 1.64 | 1.27 | 1.21 | … | … | … | … | … | 1.6 | … |
| 47 | Ag | 0.94 | 0.94 | … | 0.94 | 0.94 | 0.94 | 0.94 | 0.94 | 0.9 | 0.9 | 0.85 | 0.67 | 0.75 | 1.12 | 1.04 | 0.14 | … | … | … | … | … | 1.5 | … |
| 48 | Cd | 1.77 | 1.77 | … | 1.77 | 1.77 | 1.77 | 1.86 | 1.86 | 2 | 2 | 1.85 | 1.97 | 1.54 | 2.64 | 1.66 | 1.46 | … | … | … | … | … | 2.7 | … |
| 49 | In | 1.66 | 1.66 | … | 1.66 | 1.66 | 1.66 | 1.66 | 1.66 | 1.7 | 1.7 | 1.65 | 1.71 | 1.45 | 1.68 | 1.28 | 1.16 | … | … | … | … | … | 0.5 | … |
| 50 | Sn | 2.00 | 2.00 | … | 2.00 | 2.00 | 2.00 | 2.00 | 2.00 | 2 | 2 | 2 | 1.71 | 1.54/ 2.05 | 1.32 | 2.05 | 1.54 | … | … | … | … | … | 1.7 | … |
| 51 | Sb | 1.00 | 1.00 | … | 1.00 | 1.00 | 1.00 | 1.00 | 1.00 | 1 | 1 | 1 | 0.75 | 1.94/ 0.42 | 1.04 | 0.42 | 1.94 | … | … | … | … | … | 1.3 | … |
| 52 | Te | … | … | … | … | … | … | … | … | … | … | … | … | … | … | … | … | … | … | … | … | … | … | … |
| 53 | I | … | … | … | … | … | … | … | … | … | … | … | … | … | … | … | … | … | … | … | … | … | … | … |
| 54 | Xe | 2.16 | 2.17 | … | 2.17 | … | … | … | … | … | … | … | … | … | … | … | … | … | … | … | … | … | … | … |
| 55 | Cs | … | … | … | … | … | 1.12 | … | … | <1.9 | <2.1 | <1.9 | <1.79 | … | … | … | … | | … | … | … | … | … | … |
| 56 | Ba | 2.13 | 2.13 | … | 2.13 | 2.13 | 2.13 | 2.13 | 2.13 | 2.1 | 2.1 | 2.09 | 2.22 | 2.1 | 2.06 | 2.5 | 2.10 | 2.38 | 2.44 | … | … | … | 3.8 | 2.1 |
| 57 | La | 1.17 | 1.17 | … | 1.17 | 1.17 | 1.22 | 1.22 | 1.22 | 1.1 | 1.1 | 1.13 | 1.11 | 2.03 | 2.03 | … | … | … | … | … | … | … | 2.3 | … |
| 58 | Ce | 1.58 | 1.58 | … | 1.58 | 1.58 | 1.55 | 1.55 | 1.55 | 1.6 | 1.6 | 1.55 | 1.62 | 1.78 | 1.98 | … | … | … | … | … | … | … | 2.9 | … |
| 59 | Pr | 0.71 | 0.71 | … | 0.71 | 0.71 | 0.71 | 0.71 | 0.71 | 0.8 | 0.8 | 0.66 | 0.78 | 1.45 | 1.59 | … | … | … | … | … | … | … | 1.1 | … |
| 60 | Nd | 1.50 | 1.50 | … | 1.50 | 1.50 | 1.50 | 1.50 | 1.34 | 1.2 | 1.2 | 1.23 | 1.44 | 1.93 | … | … | … | … | … | … | … | … | 2.5 | … |
| 61 | Pm | … | … | … | … | … | … | … | … | … | … | … | … | … | … | … | … | … | … | … | … | … | … | … |
| 62 | Sm | 1.01 | 1.01 | … | 1.01 | 1.01 | 1.01 | 1.00 | 0.80 | 0.7 | 0.7 | 0.72 | 0.91 | 1.62 | … | … | … | … | … | … | … | … | 2 | … |
| 63 | Eu | 0.51 | 0.51 | … | 0.51 | 0.51 | 0.51 | 0.51 | 0.51 | 0.7 | 0.7 | 0.7 | 0.51 | 0.96 | … | … | … | … | … | … | … | … | 1.9 | … |
| 64 | Gd | 1.12 | 1.12 | … | 1.12 | 1.12 | 1.12 | 1.12 | 1.12 | 1.1 | 1.1 | 1.12 | 1.15 | 1.13 | … | … | … | … | … | … | … | … | 1.6 | … |
| 65 | Tb | -0.1 | -0.1 | … | -0.1 | -0.1 | -0.1 | -0.1 | 0.2 | … | … | … | 0.42 | … | … | … | … | … | … | … | … | … | … | … |
| 66 | Dy | 1.14 | 1.14 | … | 1.14 | 1.14 | 1.14 | 1.10 | 1.10 | 1.1 | 1.1 | 1.06 | 1.11 | 1 | … | … | … | … | … | … | … | … | 2.1 | … |
| 67 | Ho | 0.26 | 0.26 | … | 0.26 | 0.26 | 0.26 | 0.26 | 0.26 | … | … | … | 0.5 | … | … | … | … | … | … | … | … | … | … | … |
| 68 | Er | 0.93 | 0.93 | … | 0.93 | 0.93 | 0.93 | 0.93 | 0.93 | 0.8 | 0.8 | 0.76 | 0.89 | … | … | … | … | … | … | … | … | … | 0.6 | … |
| 69 | Tm | 0.00 | 0.00 | … | 0.00 | 0.00 | 0.00 | 0.00 | 0.00 | 0.3 | 0.3 | 0.26 | 0.09 | … | … | … | … | … | … | … | … | … | … | … |
| 70 | Yb | 1.08 | 1.08 | … | 1.08 | 1.08 | 1.08 | 1.08 | 1.08 | 0.8 | 0.2 | 0.9 | 0.87 | 1.53 | 2.33(I), 0.49(II) | … | … | … | … | … | … | … | … | … |
| 71 | Lu | 0.06 | 0.06 | … | 0.06 | 0.76 | 0.76 | 0.76 | 0.76 | 0.8 | 0.8 | 0.76 | 0.09 | … | … | … | … | … | … | … | … | … | … | … |
| 72 | Hf | 0.88 | 0.88 | … | 0.88 | 0.88 | 0.88 | 0.88 | 0.88 | 0.9 | 0.9 | 0.8 | … | … | … | … | … | … | … | … | … | … | 0.9 | … |
| 73 | Ta | … | … | … | … | … | … | … | … | … | … | … | … | … | … | … | … | … | … | … | … | … | … | … |
| 74 | W | 1.11 | 1.11 | … | 1.11 | 1.11 | 1.11 | 1.11 | 1.11 | 0.8 | 0.8 | 1.7 | … | … | 2.57 | … | … | … | … | … | … | … | 0.7 | … |
| 75 | Re | | … | … | … | … | … | … | … | <-0.3 | <-0.3 | <-0.3 | 0.25 | … | … | … | … | … | … | … | … | … | … | … |
| 76 | Os | 1.45 | 1.45 | … | 1.45 | 1.45 | 1.45 | 1.45 | 1.45 | 0.7 | 0.7 | 0.7 | 1.34 | … | 0.75 | … | … | … | … | … | … | … | … | … |
| 77 | Ir | 1.35 | 1.35 | … | 1.35 | 1.35 | 1.35 | 1.35 | 1.35 | 2.2 | 0.8 | 0.85 | 1.24 | … | 2.21 | … | … | … | … | … | … | … | 0.3 | … |
| 78 | Pt | 1.80 | 1.80 | … | 1.80 | 1.80 | 1.80 | 1.80 | 1.80 | 1.8 | 1.8 | 1.75 | 1.75 | … | … | … | … | … | … | … | … | … | 2.1 | … |
| 79 | Au | 1.01 | 1.01 | … | 1.01 | 1.01 | 1.01 | 1.01 | 1.13 | 0.1 | 0.8 | 0.75 | 0.91 | … | 0.8 | … | … | … | … | … | … | … | … | … |
| 80 | Hg | | … | … | … | … | … | … | … | <2.1 | <2.1 | <2.1 | 2.61 | … | 1.13? | … | … | … | … | … | … | … | … | … |
| 81 | Tl | 0.90 | 0.90 | … | 0.90 | 0.90 | 0.90 | 0.90 | 0.90 | 0.9 | 0.9 | 0.9 | 0.81 | … | … | … | … | … | … | … | … | … | 1.9 | … |
| 82 | Pb | 2.00 | 1.95 | … | 1.95 | 1.95 | 1.95 | 1.85 | 1.9 | 1.9 | 1.9 | 1.93 | 1.75 | 1.63 | 1.66 | 1.63 | 1.33 | … | 2.04 | … | … | … | 1.7 | … |
| 83 | Bi | … | … | … | … | … | … | … | … | <1.9 | <1.9 | <1.9 | 0.78 | … | 0.81 | … | … | … | … | … | … | … | … | … |
| 90 | Th | 0.12 | … | … | … | … | … | 0.12 | 0.02 | 0.2 | 0.2 | 0.2 | -0.02 | … | … | … | … | … | … | … | … | … | … | … |
| 92 | U | < -0.47 | < -0.47 | … | <-0.47 | <-0.47 | <-0.47 | <-0.47 | <-0.47 | <0.6 | <0.6 | <0.6 | -0.5 | … | … | … | … | … | … | … | … | … | … | … |

**Footnote for Table 4:** Abundances from photospheric data unless indicated: "in" = indirect determination. "sp" sunspot.
Helium from helioseismology, see text. Mass fractions for recommended 3D composition: X = 0.7389 (Basu & Antia 2004), Y = 0.2462 for Z = 0.0149 and Z/X = 0.0201. Mass fractions for recommended 1D composition: X = 0.7389 (Basu & Antia 2004), Y = 0.2458 for Z = 0.0153 and Z/X = 0.0207. Neon estimated from solar wind and coronal Ne/O ratios from Young (2018) using oxygen abundances recommended here, Ne abundance is still uncertain and debatable. Argon from solar wind, B-stars and other considerations, see Lodders (2008). Krypton and Xenon estimated from nuclear systematics, see text. Recommended 1D: largely from SL+, SSG15, Z16. Recommended 3D largely from C11 and SSG15, plus Amarsi et al. 2018 (O), Caffau et al. 2015 (O), Amarsi & Asplund 2017 (Si), Armarsi et al. 2019 (C), Caffau et al. 2011 (C ), Bergemann et al. 2017 (Mg), -- 2019 (Mn), Lind et al. 2017 (Fe), Nordlander et al. 2017 (Al). Recommended 1D largely from H01, SSG15, Z16, Alexeeva & Mashonkina 2015 (C). Sunspot data: Maas et al. 2016 (Cl), Maiorca et al. 2014 (F), Vitas et al. 2008 (In).
Reference codes in the Table: [AG89] Anders & Grevesse 1989. [AGS05] Asplund, Grevesse, Sauval 2005. [AGSS09] Asplund et al. 2009. [C11] Caffau et a. 2011. [C51] Claas 1951. [E77] Envold 1977. [G84] Grevesse 1984. [GAS07] Grevesse et al. 2007. [GBB68] Grevesse et al. 1968. [GMA60] Goldberg et al. 1961. [GNS96] Grevesse, et al. 1996. [GS02] Grevesse & Sauval 2002. [GS93] Grevesse & Noels 1993. [GS98] Grevesse & Sauval 1998. [H01] Holweger 2001. [L03] Lodders 2003. [LPG09] Lodders et al. 2009. [M65] Müller 1965. [M68] Müller 1968. [Me43] Menzel 1943. [P76] Pagel 1979. [PJ03] Palme & Jones 2003. [PLJ14] Palme et al. 2014. [P25] Payne 1925a,b, 1926; stellar abundances. [R29] Russell 1929. [RA76] Ross &Aller 1976. [SSG15]Scott et al. 2015a,b, Grevesse et al. 2015. [SL+] Sneden et al. 2009, 2016, Lawler et al. 2014, 2017, 2018, 2019. [U45] Unsöld 1945. [U47] Unsöld 1947. [U48] Unsöld 1948. [W71] Withbroe 1971. [Z16] Zhao et al. 2016.

## Coronal Abundances

The chromosphere is the transition zone between the photosphere (with a brightness temperature of around 5800K) and the tenuous outer corona (with a kinetic gas temperature around 2 million K). The chromosphere has a large temperature gradient and is heterogeneous. Chromospheric emission spectra are obtained during solar eclipses show the presence of neutral and singly-ionized atoms, whereas quiet coronal emissions show the forbidden lines of highly ionized atoms in the visible regions of the spectra, and permitted lines in the extreme ultra violet (EUV) and X-ray wavelengths. Analyses of chromospheric plasma spectra require r atomic properties, including ionization equilibria and deviations from them, electron densities, and line excitation mechanisms. Elemental abundances for the corona are in Table 5.

Coronal abundances are not easy to quantify because atomic data for radiative transition probabilities and cross-sections for excitation, ionization and recombination processes must be available from laboratory measurements. First, coronal abundances were derived from observations of forbidden lines and later from extreme ultraviolet observations. The first spectroscopic analyses for 20 elements in the solar corona was achieved by Menzel (1931), later abundances for the corona from optical observations are by e.g., Woolley & Allen (1948), Pottasch (1964a); see Feldman & Widing (2003) for more studies. In 1946 the US Naval Research Laboratory mounted a spectrograph on the first V-2 rocket launched by the US to obtain the first EUV spectra of the sun (see Trousey 1967 for a review of early solar UV research). Subsequently Pottasch (1964b,c,1967) analyzed solar EUV spectra and found iron was 40 times higher in the corona than in the photosphere but closer to meteoritic values, which was one of the indications of the erroneous photospheric iron abundance (see above).

EUV and X-ray spectroscopy with instruments on board of several spacecrafts recently provided measurements of coronal matter in solar flares, e.g., the SOHO Coronal Diagnostic Spectrometer (CDS, Mason et al., 1999), the Hinode EUV Imaging Spectrometer (EIS, O'Dwyer et al. 2011), the Solar X-ray spectra from the REntgenovsky Spektrometr s Izognutymi Kristalami (RESIK) crystal spectrometer on the CORONAS-F spacecraft (Sylvester et al. 2015) and Solar Assembly for X-rays (SAX) on the Mercury MESSENGER spacecraft (Dennis et al. 2015). For EUV and X-ray spectroscopic observations of the sun by an armada of spacecraft see Milligan & Ireland (2018).

## Solar Wind Abundances

The solar wind stems from mass ejections originating from the photosphere and the corona. The solar wind is an interplanetary presence of solar material and the term heliosphere is used to describe where solar wind and the solar magnetic field dominate over the interstellar particle flow (i.e., galactic cosmic rays) and the interstellar magnetic field. The solar wind as corpuscular irradiation was first recognized in order to explain why cometary tails always flow outwards with motion perpendicular from the Sun's sphere (Biermann 1951, Parker 1958). Implanted solar wind was first detected in noble-gas rich meteorites, later in lunar rocks and soil that had been exposed to the solar wind, and directly captured in the foils posted on the lunar surface during the Apollo missions (Geiss et al. 2004). There are limitations to obtain abundances for many elements from surface-exposed rocks due to the ubiquitous presence of the same elements in the lunar and meteoritic rocks themselves. Useful elemental and isotope data for solar wind can only be obtained for the noble gases, and nitrogen which are typically not part of indigenous minerals in meteoritic and lunar rocks (see Pillinger 1979 for a review). The foil experiments may not suffer from the problems of ascribing an element occurrence to the solar wind, however, the exposure times of the foils were relatively short and only major abundant elements could be collected. Space missions provided in-situ measurements of the solar wind compositions in the heliosphere with energy/charge spectrometers, these include the Advanced Composition Explorer (ACE) and Ulysses (see, e.g., Gloeckler et al., 1998, Gloeckler & Geiss 2007). Elemental abundances from these experiments are given in Table 5. The most ambitious mission with sample return and laboratory analyses of solar wind samples is the Genesis mission (Burnett et al. 2011,2019). The analyses have already provided well characterized noble gas abundances (e.g., Crowther & Gilmour 2013, Heber et al. 2009, 2012, Pepin et al. 2009, 2012, Vogel et al. 2011, Meshik et al. 2014) and C,N, and O abundances (e.g., Heber et al. 2013, Laming et al. 2017) whereas the analyses of rock-forming elements (e.g., Jurewicz et al. 2011, 2019, Heber et al. 2014, Rieck et al. 2016, Burnett et al. 2017, 2019, Westphal et al. 2019) are ongoing and for these preliminary abundances are listed in Table 5.

Figures 3a,b show coronal and solar wind abundances. The solar wind measurements pertain to ejected solar matter. The solar wind consists of a plasma with highly ionized atoms. The ionization state changes as plasma is heated and accelerated, and slow and fast plasma streamers, and coronal mass ejections develop. The slow and fast solar wind and solar energetic particles are related to these phenomena. The elemental and isotopic abundances in the corona and solar wind are fractionated from photospheric abundances due to hydromagnetic processes that depend on the charge state or charge-to-mass ratios of individual atoms.

The fractionation occurs when plasma flows from the photosphere to the chromosphere into the heliosphere and appears to be a function of the first ionization potential (FIP) of the elements. As shown in Figure 3a, elements with low FIP (below 10 eV, including Na, K, Mg, Al, Si, Fe) have more or less uniformly higher abundances in coronal and solar wind matter than observed in the photosphere, in contrast to elements with high FIP above 10eV, including C, N, O, noble gases (e.g., Reisenfeld et al. 2013, von Steiger & Zurbuchen 2016). Fast solar wind originates from coronal holes and is closer to photospheric abundances but elements with low and high FIP are still fractionated, but not as much as the slow solar wind which makes it difficult to use these solar wind abundances to derive a representative solar composition.

More energetic forms (> 1 MeV/nucleon) of the solar wind present within the interplanetary medium of the heliosphere are so-called Solar Energetic Particles, SEPs, which occur as either “gradual” or “impulsive” types (Desai & Giacalone 2016). The gradual SEPs are created when the slow solar wind is accelerated by shocks from coronal mass ejections as matter passes through the corona. These SEP abundances are sometimes taken as representative for the corona. In contrast, impulsive solar flares or impulsive SEP events resulting from flare events, are compositionally more variable and could be regarded as extreme forms of the solar wind. Elemental composition for the SEPs have been measured e.g., with the cosmic ray subsystem (CRS) aboard the Voyager 1 and 2 spacecraft (see, Desai & Giacalone 2016, Reames 2018). Interplanetary Coronal Mass Ejections (ICME) are highly energetic events and solar material in them is highly fractionated from photospheric abundances. Zurbuchen et al. (2016) analyzed data from the Solar Wind Ion Composition Spectrometer (SWICS) on board the Advanced Composition Explorer (ACE). Elemental abundances for these solar wind components are included in Table 5.

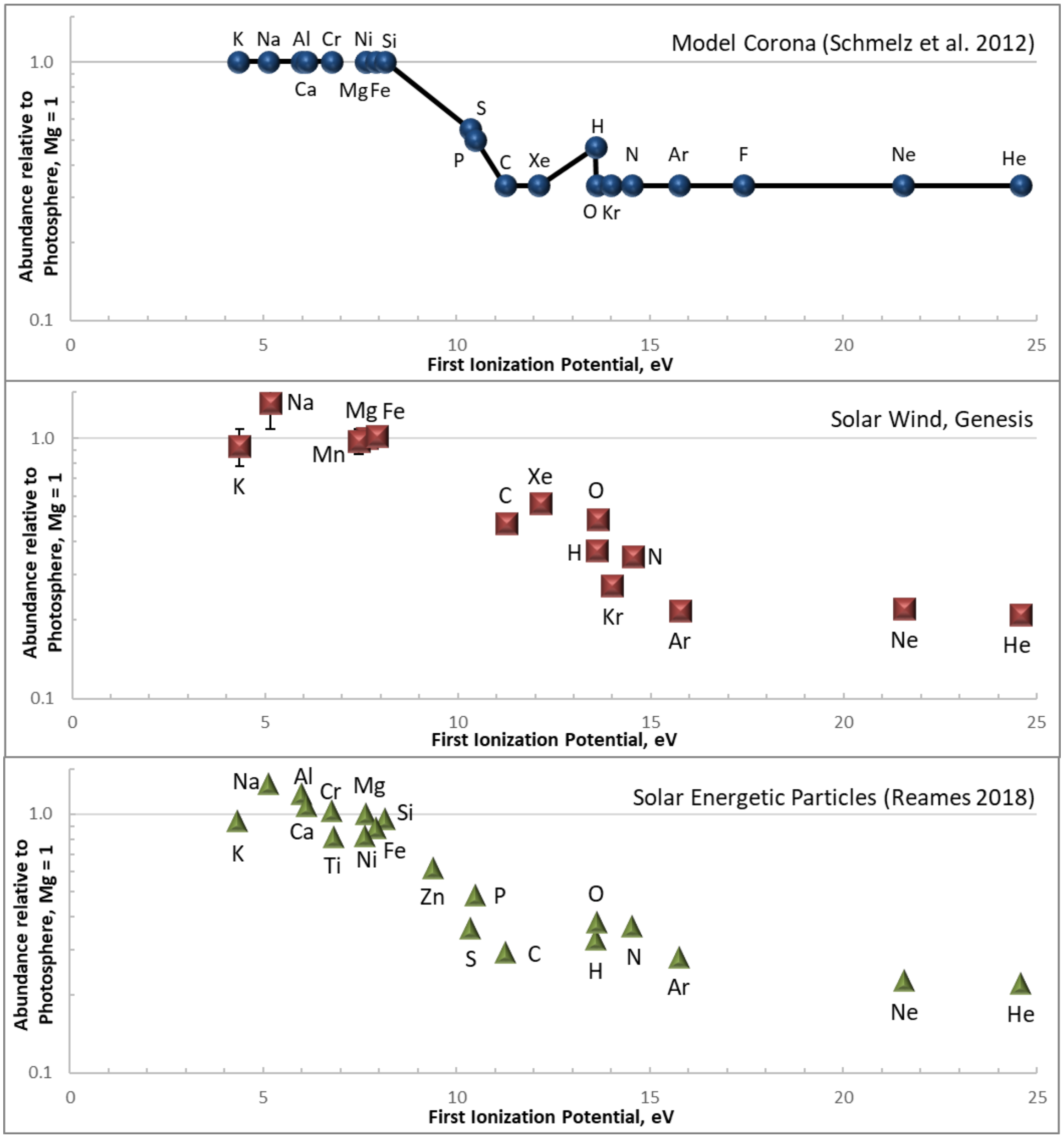

**Figure 3a**. Elemental abundances in the model corona (Schmelz et al. 2012), the solar wind as measured by the Genesis mission (see Table 5 for detailed references) and in solar energetic particles, SEPs (Reames 2018) normalized to Mg and photospheric abundances (3D from Table 4), as a function of the first ionization potential (FIP) of the elements. Error bars include the uncertainties from the photospheric element/Mg ratios used for normalization. See Figure 3b for the same data plotted as a function of first ionization time and text for explanations.

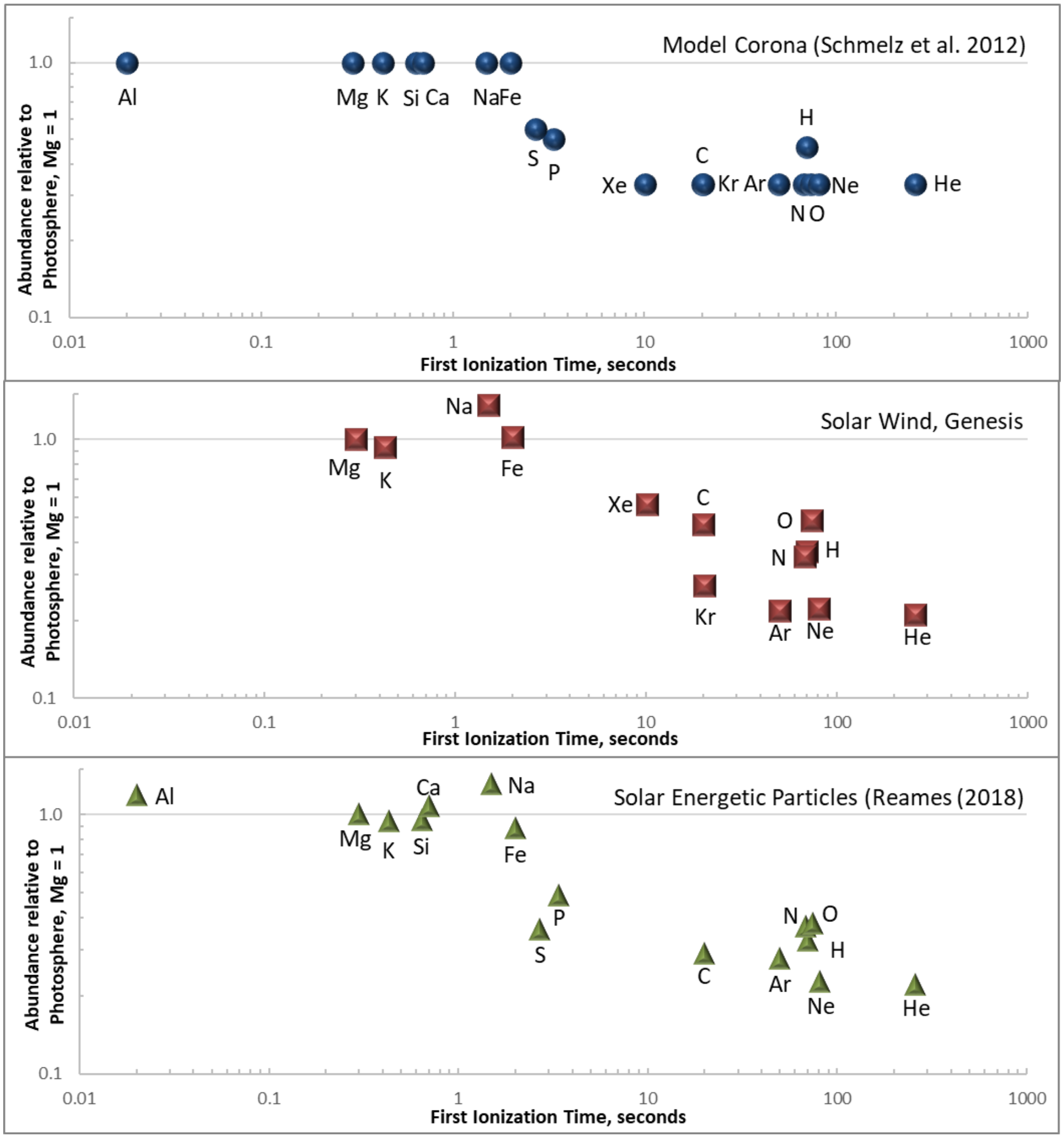

**Figure 3b**. Same as Figure 3a, but normalized abundances are plotted versus the First Ionization Time (FIT). The FIT data are from Marsch et al. (1995), Geiss (1998) and Giammenco et al. (2008); note that FIT values for Mg and Fe were estimated based on FIP.

Enrichments of elements with low FIP may arise from Alfenic waves causing ponderomotive forces (forces in spatially inhomogeneous, electro-magnetic fields acting on moving charged particles) that accelerate chromospheric ions (see, e.g. Laming 2015, Dahlburg et al. 2016, Reames 2018). However, the compositional differences between the fast and slow solar wind are not yet fully understood with existing models on the origins of the various solar wind regimes. In addition to the dependence on FIP, there is an apparent correlation of the elemental enrichments with First Ionization Time (Geiss & Bochsler 1986, Widing & Feldman 2001).

Elements with low-FIP are 2-4 times more abundant in the corona than in the photosphere whereas elements with high FIP have comparable abundances when the ratio corona to photosphere is also normalized to a high-FIP element such as H, or O. Conversely, when the abundance ratio is normalized to a low FIP element (such as Mg as done in Figure 3a,b), all low-FIP elements have close to photospheric abundances around unity whereas the high FIP elements are below unity indicating an apparent depletion. The "hybrid" abundances for the corona from Schmelz et al. (2012) are modelled by assuming low FIP elements (Mg, Fe etc.) have essentially constant element/hydrogen ratios of 2.14 times the photospheric ratios, and high FIP elements (C,N,O, noble gases) ratios of 0.71 respectively. Elements with intermediate FIP (P, S, Zn) are in-between. This approach places hydrogen, a high FIP element, at unity (the photospheric value); above all other high FIP elements at 0.71 times the photospheric value by Schmelz et al. (2012). This is peculiar because H and O have similar FIPs, and therefore H and O should show similar abundances relative to the photosphere if the FIP is the underlying cause for the element fractionations between the photosphere and the corona. The coronal abundances by Feldman & Widing (2003) place the low FIP elements at 2.1 times the photospheric abundance ratios, and the high FIP elements, including hydrogen, at the photospheric level.

The FIP may not be the main governing parameter and the First Ionization Time (FIT) may be more important. The FIT describes the time after which 50% of an element has reached a certain ionization stage and the effectiveness with which atoms can be fractionated in the competition between photoionization and dynamic atom-ion processes. It is not a fixed atomic property but a model-dependent and time-variable convolution of the ionization cross-section with the solar EUV spectrum (e.g., Geiss et al. 1994, Geiss 1998, von Steiger & Schwadron 2000). For example, the FIT might be a better ordering parameter for elements with similar FIP, if they show abundance fractionations. The same data from Figure 3a are plotted as a function of FIT in Figure 3b (see also Vogel et al. (2011) for similar figures for the solar wind). Overall, the two Figures show similar trends, and qualitatively, the correlation with FIP is more like a step-function than a steady correlation with FIT. The correlation with FIT shows more scatter which is in part due to the difficulty in deriving estimates for the FIT, which can vary greatly (see Marsch et al. 1995, Geiss 1998, Giammenco et al. 2008). The uncertainties in some abundance measurements remain relatively large for various solar wind regimes (see e.g., similar diagrams in Vogel et al. 2011) but also for photospheric values used for normalization (Table 4). Overall, the existing and anticipated high-quality results of the Genesis mission should help to clarify the issue whether FIP, FIT or yet another mechanism is a key parameter of the fractionation of the solar wind from photospheric material.

**Table 5. Elemental Abundances in the Solar Corona and the Solar Wind Normalized to Mg = 100 Atoms**

| Z | E | FIP, eV | Photo-sphere [Table 4] | Corona (model) [Sch12] | Corona [M85ab] | Corona [W71] | Corona (optical) [FW03] | Corona (EUV) [W75] | Corona (X-ray) [VP81] | Gradual SEP [R18] | Gradual SEP [M85ab] | $^3$He-rich SEP [Mas04] | Slow Solar Wind [SSWave] | Fast Solar Wind FSWave] | Coronal Mass Ejections [CMEave] | Solar Wind [SWBulk] | Solar Wind [Genesis] |
|---|---|---|---|---|---|---|---|---|---|---|---|---|---|---|---|---|---|
| 2 | H | 13.598 | 2.75E+6 | 1.35E+6 | 2.68E+6 | 2.24E+6 | * | 3.16E+6 | 2.24E+6 | 1.00E+6 | … | … | 1.68E+6 | 1.81E+6 | 9.76E+5 | 1.51E+6 | 1.02E+6 |
| 2 | He | 24.587 | 2.34E+5 | 8.05E+4 | 2.63E+5 | … | * | … | … | 5.11E+4 | 3.09E+4 | 1.46E+4 | (6.06±0.03)E+4 | (7.10±0.03)E+4 | (4.79±0.09)E+4 | 5.94E+4 | (4.87±0.27)E+4 |
| 6 | C | 11.260 | 813 | 303 | 632 | 1260 | * | 1200 | … | 236 | 236 | 87 | 450±49 | 635±27 | 303±7 | 428 | 383±19 |
| 7 | N | 14.534 | 195 | 69 | 105 | 251 | * | 141 | … | 72 | 66 | 35 | 71±25 | 126±13 | 74±4 | 103 | 69±5 |
| 8 | O | 13.618 | 1413 | 514 | 663 | 1259 | * | 660 | 631 | 562 | 528 | 270 | 693±49 | 918±50 | 531±22 | 687 | 723±47 |
| 10 | Ne | 21.565 | 389 | 107 | 95 | 79 | * | 51 | 100 | 88 | 68 | 70 | 77±13 | 96±34 | 88 | 75 | 86±5 |
| 11 | Na | 5.139 | 4.5 | 5.8 | 7.4 | 6.3 | 5.5 | 5.8 | 7.1 | 5.8 | 6.9 | … | 6.1 | 4.8 | … | … | 6.1±1.2 |
| 12 | Mg | 7.646 | 100 | 100 | 100 | 100 | 100 | 100 | 100 | 100 | 100 | 100 | 100 | 100 | 100 | 100 | 100±6 |
| 13 | Al | 5.986 | 7.4 | 8.5 | 7.4 | 6.3 | 8.1 | 5.6 | … | 8.8 | 7.2 | … | 8.1 | 7.6 | … | … | … |
| 14 | Si | 8.152 | 89 | 95 | 105 | 100 | 95 | 100 | 105 | 85 | 81 | 111 | 109±13 | 104±27 | 94±3 | 118 | … |
| 15 | P | 10.487 | 0.71 | 0.4 | … | 0.60 | * | 0.8 | … | 0.37 | … | 1 | 0.95 | … | … | … | … |
| 16 | S | 10.360 | 39 | 23 | 23 | 32 | * | 35 | 18 | 14 | 16 | 32 | 39±6 | 41±11 | 35±2 | 38 | … |
| 18 | Ar | 15.760 | 8.7 | 3.0 | 5.7 | 13 | * | 17 | 5.4 | 2.4 | 3.1 | … | 2.1 | 2.3±0.8 | … | … | 1.90±0.09 |
| 19 | K | 4.341 | 0.33 | 0.37 | … | 1.3 | * | … | … | 0.31 | … | … | … | … | … | … | 0.31±0.05 |
| 20 | Ca | 6.113 | 5.8 | 6.2 | 7.9 | 5.6 | 5.9 | 3 | 7.2 | 6.2 | 6.2 | 16 | 5.5 | 5.7±1 | … | … | … |
| 24 | Cr | 6.767 | 1.15 | 1.26 | … | 1.60 | * | … | … | 1.2 | 1.8 | … | 1.4 | 1.4 | … | … | … |
| 25 | Mn | 7.434 | 0.91 | 0.7 | … | 0.80 | * | … | … | … | … | … | … | … | … | … | 0.89±0.10 |
| 26 | Fe | 7.902 | 83 | 87 | 105 | 105 | 83 | 85 | 100 | 74 | 80 | 260 | 90±15 | 73±10 | 85±9 | 88 | 85±6 |
| 28 | Ni | 7.640 | 4.40 | 4.9 | 5.8 | 7.6 | 4.7 | 6.3 | 5.6 | 3.6 | 3.7 | … | 4.4 | … | … | … | … |
| 36 | Kr | 14.000 | 4.57E-3 | 1.82E-3 | … | … | … | … | … | … | … | … | … | … | … | … | (1.24±0.10)E-3 |
| 54 | Xe | 12.130 | 4.90E-4 | 1.78E-4 | … | … | … | … | … | … | … | … | … | … | … | … | (2.76±0.15)E-4 |

**Footnotes to Table 5:** [FW03] Feldman & Widing 2003, optical spectra/forbidden lines. Entries marked * indicate that photospheric abundances were assumed in their table. [M85ab] Meyer 1985a,b as recommended by the author. [Mas04] Mason et al. 2004. $^3$He-rich SEP, impulsive flares. [R18] Reames 2018, gradual SEP. [Sch12] Schmelz et al. 2012. Calculated from the adopted photospheric composition here and their relative enrichments/depletion factors, see text. [VP81] Veck & Parkinson 1981, OSO-8 X-ray spectrometer. [W71] Withbroe 1971 as recommended by the author. [W75] Withbroe 1975, EUV spectra. [CMEave] Coronal Mass Ejections from ACE spacecraft measurements, uncertainty given here probably mainly reflects real temporal variations, Reisenfeld et al. 2013, Zurbuchen et al. 2016. [FSWave] fast solar wind average (originating from coronal holes) from ACE and Ulysses spacecraft measurements, uncertainty given here probably mainly reflects real temporal variations, Bochsler 2007, 2009, Gloeckler & Geiss 2007, von Steiger et al. 2000, von Steiger & Zurbuchen 2016, Reisenfeld et al. 2013. [SSWave] Slow solar wind (interstream) average from ACE and Ulysses spacecraft measurements, uncertainty given probably mainly reflects real temporal variations. Bochsler 2007, 2009, von Steiger et al. 2000, von Steiger & Zurbuchen 2016, Reisenfeld et al. 2013, Pilleri et al. 2015. [SWBulk] Bulk solar wind, Pilleri et al. 2015, Reisenfeld et al. 2013; see comment above. [Genesis] Burnett et al. 2017, Rieck et al. 2019 (Mg) Heber et al. 2013 (C,N,O). Jurewicz et al. 2011, 2019 (Fe, Mg). Koeman-Shields et al. 2016 (H). Crowther & Gilmour 2013, Heber et al. 2009, Meshik et al. 2014, Vogel et al. 2011, (Ar, Kr, Xe). Pepin et al. 2009, 2012 (Ne, Ar). Rieck et al. 2016, (Na,K). Reisenfeld et al. 2013 (H, He). Westphal et al. 2019 (Mn).

## Other Sources for Solar System Abundances

For some elements, obtaining their abundances by measurements of meteorites or solar environs remain elusive. The noble gases are highly depleted in meteorites and geochemical analyses from meteorites cannot provide their abundances.

Solar measurements are problematic because spectral lines are absent, weak, or very difficult to model, and physical fractionations as noted above for the solar wind make abundance data difficult to interpret without any baseline for reference as described above. For helium, results from helioseismology are used as described above.

For neon and argon, spectroscopic observations of the much hotter B-type dwarf stars can provide abundances. These stars have similar relative abundances for elements observable in the solar photosphere and therefore, the Ne and Ar abundances can be derived by analogy. However, B-stars are younger than the sun and they could have formed from a different mix of elements because of continuing nucleosynthesis in stars and element release into the interstellar medium. Galactic chemical evolution models consider yields of nucleosynthesis models and for elements with similar production mechanism, the element abundance ratios should vary in step (just like elements with similar chemical properties fractionate in a similar manner during chemical processing so that element abundance ratios remain about constant).

Another approach to estimate elemental abundances is to consider elemental and isotopic variations in the solar abundance distributions. This type of work is directly related to understanding nuclear structure and nucleosynthetic mechanisms, which was one of the reasons why researchers began to look at elemental abundances to begin with. Starting with Oddo's (1914) and Harkins' (1917) discovery that elements with odd atomic numbers are less abundant than their even-numbered neighbors, better abundance analyses eventually revealed well-defined structures in elemental abundances, and later in the isotopic abundance distribution curves (e.g., Goldschmidt 1937). Figure 4, for which the data are described in the following section, shows the abundance curve of the elements as a function of atomic number and illustrates Harkin's discovery. This, and more importantly, the isotopic composition of the element and isotopic (or nuclide) distributions led to the recognition of nuclide abundance rules by Suess (1947a,b) which then enabled him to predict nuclide abundances, and elemental abundances. As already done by Goldschmidt (1937), the abundances of Kr and Xe still rely on interpolations from abundances of their neighboring elements and nuclides, but are based on much better understood underlying physics about nuclear structure and nucleosynthesis (see, e.g., Cameron 1982, Anders & Grevesse 1989, Palme & Beer 1993, Lodders 2003). Earlier abundance compilations often used graphical interpolations between neighboring elements (or isotopes) to obtain an abundance estimate for "missing" elements (or isotopes) from plots of elemental abundances vs atomic number (such as Figure 4) or isotope abundances as a function of mass number (such as Figure 5). When employing this method, the best approach is to only use the abundances for isotopes with odd mass numbers because there is usually only one odd-numbered isotope per mass number, and the abundance curve for odd-numbered isotopes is particularly smooth as noted long ago (Suess 1947a,b). If the isotopic composition of an element is known, and the abundance of one (or more) of its isotopes can be estimated from the abundance curve, the total elemental abundance is obtained. The same principle, based on better underlying physics, uses nucleosynthesis network calculations to predict the relative contributions from the different nucleosynthetic production process operating in stellar sources. Most successful abundance estimates come from models for the slow-neutron capture process ("s-process") that operates in evolved low to intermediate mass stars (2-5 solar masses), and produces many isotopes of the elements heavier than iron (e.g., Arlandini et al. 1999, Sneden et al. 2008, Bisterzo et al. 2017). If an element has one or more isotopes only or largely produced by the s-process (which is the case for e.g., $^{82}$Kr and $^{128,130}$Xe), the calculated isotopic abundances from the s-process network can serve to obtain the total elemental abundance, again if the overall isotopic composition of the element is known. Because there are only elements heavier than Fe in the s-process, the anchoring to the solar composition scale relative to $10^6$ silicon atoms is usually done via the s-process-only isotope $^{150}$Sm, because its abundance is independently known on the solar abundance scale. Re-normalizing the computed s-process abundances to the same $^{150}$Sm abundance value on the existing abundance curve gives the means to derive the "missing" elements. This modelling with the "main" s-process yields works well for the heavier elements such as Xe. For the lighter elements such as Kr, additional contributions from the weak s-process operating in more massive stars must be considered (e.g., see reviews by Busso et al. 2999, Sneden et al. 2008). The recommended abundances of Kr and Xe in Tables 4,6, and 8 are based on interpolations with the elemental nuclide abundances resulting from the adopted compositions and s-process systematics, and from interpolation of their odd-numbered nuclides in the overall distribution curve for odd numbered nuclides.

## Solar System Abundances

The proto-solar, or solar system abundances were traditionally derived from photospheric, meteoritic, and for some elements, theoretical considerations. Table 6 lists the solar system abundances published over time on a scale relative to $10^6$ silicon atoms. Sometimes these abundances are referred to as "solar", "cosmic", or "local galactic". In many cases the data are derived from several sources as indicated in the column headers "ph" for photosphere and "mets" for meteorite data. An "o" means other considerations than meteoritic or solar abundances were included to larger extent.

An important difference between "present" and "proto-solar" results from element diffusion and gravitational settling from the solar convection zone which lowered the abundances of elements heavier than H (see above and Table 7). Radioactive decay of "long-lived" isotopes (which are still present since the time the oldest solids formed in the solar nebula 4.567 Ga ago) decreased abundances of the radioactive parent isotopes and increased abundances of the daughter isotopes which has (usually small) effects on the total elemental abundances. Table 6 therefore distinguishes the point in time the abundances are relevant ("present" or "proto-solar"). However, the element settling from the outer convection zone was only considered since the late 1980s for He and for heavy elements since the late 1990s, whereas older compilations do not distinguish solar photospheric from proto-solar composition (except for Li and isotopes of H and He).

**Table 6. Solar System Abundances Relative to Si= $10^6$ atoms**
Part 1.

| Z | E | Proto-Solar Sun+CI+o Recomm. | ±σ | Present Sun+CI+o Recomm. | ±σ | CI-chond. [PLJ14] | Present Sun+CI+o [LPG09] | Proto-Solar Sun+CI+o [L03] | Present Sun+CI+o [L03] | CI-chond. [PJ03] | CI-chond [PB93] | Present Sun+CI+o [AG89] | Present Sun+CI+o [P88] | Present Sun [M85b] | Present ph+CI+o [AE82] | Proto-solar ph+CI+o [Cam82] | Present ph+CI+o [PSZ81] | Proto-solar ph+CI+o [H79] | Chond. [SZ73] |
|---|---|---|---|---|---|---|---|---|---|---|---|---|---|---|---|---|---|---|---|
| 1 | H | 2.52E+10 | | 3.09E+10 | | [5.13E+6] | 2.93E+10 | 2.431E+10 | 2.884E+10 | [6.27E+6] | [6.27E+6] | 2.79E+10 | 2.82E+10 | 2.71E+10 | 2.72E+10 | 2.66E+10 | 2.50E+10 | 3.10E+10 | … |
| 2 | He | 2.51E+9 | 1.2E+8 | 2.59E+9 | 1.2E+8 | [0.601] | 2.47E+9 | 2.343E+9 | 2.288E+9 | | | 2.72E+9 | 2.82E+9 | 2.60E+9 | 2.18E+9 | 1.80E+9 | 2.00E+9 | 3.10E+9 | … |
| 3 | Li | 56.9 | 3.4 | 0.339 (sun) | 0.088 | 54.8 | 55.6 | 55.47 | 55.47 | 56.5 | 56.5 | 57.1 | 55.0 | | 60.0 | 60.0 | 55 | 60.0 | … |
| 4 | Be | 0.637 | 0.044 | 0.637 | 0.044 | 0.638 | 0.612 | 0.7374 | 0.7374 | 0.727 | 0.727 | 0.73 | 0.73 | | 0.78 | 1.2 | 0.73 | 0.81 | … |
| 5 | B | 18.0 | 1.3 | 18.0 | 1.3 | 18.8 | 18.8 | 17.32 | 17.32 | 16.8 | 21.2 | 21.2 | 6.6 | | 24 | 9.0 | 6.6 | 44.0 | … |
| 6 | C | 9.12E+6 | 1.80E+5 | 9.12E+6 | 1.80E+5 | [7.60E+5] | 7.19E+6 | 7.079E+6 | 7.079E+6 | [7.05E+5] | [7.05E+5] | 1.01E+7 | 1.38E+7 | 1.26E+7 | 1.21E+7 | 1.11E+7 | 7.9E+6 | 1.50E+7 | … |
| 7 | N | 2.19E+6 | 9700 | 2.19E+6 | 9700 | [5.53E+4] | 2.12E+6 | 1.950E+6 | 1.950E+6 | [5.97E+4] | [5.97E+4] | 3.13E+6 | 2.75E+6 | 2.25E+6 | 2.48E+6 | 2.31E+6 | 2.1E+6 | 3.10E+6 | … |
| 8 | O | 1.66E+7 | 1.30E+5 | 1.66E+7 | 1.30E+5 | [7.53E6] | 1.57E+7 | 1.413E+7 | 1.413E+7 | [7.64E+6] | [7.64E+6] | 2.38E+7 | 2.29E+7 | 2.25E+7 | 2.01E+7 | 1.84E+7 | 1.7E+7 | 2.60E+7 | … |
| 9 | F | 1270 | 275 | 1270 | 275 | 804 | 804 | 841.1 | 841.1 | 806 | [59700] | 843 | 710 | 930 | 843 | 780 | 710 | 1000 | … |
| 10 | Ne | 4.37E+6 | 1.13E+6 | 4.37E+6 | 1.13E+6 | [0.0023] | 3.29E+6 | 2.148E+6 | 2.148E+6 | … | … | 3.44E+6 | 2.82E+6 | 3.25E+6 | 3.76E+6 | 2.60E+6 | 2.50E+6 | 1.60E+6 | … |
| 11 | Na | 57800 | 4700 | 57800 | 4700 | 5.67E+4 | 5.77E+4 | 5.751E+4 | 5.751E+4 | 5.70E+4 | 5.70E+4 | 5.74E+4 | 5.70E+4 | 5.50E+4 | 5.70E+4 | 6.00E+4 | 5.7E+4 | 6.00E+4 | … |
| 12 | Mg | 1.03E+6 | 4.40E+4 | 1.03E+6 | 4.40E+4 | 1.03E+6 | 1.03E+6 | 1.020E+6 | 1.020E+6 | 1.04E+6 | 1.04E+6 | 1.07E+6 | 1.01E+6 | 1.05E+6 | 1.08E+6 | 1.06E+6 | 1.01E+6 | 1.06E+6 | … |
| 13 | Al | 81820 | 6110 | 81820 | 6110 | 81700 | 8.46E+4 | 8.410E+4 | 8.410E+4 | 8.27E+4 | 8.78E+4 | 8.49E+4 | 8.00E+4 | 8.40E+4 | 8.49E+4 | 8.50E+4 | 8.00E+4 | 8.50E+4 | … |
| 14 | Si | 1.00E+6 | 3.40E+4 | 1.00E+6 | 3.40E+4 | 1.00E+6 | 1.00E+6 | 1.000E+6 | 1.000E+6 | 1.00E+6 | 1.00E+6 | 1.00E+6 | 1.00E+6 | 1.00E+6 | 1.00E+6 | 1.00E+6 | 1.00E+6 | 1.00E+6 | … |
| 15 | P | 8260 | 530 | 8260 | 530 | 8340 | 8300 | 8373 | 8373 | 7860 | 9380 | 10400 | 8580 | 9400 | 10400 | 6500 | 8580 | 7000 | … |
| 16 | S | 4.37E+5 | 2.60E+4 | 4.37E+5 | 2.60E+4 | 4.38E+5 | 4.21E+5 | 4.449E+5 | 4.449E+5 | 4.44E+5 | 1.49E+6 | 5.15E+5 | 4.80E+5 | 4.30E+5 | 5.15E+5 | 5.00E+5 | 4.8E+5 | 5.02E+5 | … |
| 17 | Cl | 5290 | 810 | 5290 | 810 | 5170 | 5170 | 5237 | 5237 | 5180 | 5180 | 5240 | 5000 | 4700 | 5240 | 4740 | 5000 | 5700 | … |
| 18 | Ar | 97700 | 25300 | 97700 | 25300 | [0.0096] | 92700 | 1.025E+5 | 1.025E+5 | … | … | 1.01E+5 | 1.07E+5 | 1.07E+5 | 2.00E+5 | 1.06E+5 | 1.00E+5 | 2.10E+5 | … |
| 19 | K | 3611 | 190 | 3606 | 190 | 3670 | 3660 | 3697 | 3692 | 3660 | 3660 | 3770 | 3480 | 3400 | 3770 | 3500 | 3480 | 3500 | … |
| 20 | Ca | 57234 | 4500 | 57239 | 4500 | 59700 | 60400 | 62870 | 62870 | 61200 | 431000 | 61100 | 59100 | 62000 | 61100 | 62500 | 59100 | 72000 | … |
| 21 | Sc | 33.7 | 2.2 | 33.7 | 2.2 | 33.9 | 34.4 | 34.20 | 34.20 | 34.5 | 34.5 | 35.5 | 35 | 35.0 | 33.8 | 31 | 35 | 31 | … |
| 22 | Ti | 2459 | 150 | 2459 | 150 | 2430 | 2470 | 2422 | 2422 | 2520 | 2420 | 2400 | 2420 | 2700 | 2400 | 2400 | 2420 | 2400 | … |
| 23 | V | 275 | 18 | 275 | 18 | 281 | 386 | 288.4 | 288.4 | 280 | 280 | 293 | 290 | 260 | 295 | 254 | 290 | 254 | … |
| 24 | Cr | 13130 | 460 | 13130 | 460 | 13200 | 13100 | 12860 | 12860 | 13400 | 13400 | 13500 | 13500 | 12900 | 13400 | 12700 | 13500 | 12700 | … |
| 25 | Mn | 9090 | 620 | 9090 | 620 | 9150 | 9220 | 9168 | 9168 | 9250 | 9250 | 9550 | 8720 | 7700 | 9510 | 9300 | 8720 | 9300 | … |
| 26 | Fe | 8.72E+5 | 3.80E+4 | 8.72E+5 | 3.80E+4 | 8.77E+5 | 8.48E+5 | 8.380E+5 | 8.380E+5 | 8.68E+5 | 8.58E+5 | 9.00E+5 | 8.60E+5 | 8.80E+5 | 9.00E+5 | 9.00E+5 | 8.60E+5 | 9.01E+5 | 9.00E+5 |
| 27 | Co | 2260 | 100 | 2260 | 100 | 2280 | 2350 | 2323 | 2323 | 2260 | 2260 | 2250 | 2240 | 2100 | 2250 | 2220 | 2240 | 2200 | 2300 |
| 28 | Ni | 48670 | 2940 | 48670 | 2940 | 4.83E+4 | 4.90E+4 | 4.780E+4 | 4.780E+4 | 4.82E+4 | 4.82E+4 | 4.93E+4 | 4.83E+4 | 4.80E+4 | 4.93E+4 | 4.78E+4 | 4.83E+4 | 4.78E+4 | 4.90E+4 |
| 29 | Cu | 535 | 50 | 535 | 50 | 549 | 541 | 527 | 527 | 542 | 542 | 522 | 450 | 520 | 514 | 540 | 450 | 540 | 540 |
| 30 | Zn | 1260 | 180 | 1260 | 180 | 1240 | 1300 | 1226 | 1226 | 1300 | 1300 | 1260 | 1400 | 980 | 1260 | 1260 | 1400 | 1260 | 1260 |
| 31 | Ga | 36.2 | 1.8 | 36.2 | 1.8 | 36.2 | 36.6 | 35.97 | 35.97 | 36.6 | 36.6 | 37.8 | 34.0 | … | 37.8 | 38 | 34.0 | 14 | 45.0 |
| 32 | Ge | 120 | 7 | 120 | 7 | 118 | 115 | 120.6 | 120.6 | 118 | 118 | 119 | 114 | … | 118 | 117 | 114 | 117 | 117 |
| 33 | As | 6.07 | 0.5 | 6.07 | 0.5 | 6.10 | 6.10 | 6.089 | 6.089 | 6.35 | 6.35 | 6.56 | 6.50 | … | 6.79 | 6.2 | 6.5 | 6.2 | 6.60 |
| 34 | Se | 67.6 | 5 | 67.6 | 5 | 67.53 | 67.5 | 65.79 | 65.79 | 71.3 | 70.9 | 62.1 | 63 | … | 62.1 | 67 | 63 | 67 | 66.0 |
| 35 | Br | 12.3 | 2.9 | 12.3 | 2.9 | 10.7 | 10.7 | 11.32 | 11.32 | 11.5 | 11.5 | 11.8 | 8.0 | … | 11.8 | 9.2 | 8 | 14 | 18.0 |
| 36 | Kr | 51.3 | 10.4 | 51.3 | 10.4 | [1.64E-4] | 55.8 | 55.15 | 55.15 | … | … | 45.0 | … | … | 45.3 | 41.3 | … | 40 | 60.0 |
| 37 | Rb | 7.17 | 0.47 | 7.04 | 0.46 | 7.13 | 7.10 | 6.694 | 6.572 | 7.14 | 7.14 | 7.09 | 6.4 | … | 7.09 | 6.10 | 6.4 | 6.0 | 6.4 |
| 38 | Sr | 23.3 | 1.3 | 23.4 | 1.3 | 23.3 | 23.4 | 23.52 | 23.64 | 21.8 | 21.8 | 23.5 | 26 | … | 23.8 | 22.9 | 26 | 27.0 | 27.0 |
| 39 | Y | 4.35 | 0.24 | 4.35 | 0.24 | 4.31 | 4.63 | 4.608 | 4.608 | 4.61 | 4.63 | 4.64 | 4.64 | … | 4.64 | 4.8 | 4.3 | 4.8 | 4.80 |
| 40 | Zr | 10.9 | 1 | 10.9 | 1 | 10.4 | 10.80 | 11.33 | 11.33 | 11.10 | 11.16 | 11.4 | 11.2 | … | 10.7 | 12 | 11 | 9.10 | 13.00 |
| 41 | Nb | 0.78 | 0.07 | 0.78 | 0.07 | 0.800 | 0.780 | 0.7554 | 0.7554 | 0.699 | 0.696 | 0.698 | 0.70 | … | 0.7 | 0.9 | 0.85 | 0.9 | 0.900 |
| 42 | Mo | 2.6 | 0.26 | 2.6 | 0.26 | 2.66 | 2.55 | 2.601 | 2.601 | 2.54 | 2.54 | 2.55 | 2.5 | … | 2.52 | 4 | 2.5 | 4.0 | 2.50 |
| 43 | Tc | … | | … | | … | … | … | … | … | … | … | … | … | … | … | … | … | … |
| 44 | Ru | 1.81 | 0.02 | 1.81 | 0.02 | 1.79 | 1.78 | 1.900 | 1.900 | 1.78 | 1.86 | 1.86 | 1.8 | … | 1.86 | 1.9 | 1.8 | 1.90 | 1.90 |
| 45 | Rh | 0.338 | 0.015 | 0.338 | 0.015 | 0.337 | 0.370 | 0.3708 | 0.3708 | 0.358 | 0.342 | 0.344 | 0.33 | … | 0.344 | 0.40 | 0.33 | 0.40 | 0.45 |
| 46 | Pd | 1.38 | 0.07 | 1.38 | 0.07 | 1.38 | 1.36 | 1.435 | 1.435 | 1.37 | 1.37 | 1.39 | 1.32 | … | 1.39 | 1.3 | 1.32 | 1.30 | 1.30 |
| 47 | Ag | 0.497 | 0.022 | 0.497 | 0.022 | 0.489 | 0.489 | 0.4913 | 0.4913 | 0.480 | 0.480 | 0.486 | 0.5 | … | 0.529 | 0.46 | 0.5 | 0.46 | 0.46 |
| 48 | Cd | 1.58 | 0.06 | 1.58 | 0.06 | 1.49 | 1.57 | 1.584 | 1.584 | 1.59 | 1.59 | 1.61 | 1.80 | … | 1.69 | 1.55 | 1.3 | 1.55 | 1.55 |
| 49 | In | 0.179 | 0.008 | 0.179 | 0.008 | 0.168 | 0.178 | 0.1810 | 0.1810 | 0.178 | 0.178 | 0.184 | 0.17 | … | 0.184 | 0.19 | 0.174 | 0.19 | 0.19 |
| 50 | Sn | 3.59 | 0.22 | 3.59 | 0.22 | 3.35 | 3.60 | 3.733 | 3.733 | 3.72 | 3.72 | 3.82 | 3.88 | … | 3.82 | 3.7 | 2.4 | 3.7 | 3.6 |
| 51 | Sb | 0.359 | 0.045 | 0.359 | 0.045 | 0.300 | 0.313 | 0.3292 | 0.3292 | 0.287 | 0.287 | 0.309 | 0.27 | … | 0.352 | 0.31 | 0.27 | 0.31 | 0.31 |
| 52 | Te | 4.72 | 0.23 | 4.72 | 0.23 | 4.56 | 4.69 | 4.815 | 4.815 | 4.68 | 4.68 | 4.81 | 4.83 | … | 4.91 | 6.5 | 4.8 | 6.5 | 6.5 |
| 53 | I | 1.59 | 0.64 | 1.59 | 0.64 | 1.05 | 1.10 | 0.9975 | 0.9975 | 0.90 | 0.90 | 0.90 | 1.16 | … | 0.90 | 1.27 | 1.16 | 1.16 | 1.20 |
| 54 | Xe | 5.50 | 1.11 | 5.50 | 1.11 | [1.74E-4] | 5.46 | 5.391 | 5.391 | … | | 4.70 | … | … | 4.35 | 5.84 | 6.1 | 6.3 | 7.8 |
| 55 | Cs | 0.368 | 0.043 | 0.368 | 0.043 | 0.371 | 0.371 | 0.3671 | 0.3671 | 0.372 | 0.372 | 0.372 | 0.37 | … | 0.372 | 0.39 | 0.37 | 0.39 | 0.38 |
| 56 | Ba | 4.55 | 0.27 | 4.55 | 0.27 | 4.63 | 4.47 | 4.351 | 4.351 | 4.61 | 4.61 | 4.49 | 4.22 | … | 4.36 | 4.8 | 4.2 | 4.8 | 4.8 |
| 57 | La | 0.459 | 0.024 | 0.459 | 0.024 | 0.4561 | 0.457 | 0.4405 | 0.4405 | 0.464 | 0.464 | 0.446 | 0.46 | … | 0.448 | 0.37 | 0.46 | 0.37 | 0.380 |
| 58 | Ce | 1.16 | 0.05 | 1.16 | 0.05 | 1.160 | 1.18 | 1.169 | 1.169 | 1.20 | 1.20 | 1.14 | 1.2 | … | 1.16 | 1.20 | 1.2 | 1.20 | 1.20 |
| 59 | Pr | 0.175 | 0.012 | 0.175 | 0.012 | 0.1749 | 0.172 | 0.1737 | 0.1737 | 0.180 | 0.180 | 0.167 | 0.18 | … | 0.174 | 0.18 | 0.18 | 0.180 | 0.18 |
| 60 | Nd | 0.864 | 0.022 | 0.865 | 0.022 | 0.8621 | 0.857 | 0.8343 | 0.8355 | 0.864 | 0.864 | 0.828 | 0.87 | … | 0.870 | 0.70 | 0.87 | 0.790 | 0.80 |
| 61 | Pm | … | | … | | … | … | … | … | … | … | … | … | … | … | … | … | … | … |
| 62 | Sm | 0.273 | 0.012 | 0.271 | 0.012 | 0.2681 | 0.265 | 0.2542 | 0.2542 | 0.269 | 0.269 | 0.258 | 0.27 | … | 0.27 | 0.24 | 0.27 | 0.24 | 0.24 |
| 63 | Eu | 0.1 | 0.005 | 0.1 | 0.005 | 0.1016 | 0.0984 | 0.09513 | 0.09513 | 0.100 | 0.100 | 0.0973 | 0 | … | 0.100 | 0.094 | 0.0972 | 0.094 | 0.094 |
| 64 | Gd | 0.346 | 0.013 | 0.346 | 0.013 | 0.3453 | 0.360 | 0.3321 | 0.3321 | 0.341 | 0.341 | 0.330 | 0.34 | … | 0.34 | 0.42 | 0.331 | 0.12 | 0.42 |
| 65 | Tb | 0.0625 | 0.0025 | 0.0625 | 0.0025 | 0.06271 | 0.0634 | 0.05907 | 0.05907 | 0.0621 | 0.0621 | 0.0603 | 0.061 | … | 0.061 | 0.076 | 0.0589 | 0.076 | 0.07 |
| 66 | Dy | 0.407 | 0.016 | 0.407 | 0.016 | 0.4132 | 0.404 | 0.3862 | 0.3862 | 0.411 | 0.411 | 0.394 | 0.41 | … | 0.41 | 0.37 | 0.398 | 0.37 | 0.37 |
| 67 | Ho | 0.0891 | 0.0035 | 0.0891 | 0.0035 | 0.08982 | 0.0910 | 0.08986 | 0.08986 | 0.0904 | 0.0904 | 0.0889 | 0.091 | … | 0.091 | 0.092 | 0.0875 | 0.092 | 0.093 |
| 68 | Er | 0.256 | 0.009 | 0.256 | 0.009 | 0.2597 | 0.262 | 0.2554 | 0.2554 | 0.261 | 0.261 | 0.251 | 0.26 | … | 0.26 | 0.23 | 0.253 | 0.23 | 0.24 |
| 69 | Tm | 0.0403 | 0.0014 | 0.0403 | 0.0014 | 0.04054 | 0.0406 | 0.03700 | 0.03700 | 0.0399 | 0.0399 | 0.0378 | 0.041 | … | 0.041 | 0.035 | 0.0386 | 0.035 | 0.036 |
| 70 | Yb | 0.252 | 0.009 | 0.252 | 0.009 | 0.2559 | 0.256 | 0.2484 | 0.2484 | 0.251 | 0.251 | 0.248 | 0.25 | … | 0.25 | 0.20 | 0.243 | 0.20 | 0.22 |
| 71 | Lu | 0.0381 | 0.0018 | 0.038 | 0.0018 | 0.03755 | 0.0380 | 0.03580 | 0.03572 | 0.0382 | 0.0382 | 0.0367 | 0.038 | … | 0.038 | 0.035 | 0.0369 | 0.035 | 0.04 |
| 72 | Hf | 0.155 | 0.011 | 0.155 | 0.011 | 0.1566 | 0.156 | 0.1698 | 0.1699 | 0.158 | 0.158 | 0.154 | 0.157 | … | 0.180 | 0.17 | 0.176 | 0.17 | 0.22 |
| 73 | Ta | 0.0215 | 0.001 | 0.0215 | 0.001 | 0.0218 | 0.0210 | 0.02099 | 0.02099 | 0.0206 | 0.0203 | 0.0207 | 0.02 | … | 0.021 | 0.02 | 0.0226 | 0.020 | 0.030 |
| 74 | W | 0.144 | 0.013 | 0.144 | 0.013 | 0.137 | 0.137 | 0.1277 | 0.1277 | 0.129 | 0.136 | 0.133 | 0.13 | … | 0.127 | 0.30 | 0.137 | 0.30 | 0.16 |
| 75 | Re | 0.0547 | 0.0042 | 0.0521 | 0.0042 | 0.0554 | 0.0554 | 0.05509 | 0.05254 | 0.0558 | 0.0541 | 0.0517 | 0.052 | … | 0.052 | 0.051 | 0.0507 | 0.051 | 0.052 |
| 76 | Os | 0.652 | 0.04 | 0.655 | 0.04 | 0.6830 | 0.680 | 0.6713 | 0.6738 | 0.699 | 0.672 | 0.675 | 0.68 | … | 0.72 | 0.69 | 0.717 | 0.69 | 0.85 |
| 77 | Ir | 0.633 | 0.029 | 0.633 | 0.029 | 0.640 | 0.672 | 0.6448 | 0.6448 | 0.657 | 0.628 | 0.661 | 0.65 | … | 0.66 | 0.72 | 0.65 | 0.72 | 0.73 |
| 78 | Pt | 1.24 | 0.1 | 1.24 | 0.1 | 1.24 | 1.27 | 1.357 | 1.357 | 1.32 | 1.34 | 1.34 | 0.14 | … | 1.37 | 1.41 | 1.42 | 1.40 | 1.40 |
| 79 | Au | 0.195 | 0.016 | 0.195 | 0.016 | 0.197 | 0.195 | 0.1955 | 0.1955 | 0.198 | 0.203 | 0.187 | 0.19 | … | 0.186 | 0.21 | 0.19 | 0.21 | 0.21 |
| 80 | Hg | 0.376 | 0.156 | 0.376 | 0.156 | 0.41 | 0.458 | 0.4128 | 0.4128 | 0.406 | 0.406 | 0.340 | 0.52 | … | 0.52 | 0.21 | 0.40 | 1.40 | 0.44 |
| 81 | Tl | 0.179 | 0.015 | 0.179 | 0.015 | 0.184 | 0.182 | 0.1845 | 0.1845 | 0.184 | 0.184 | 0.184 | 0.17 | … | 0.184 | 0.19 | 0.17 | 0.19 | 0.19 |
| 82 | Pb | 3.31 | 0.2 | 3.33 | 0.2 | 3.32 | 3.33 | 3.234 | 3.258 | 3.21 | 3.21 | 3.15 | 3.09 | … | 3.15 | 2.6 | 3.1 | 2.6 | 4.0 |
| 83 | Bi | 0.141 | 0.01 | 0.141 | 0.01 | 0.138 | 0.138 | 0.1388 | 0.1388 | 0.140 | 0.140 | 0.144 | 0.14 | … | 0.144 | 0.14 | 0.136 | 0.14 | 0.14 |
| 90 | Th | 0.0421 | 0.0021 | 0.0336 | 0.0017 | 0.0339 | 0.0351 | 0.04399 | 0.03512 | 0.0338 | 0.0338 | 0.0335 | 0.032 | … | 0.0335 | 0.045 | 0.032 | 0.057 | 0.0105 |
| 92 | U | 0.02389 | 0.0017 | 0.00897 | 0.00064 | 0.00893 | 0.00893 | 0.024631 | 0.009306 | 0.00860 | 0.0086 | 0.0090 | 0.0091 | … | 0.0090 | 0.027 | 0.0091 | 0.0270 | ... |

## Table 6. Solar System Abundances Relative to Si= $10^6$ atoms

Part 2.

| Z | E | ph+CI+o [Cam73a] | ph+CI+o [U72] | ph+CM2+o [Go69] | ph+CI+o [Cam67] | Photo +chon+o [Cam63] | Sun [All61] | Chond. [E61] | Sun +mets+o [Cam59] | Sun +mets+o [Ran56] | Sun +mets+o [SU56] | Chond. [U52] | Sun +mets+o [S49] | Sun +mets+o [Bro49] | Sun +mets+o [S47b] | Sun +mets+o [Gs37] | Mets. [NN34] | Mets. [NN30] | Sun [R29] |
|---|---|---|---|---|---|---|---|---|---|---|---|---|---|---|---|---|---|---|---|
| 1 | H | 3.18E+10 | 2.25E+10 | 4.80E+10 | 2.60E+10 | 3.20E+10 | 2.94E+10 | … | 2.50E+10 | 3.50E+10 | 4.00E+10 | … | 1.62E+10 | 3.50E+10 | 2.00E+10 | 2.30E+9 | [7.5E+4] | … | 1.6E+10 |
| 2 | He | 2.21E+9 | 2.10E+9 | 3.90E+9 | 2.10E+9 | 5.00E+9 | 4.05E+9 | … | 3.80E+9 | 3.50E+9 | 3.08E+9 | … | 2.88E+9 | 3.50E+9 | 3.16E+9 | 9.0E+7 | … | … | 5.0E+8 |
| 3 | Li | 49.5 | 49.5 | 16.0 | 45.0 | 38 | 0.6 (sun) | 38.0 | 100 | 61 | 100 | 100 | 100 | … | 100 | 100 | 66 | 90 | 5 (sun) |
| 4 | Be | 0.81 | … | 0.81 | 0.690 | 7 | 1.000 | … | 20 | 16 | 20 | 16 | 20 | … | 19 | 19 | 15 | 140 | 3.2 |
| 5 | B | 350 [sic] | 135 | 6 | 6.2 | 6 | 1580 | … | 24 | 20 | 24 | 20 | 23 | … | 22.9 | 23 | 38 | | 0.1 |
| 6 | C | 1.18E+7 | 1.37E+7 | 1.70E+7 | 1.35E+7 | 1.66E+7 | 2.70E+6 | … | 9.30E+6 | 8.00E+6 | 3.50E+6 | … | 5.50E+6 | 8.00E+6 | 8.71E+6 | 2.25E+7 | [1.77E+4] | [6300] | 1.3E+6 |
| 7 | N | 3.74E+6 | 2.44E+6 | 4.60E+6 | 2.44E+6 | 3.00E+6 | 4.90E+6 | … | 2.40E+6 | 1.60E+7 | 6.60E+6 | … | 1.15E+7 | 1.60E+7 | 1.45E+7 | 7.50E+6 | [10] | [63] | 2.0E+6 |
| 8 | O | 2.15E+7 | 2.36E+7 | 4.40E+7 | 2.36E+7 | 2.90E+7 | 1.58E+7 | … | 2.50E+7 | 2.20E+7 | 2.15E+7 | … | 1.51E+7 | 2.20E+7 | 1.91E+7 | 7.50E+7 | [3.45E+6] | [3.33e+6] | 5.0E+7 |
| 9 | F | 2450 | 3630 | 2500 | 3630 | 10000 | … | … | 1600 | 300 | 1600 | 300 | 1510 | 9000 | 1510 | 1500 | 290 | … | … |
| 10 | Ne | 3.44E+6 | 2.36E+6 | 4.40E+6 | 2.36E+6 | 1.70E+7 | … | … | 8.00E+5 | 9E+5 to 2.4E+7 | 8.60E+6 | … | 1.82E+7 | 3.70E+6 | 2.29E+7 | 2.50E+6 | … | … | … |
| 11 | Na | 6.00E+4 | 5.93E+4 | 3.50E+4 | 6.32E+4 | 4.18E+4 | 7.70E+4 | … | 4.38E+4 | 5.10E+4 | 4.38E+4 | 5.10E+4 | 4.47E+4 | 4.62E+4 | 4.37E+4 | 4.42E+4 | 4.3E+4 | 4.00E+4 | 7.9E+5 |
| 12 | Mg | 1.061E+6 | 1.05E+6 | 1.04E+6 | 1.05E+6 | 1.05E+6 | 1.78E+6 | … | 9.12E+5 | 8.70E+5 | 9.12E+5 | 8.70E+5 | 8.71E+5 | 8.87E+5 | 8.71E+5 | 8.72E+5 | 8.6E+5 | 8.33E+5 | 3.2E+6 |
| 13 | Al | 8.50E+4 | 8.50E+4 | 8.40E+4 | 8.51E+4 | 8.93E+4 | 7.40E+4 | … | 9.48E+4 | 8.20E+4 | 9.48E+4 | 8.20E+4 | 8.71E+4 | 8.82E+4 | 8.71E+4 | 8.79E+4 | 8.7E+4 | 8.00E+4 | 1.3E+5 |
| 14 | Si | 1.00E+6 | 1.00E+6 | 1.00E+6 | 1.00E+6 | 1.00E+6 | 1.00E+6 | … | 1.00E+6 | 1.00E+6 | 1.00E+6 | 1.00E+6 | 1.00E+6 | 1.00E+6 | 1.00E+6 | 1.00E+6 | 1.00E+6 | 1.00E+6 | 1.00E+6 |
| 15 | P | 9600 | 12700 | 8100 | 12700 | 9320 | 19000 | … | 10000 | 7500 | 10000 | 7500 | 5900 | 13000 | 5900 | 5900 | 6500 | 7300 | … |
| 16 | S | 5.00E+5 | 5.05E+5 | 8.00E+5 | 5.06E+5 | 6.00E+5 | 5.20E+5 | … | 3.75E+5 | 9.80E+4 | 3.75E+5 | 9.80E+4 | 1.15E+5 | 3.50E+5 | 1.15E+5 | 1.14E+5 | 8.73E+4 | 1.33E+4 | 2.5E+4 |
| 17 | Cl | 5700 | 2000 | 2100 | 1970 | 1836 | 300000 | … | 2770 | 2100 | 8850 | 2100 | 3890 | 17000 | 3890 | 3900 | 2000 | 3300 | … |
| 18 | Ar | 1.17E+5 | 1.00E+5 | 3.40E+5 | 2.28E+5 | 2.40E+5 | 1.00E+5 | … | 1.50E+5 | 5.30E+4 | 1.50E+5 | 2.50E+4 | 1.58E+5 | 1.00E+5 | 1.26E+5 | 1.00E+6 | … | … | … |
| 19 | K | 4200 | 3200 | 2100 | 3240 | 2970 | 3900 | 3290 | 3160 | 3500 | 3160 | 3160 | 6920 | 6930 | 6900 | 6900 | 6560 | 8700 | 316000 |
| 20 | Ca | 72100 | 66600 | 72000 | 73600 | 72800 | 83000 | 45000 | 49000 | 56000 | 49000 | 56000 | 57500 | 67000 | 57500 | 57100 | 6.5E+5 | 63300 | 2.5E+5 |
| 21 | Sc | 35 | 35 | 35 | 33 | 29 | 42 | 32 | 28 | 21 | 28 | 17 | 15 | 18 | 15 | 15 | 2 | 320 | 200 |
| 22 | Ti | 2775 | 2500 | 2400 | 2300 | 3140 | 1800 | 2090 | 1680 | 1800 | 2240 | 1800 | 4680 | 2600 | 4700 | 4700 | 4700 | 5700 | 7900 |
| 23 | V | 262 | 298 | 590 | 900 | 590 | 300 | … | 220 | 150 | 220 | 150 | 129 | 250 | 130 | 130 | 270 | 770 | 5000 |
| 24 | Cr | 12700 | 11900 | 12400 | 12400 | 12000 | 1900 | 6400 | 7800 | 8200 | 7800 | 8200 | 11200 | 9500 | 11200 | 11300 | 52400 | 13000 | 25100 |
| 25 | Mn | 9300 | 9300 | 6200 | 8800 | 6320 | 5600 | 7200 | 6850 | 6800 | 6850 | 6800 | 6600 | 7700 | 6600 | 6600 | 6600 | 5300 | 40000 |
| 26 | Fe | 8.30E+5 | 8.90E+5 | 2.50E+5 | 8.90E+5 | 8.42E+5 | 4.80E+5 | … | 8.50E+5 | 6.70E+5 | 6.00E+5 | 6.70E+5 | 8.91E+5 | 1.83E+6 | 1.15E+6 | 8.90E+5 | 6.11E+5 | 1.97E+6 | 7.9E+5 |
| 27 | Co | 2210 | 2300 | 1900 | 2300 | 2290 | 2200 | 1190 | 1800 | 2900 | 1800 | 2900 | 3470 | 9900 | 3500 | 3500 | 3000 | 9000 | 20000 |
| 28 | Ni | 4.80E+4 | 4.57E+4 | 4.50E+4 | 4.57E+4 | 4.44E+4 | 4.40E+4 | 2.48E+4 | 2.74E+4 | 3.90E+4 | 2.74E+4 | 3.90E+4 | 4.57E+4 | 1.34E+5 | 4.57E+4 | 4.6E+4 | 3.9E+4 | 1.4E+5 | 5.0E+4 |
| 29 | Cu | 540 | 920 | 420 | 919 | 861 | 932 | 186 | 212 | 420 | 212 | 420 | 230 | 460 | 230 | 460 | 500 | 1300 | 5000 |
| 30 | Zn | 1244 | 1350 | 630 | 1500 | 930 | 2880 | … | 202 | 180 | 486 | 180 | 1510 | 160 | 1510 | 360 | 180 | 430 | 4000 |
| 31 | Ga | 48 | 51 | 28 | 45.5 | 39 | 2.5 | … | 9.05 | 10 | 11.4 | 10 | 62 | 65 | 66 | 8 | 10 | 17 | 5 |
| 32 | Ge | 115 | 135 | 76 | 126 | 134 | 25 | 18.7 | 25.3 | 110 | 50.5 | 110 | 186 | 250 | 324 | 188 | 65 | 570 | 50 |
| 33 | As | 6.60 | 6.40 | 3.80 | 7.2 | 4.4 | … | … | 1.70 | 38 | 4.0 | 38 | 34 | 480 | 18 | 19 | 36 | 6.3 | 0.2 |
| 34 | Se | 67.2 | 89 | 27 | 70.1 | 18.8 | … | 18.8 | 18.8 | 13 | 67.6 | 13 | 257 | 25 | 234 | 15 | 15 | 140 | … |
| 35 | Br | 13.5 | 14.5 | 5.4 | 20.6 | 3.95 | … | … | 3.95 | 49 | 13.4 | 49 | 43 | 42 | 43 | 43 | 2.6 | 1.7 | … |
| 36 | Kr | 46.8 | 54.5 | 25 | 64.4 | 20 | … | … | 42.0 | 270 | 51.3 | ... | 151.0 | 87 | 190 | 50 | … | … | … |
| 37 | Rb | 5.88 | 6.5 | 4.1 | 5.95 | 5 | … | 4.6; 5.8 | 6.50 | 15 | 6.50 | 15 | 17 | 7.1 | 24 | 6.8 | 12 | 6.7 | 2.5 |
| 38 | Sr | 26.9 | 24 | 25 | 58.4 | 21 | … | … | 61.0 | 20 | 18.9 | 41 | 46 | 41 | 58 | 37 | 100 | 107 | 100 |
| 39 | Y | 4.80 | 4.60 | 4.70 | 4.60 | 3.60 | … | … | 8.90 | 9.7 | 8.90 | 9.7 | 9.8 | 10 | 9.8 | 9.7 | 9.7 | 50 | 20 |
| 40 | Zr | 28 | 26 | 23 | 30 | 21.00 | … | … | 14.20 | 57 | 54.50 | 140 | 63 | 150 | 57 | 140 | 150 | 140 | 16 |
| 41 | Nb | 1.4 | 1.3 | 0.9 | 1.2 | 0.810 | … | … | 0.81 | 0.7 | 1.00 | 0.70 | 5.60 | 0.90 | 3.89 | 7.0 | 0.16 | 3.7 | 0.5 |
| 42 | Mo | 4.0 | 2.20 | 2.50 | 2.52 | 2.42 | … | … | 2.42 | 5.9 | 2.42 | 5.90 | 26.00 | 19 | 9.55 | 9.5 | 6.5 | 20 | 1.3 |
| 43 | Tc | … | … | … | … | … | … | … | … | 0 | … | … | … | … | … | … | 0.002 | … | … |
| 44 | Ru | 1.9 | 1.85 | 1.83 | 1.60 | 1.58 | … | 1.20 | 0.87 | 2.1 | 1.49 | 2.10 | 5.10 | 9.30 | 3.63 | 3.6 | 1.9 | 20 | 2.5 |
| 45 | Rh | 0.40 | 0.29 | 0.33 | 0.33 | 0.26 | … | 0.27 | 0.15 | 0.71 | 0.214 | 0.71 | 1.40 | 3.5 | 0.96 | 1.3 | 0.71 | 4.3 | 0.16 |
| 46 | Pd | 1.3 | 1.28 | 1.33 | 1.50 | 1.0 | … | … | 0.68 | 1.3 | 0.68 | 1.3 | 5.6 | 3.2 | 3.39 | 2.5 | 1.5 | 15 | 0.63 |
| 47 | Ag | 0.45 | 0.96 | 0.33 | 0.50 | 0.26 | … | 0.131 | 0.26 | 1.9 | 0.26 | 0.90 | 2.30 | 2.7 | 1.62 | 3.2 | 0.55 | 5 | 0.5 |
| 48 | Cd | 1.48 | 2.40 | 1.20 | 2.12 | 0.89 | … | … | 0.89 | 2.2 | 0.89 | 2.2 | 8.7 | 3 | 4.79 | 2.6 | 2.2 | 9.7 | 7.90 |
| 49 | In | 0.189 | 0.217 | 0.100 | 0.217 | 0.11 | … | 0.0013 | 0.11 | 0.27 | 0.11 | 0.27 | 1.30 | 1.0 | 0.71 | 0.23 | 0.31 | 0.09 | 0.05 |
| 50 | Sn | 3.6 | 4.2 | 1.7 | 4.22 | 1.33 | … | … | 1.33 | 18 | 1.33 | 18 | 16 | 62 | 9.12 | 29 | 50 | 240 | 0.8 |
| 51 | Sb | 0.316 | 0.380 | 0.200 | 0.381 | 0.150 | … | … | 0.227 | 0.79 | 0.246 | 0.79 | 0.72 | 1.7 | 0.72 | 0.73 | 0.79 | 2.4 | 0.32 |
| 52 | Te | 6.42 | 6.80 | 3.10 | 6.76 | 3.0 | … | (0.50-4.0) | 2.91 | 0.16 | 4.67 | 0.16 | 14.00 | ? | 7.94 | 0.2 | 0.09 | 1.7 | … |
| 53 | I | 1.09 | 1.40 | 0.41 | 1.41 | 0.46 | … | (0.044-0.67) | 0.6 | 1.5 | 0.8 | 1.5 | 1.3 | 1.8 | 8.51 | 1.4 | 0.04 | … | … |
| 54 | Xe | 5.38 | 6.62 | 3.00 | 7.10 | 3.15 | … | … | 3.35 | 2.5 | 4.0 | … | 6.3 | 1.5 | 3.72 | 5 | … | … | … |
| 55 | Cs | 0.387 | 0.370 | 0.210 | 0.367 | 0.25 | … | 0.100 | 0.456 | 0.1 | 0.456 | 1.3 | 0.4 | 0.1 | 0.35 | 0.1 | 1.3 | 0.10 | … |
| 56 | Ba | 4.8 | 4.67 | 5.00 | 4.7 | 4.0 | … | 3.96 | 3.66 | 9 | 3.66 | 3.3 | 8.3 | 3.9 | 7.76 | 2.1 | 75 | 19 | 100 |
| 57 | La | 0.445 | 0.36 | 0.47 | 0.36 | 0.380 | … | 0.400 | 0.5 | 2.1 | 2.0 | 2.1 | 2.1 | 2.1 | 2.09 | 2.1 | 2.1 | … | 3.2 |
| 58 | Ce | 1.18 | 1.17 | 1.38 | 1.17 | 1.08 | … | 0.62 | 0.58 | 2.3 | 2.26 | 2.3 | 5.2 | 2.3 | 5.25 | 5.2 | 2.3 | 3.67 | 13 |
| 59 | Pr | 0.149 | 0.17 | 0.19 | 0.17 | 0.160 | … | 0.150 | 0.23 | 0.96 | 0.40 | 0.96 | 0.95 | 0.96 | 0.95 | 0.96 | 0.97 | … | 0.2 |
| 60 | Nd | 0.78 | 0.77 | 0.88 | 0.77 | 0.690 | … | 0.740 | 0.874 | 3.3 | 1.44 | 3.3 | 3.3 | 3.3 | 3.3 | 3.3 | 3.3 | 2.6 | 5 |
| 61 | Pm | … | … | … | … | … | … | … | … | … | … | … | … | … | … | … | … | … | … |
| 62 | Sm | 0.226 | 0.23 | 0.28 | 0.23 | 0.24 | … | 0.25 | 0.238 | 1.1 | 0.664 | 1.1 | 1.1 | 1.2 | 1.1 | 1.2 | 1.2 | 2.6 | 1.6 |
| 63 | Eu | 0.085 | 0.091 | 0.100 | 0.091 | 0.083 | … | 0.097 | 0.115 | 0.28 | 0.187 | 0.28 | 0.28 | 0.28 | 0.28 | 0.28 | 0.28 | … | 1.3 |
| 64 | Gd | 0.297 | 0.40 | 0.43 | 0.34 | 0.33 | … | 0.36 | 0.516 | 1.6 | 0.684 | 1.6 | 1.7 | 1.7 | 1.7 | 1.65 | 1.6 | … | 0.63 |
| 65 | Tb | 0.055 | 0.074 | 0.061 | 0.052 | 0.054 | … | 0.056 | 0.09 | 0.52 | 0.0956 | 0.52 | 0.37 | 0.52 | 0.52 | 0.52 | 0.52 | … | … |
| 66 | Dy | 0.36 | 0.36 | 0.45 | 0.36 | 0.33 | … | 0.39 | 0.665 | 2 | 0.556 | 2 | 2 | 2.0 | 2.0 | 2.0 | 2.0 | … | 2.0 |
| 67 | Ho | 0.079 | 0.090 | 0.093 | 0.090 | 0.076 | … | 0.078 | 0.18 | 0.57 | 0.118 | 0.57 | 0.44 | 0.57 | 0.58 | 0.57 | 0.57 | … | … |
| 68 | Er | 0.225 | 0.22 | 0.28 | 0.22 | 0.21 | … | 0.21 | 0.583 | 1.6 | 0.316 | 1.6 | 1.6 | 1.6 | 1.6 | 1.6 | 1.6 | … | 0.06 |
| 69 | Tm | 0.034 | 0.035 | 0.041 | 0.035 | 0.032 | … | 0.039 | 0.09 | 0.29 | 0.0318 | 0.29 | 0.29 | 0.29 | 0.29 | 0.29 | 0.29 | … | … |
| 70 | Yb | 0.216 | 0.21 | 0.22 | 0.21 | 0.18 | … | 0.19 | 0.393 | 1.5 | 0.220 | 1.5 | 1.8 | 1.5 | 1.5 | 1.5 | 1.5 | … | … |
| 71 | Lu | 0.036 | 0.035 | 0.036 | 0.035 | 0.031 | … | 0.036 | 0.0358 | 0.48 | 0.0500 | 0.48 | 0.49 | 0.48 | 0.49 | 0.490 | 0.48 | … | … |
| 72 | Hf | 0.21 | 0.47 | 0.31 | 0.16 | 0.16 | … | … | 0.113 | 1.4 | 0.438 | 1.4 | 2.4 | 0.7 | 3.0 | 1.5 | 1.7 | 0.7 | 0.1 |
| 73 | Ta | 0.021 | 0.025 | 0.019 | 0.022 | 0.021 | … | 0.019 | 0.015 | 0.26 | 0.0650 | 0.26 | 0.40 | 0.31 | 0.45 | 2.8 | 0.07 | 1 | … |
| 74 | W | 0.16 | 0.16 | 0.16 | 0.16 | 0.11 | … | 0.10 | 0.105 | 13? | 0.490 | 0.13 | 10.00 | 17 | 5 | 15 | 13 | 0.16 | 0.08 |
| 75 | Re | 0.053 | 0.052 | 0.059 | 0.055 | 0.054 | … | … | 0.054 | 0.07 | 0.1350 | 0.07 | 0.22 | 0.41 | 0.22 | 0.002 | 0.002 | 0.004 | … |
| 76 | Os | 0.75 | 0.70 | 0.86 | 0.71 | 0.73 | … | 0.60 | 0.64 | 0.97 | 1.000 | 0.97 | 1.70 | 3.5 | 1.7 | 1.7 | 0.72 | 4.7 | … |
| 77 | Ir | 0.717 | 0.73 | 0.96 | 0.43 | 0.50 | … | 0.38 | 0.494 | 0.31 | 0.821 | 0.31 | 1.00 | 1.4 | 1.0 | 0.58 | 0.28 | 1.1 | 0.03 |
| 78 | Pt | 1.40 | 1.30 | 1.40 | 1.13 | 1.16 | … | … | 1.28 | 1.5 | 1.63 | 1.5 | 2.9 | 8.7 | 2.9 | 2.9 | 1.8 | 7.7 | 2.0 |
| 79 | Au | 0.202 | 0.19 | 0.18 | 0.20 | 0.13 | … | 0.13 | 0.145 | 0.21 | 0.145 | 0.21 | 0.58 | 0.82 | 0.58 | 0.57 | 0.21 | 0.63 | … |
| 80 | Hg | 0.40 | 0.472 | 0.600 | 0.75 | 0.27 | … | [0.076] | 0.408 | < 0.006 | 0.284 | … | 0.8 | … | 1.91 | 0.3 | … | … | ... |
| 81 | Tl | 0.192 | 0.185 | 0.130 | 0.182 | 0.11 | … | 0.0010-0.00038 | 0.31 | 0.11 | 0.108 | 0.110 | 0.680 | … | 1.1 | 0.17 | 0.11 | 0.02 | 1.3 |
| 82 | Pb | 4 | 3.8 | 1.3 | 2.90 | 2.2 | … | 0.05 - 0.28 | 21.7 | < 2 | 0.47 | … | 9.1 | 27.00 | 0.58 | 9.2 | 13 | 70 | 0.8 |
| 83 | Bi | 0.143 | 0.16 | 0.19 | 0.164 | 0.14 | … | 0.0016 | 0.3 | 0.014 | 0.144 | 0.001 | 0.112 | 0.210 | 0.12 | 0.11 | 0.14 | 0.33 | … |
| 90 | Th | 0.058 | 0.053 | 0.04 | 0.034 | 0.069 | … | 0.026 | 0.027 | … | … | … | 0.630 | 1.2 | … | 0.59 | 2.8 | 1.1 | … |
| 92 | U | 0.0262 | 0.0334 | 0.010 | 0.0234 | 0.042 | … | 0.0072 | 0.0078 | … | … | … | 0.3 | 0.26 | … | 0.23 | 1.3 | … | … |

**Footnote for Table 6:** Values in square brackets are not representative for the solar system. The following all use combinations of abundances from the photosphere “Sun”, CI-chondrites “CI”, and other methods “o” (interpolation, nuclear systematics): [Cam59],[Cam63], [Cam67] = Cameron 1959, 1963,1967, all for proto-solar. [H79] Holweger 1979. [PSZ81] Palme, Suess, Zeh 1981. [AE82] Anders & Ebihara 1982. [Cam82] Cameron 1982, proto solar. [P88] Palme 1988. [AG89] Anders & Grevesse 1989. [PB93] Palme & Beer 1993. [L03] Lodders 2003. [PJ05] Palme & Jones 1993. [LPG09] Lodders, Palme, Gail 2009. [PLJ14] Palme, Lodders, Jones 2014. References to other abundances: [R29] Russell 1929, solar photosphere. [NN30] Noddack & Noddack 1930 meteoritic based on stony meteorite averages (see also Table 2). [NN34] Noddack & Noddack 1934, 1935, meteoritic based on weighted meteorite phase averages (see also Table 2). [Gs37] Goldschmidt 1937, photosphere (H,C,N,O, noble gases), meteoritic based on weighted meteorite phase averages (see also Table 2). [S47b] Suess 1947b, photosphere, meteoritic, interpolation. [Bro49] Brown 1949, photospheric (H,C,N,O, noble gases), meteoritic based on weighted meteorite phase averages to match Earth’s density (see Table 2). [S49] Suess 1949, as in [S47b]. [U52] Urey 1952, photospheric, direct chondrite compositions and weighted average of meteorite phases as observed in meteorites by Prior 1916. [SU56] Suess & Urey 1956, as in [U52], nuclear systematics and interpolation. [Ran56] Rankama 1956, photosphere, meteoritic from various sources including [Gs37], [Bro49], [U52]. [SU56], meteorites, heavier on nuclear systematics, interpolation. [All61] Aller 1961, photosphere. [E61] Ehman 1961, chondrites. [Go69] Goles 1969, photosphere, CM2-chondrites. [U72] Urey 1972, photosphere, CI- and CM2-chondrites, interpolation. [SZ73] Suess & Zeh 1973, chondrites and interpolation. [M85b] Meyer 1985b, “local galactic”, photosphere.

The calculation of the mass fractions X (for H), Y (He) and Z (for heavy elements Li-U) from the abundances on the atomic scale are done by multiplying the abundance of an element by its atomic weight and re-normalizing the mass abundances to unity (or 100 percent). The atomic weights for several elements are different in the sun than on Earth because of their different isotopic compositions (e.g.; for C,N,O, Ar, see isotope Table 9). Therefore atomic weights derived from the adopted isotopic compositions of the elements were used. Calculating mass-fractions requires the knowledge of abundances of all (major) elements and can be done for the present-day compositions. If the He abundance is unknown, only the mass fraction ratio (Z/X) of hydrogen (X) and that of all other elements (Z) can be calculated; this ratio is an important parameter in solar models from which the He abundances is actually derived. The recommended atomic abundance of He in Table 6 and 8 comes from the helioseismically determined mass fraction of Y (Table 7).

The mass fractions are related such as X = (1 – Y) /[ 1 + Z/X ] and Z = 1 – X – Y. The atomic ratio n(He)/n(H) and the He abundance $A(He) = \log_{10}\{ n(He)/n(H)\} +12$ on the atomic scale depend in Y and Z/X because

$$\frac{n(He)}{n(H)} = \frac{Y}{X} = \frac{Y(1+X/Z)}{4(1-Y)}$$

Where the "4" is shorthand for the ratio of the atomic weights of He and H.

The proto-solar mass fractions $X_o$, $Y_o$, and $Z_o$ (indicated by subscript "0") are derived from the present-day mass fraction by applying the settling factors SF(He) and SF(Li-U) to the present day mass ratios Y/X and Z/X. The settling factors should never be applied to the individual fractions Y or Z. From the definition of the settling factor we have

$$SF(He) = (Y/X) / (Y_o/X_o)$$

The proto-solar ratio $Y_o/X_o = (Y/X)/SF(He)$ follows from the present-day mass fraction ratio. The same applies to $Z_o/X_o$ using the settling factor SF(Li-U). With the mass-balance relations, $Xo = 1/(1 + Y_o/X_o + Z_o/X_o)$, all proto solar mass fractions are derived.

Note the atomic weights cancel out in the definition of the settling factors and the individual protosolar abundances on the atomic scale relative to hydrogen are easily obtained. The settling correction increases the present-day abundances on the abundances scale normalized to hydrogen are by a factor "SF" which means adding a constant $\log_{10}SF$ in the logarithmic scale with A(H) = 12:

$$\begin{aligned} A(E)_{proto\text{-}solar} &= 12 + \log_{10}\{n(E)/n(H)\}_{proto\text{-}solar} \\ &= 12 + A(E)_{present} + \log_{10}SF \\ &= 12 + \log_{10}\{n(E)/n(H)\}_{present} + \log_{10}SF \end{aligned}$$

The settling factors relate to the mass fraction ratios as $\log_{10}SF(He) = \log_{10}(Y/X)/(Y_o/X_o)$ for the helium/hydrogen ratio and $\log_{10}SF(Li\text{-}U) = \log_{10}(Z/X)/(Z_o/X_o)$ for the mass fraction ratio of all other elements. Table 7 summarizes the mass fractions X,Y.Z, their ratios, and settling factors of the recommended compositions (in Table 8) and those of other compilations and standard solar models using these compositions. The settling corrections affect the He and heavy element abundances on the logarithmic atomic scale normalized to hydrogen. However, on the cosmochemical abundance scale, only the H and He abundances are affected because this scale is relative to $10^6$ Si.

## Table 7. Mass Fractions of Hydrogen (X), Helium (Y) and All Other Heavier Elements (Z) in the Solar System

| | X | Y | Z | Z/X | Y/X | Xo | Yo | Zo | Zo/Xo | Yo/Xo | log SF(He) | log SF(Li-U) |
|---|---|---|---|---|---|---|---|---|---|---|---|---|
| *Solar System based on:* | | | | | | | | | | | | |
| Photosphere (3D) and CI-chondrites | ≡0.7389 | 0.2463 | 0.0148 | 0.0200 | 0.3334 | 0.7061 | 0.2766 | 0.0173 | 0.0245 | 0.3917 | ≡0.07 | ≡0.088 |
| Photosphere (3D) | ≡0.7389 | 0.2462 | 0.0149 | 0.0201 | 0.3333 | 0.7061 | 0.2765 | 0.0174 | 0.0246 | 0.3916 | ≡0.07 | ≡0.088 |
| Photosphere (1D) | ≡0.7389 | 0.2458 | 0.0153 | 0.0207 | 0.3327 | 0.7061 | 0.2760 | 0.0179 | 0.0253 | 0.3908 | ≡0.07 | ≡0.088 |
| *Helioseismic Measurements, Convection Zone Sun (Photosphere)* | | | | | | | | | | | | |
| Basu & Antia 2004, | 0.7389±0.0035 | 0.2485±0.0035 | 0.0126 | ≡0.0171 | 0.3318 | ... | ... | ... | ... | ... | ... | |
| Basu & Antia 2004 | 0.7389±0.0035 | 0.2450±0.0035 | 0.0161 | ≡0.0218 | 0.3316 | ... | ... | ... | ... | ... | ... | |
| Bahcall et al. 2006 | ... | 0.249±0.003 | ... | ... | ... | ... | ... | ... | ... | ... | ... | |
| Asplund et al. 2009 (AGSS09) composition | 0.7381 | ≡0.2485 | 0.0134 | 0.0181 | 0.3367 | 0.7154 | 0.2703 | 0.0142 | 0.0199 | 0.3778 | ≡0.05 | ≡0.04 |
| *Standard Solar Models based on Z/X by AGSS09:* | | | | | | | | | | | | |
| Zhang et al. 2019 | 0.7484 | 0.2381 | 0.0135 | ≡0.0181 | 0.3182 | 0.7190 | 0.2662 | 0.0148 | 0.0206 | 0.3703 | 0.066 | 0.055 |
| Yang 2016 | 0.7507 | 0.2351 | 0.0142 | ≡0.0189 | 0.3131 | 0.7177 | 0.2664 | 0.0159 | 0.0221 | 0.3712 | 0.075 | 0.069 |
| Basu et al. 2015 | 0.7513 | 0.2352 | 0.0135 | ≡0.0180 | 0.3131 | ... | 0.2650 | ... | ... | ... | ... | ... |
| Serenelli et al. 2011 | 0.7547 | 0.2319 | 0.0134 | ≡0.0178 | 0.3073 | 0.7231 | 0.2620 | 0.0149 | 0.0206 | 0.3623 | 0.072 | 0.064 |
| Asplund et al. 2005 (A05) composition | 0..7392 | ≡0.2486 | 0.0122 | 0.0165 | 0.3363 | 0.7135 | 0.2735 | 0.0132 | 0.0185 | 0.3833 | ≡0.05 | ≡0.04 |
| *Standard Solar Models based on Z/X by A05* | | | | | | | | | | | | |
| Basu et al. 2015 | 0.7589 | 0.2286 | 0.0125 | ≡0.0165 | 0.3012 | | 0.2586 | | | | | |
| Castro et al. 2007 | 0.7644 | 0.2330 | 0.0125 | ≡0.0164 | 0.2917 | 0.7304 | 0.2562 | 0.0134 | 0.0183 | 0.3508 | 0.080 | 0.048 |
| Bahcall et al. 2006 | 0.7583 | 0.2291 | 0.0126 | ≡0.0166 | 0.3021 | 0.7259 | 0.2600 | 0.0141 | 0.0194 | 0.3582 | 0.074 | 0.068 |
| Turck-Cheize et al. 2004 | 0.7518 | 0.2353 | 0.0129 | ≡0.0172 | 0.3130 | 0.7195 | 0.2664 | 0.0141 | 0.0196 | 0.3703 | 0.073 | 0.057 |
| Grevesse & Sauval 1998 (GS98) composition | 0.7347 | ≡0.2483 | 0.0169 | 0.0231 | 0.3380 | 0.7086 | ≡0.275 | 0.0163 | 0.0230 | 0.3881 | ≡0.060 | ≡0 |
| *Standard Solar Models based on Z/X by GS98* | | | | | | | | | | | | |
| Zhang et al. 2019 | 0.7378 | 0.2453 | 0.0169 | ≡0.0229 | 0.03325 | 0.7089 | 0.2727 | 0.01837 | 0.0259 | 0.3847 | 0.063 | 0.054 |
| Yang 2019 | 0.7364 | 0.2461 | 0.0175 | ≡0.0238 | 0.3342 | 0.7038 | 0.2767 | 0.195 | 0.0277 | 0.3932 | 0.071 | 0.066 |
| Basu et al. 2015 | 0.7374 | 0.2456 | 0.0170 | ≡0.0230 | 0.3338 | ... | 0.2755 | ... | ... | ... | ... | ... |
| Serenelli et al. 2011 | 0.7401 | 0.2429 | 0.0170 | ≡0.0229 | 0.3282 | 0.7089 | 0.2724 | 0.0187 | 0.0264 | 0.3843 | 0.0685 | 0.061 |
| Bahcall et al. 2006 | 0.7404 | 0.2426 | 0.0170 | ≡0.02292 | 0.3276 | 0.7087 | 0.2725 | 0.01884 | 0.0266 | 0.3845 | 0.070 | 0.064 |

**Footnote for Table 7:** Entries with ≡ are adopted quantities.

Using solar evolution models from results of Boothroyd & Sackmann (2003) and the dependence of the helium mass-fraction (Y) to hydrogen (X) ratio (Z/X) and of all other elements (Z/X), Lodders (2003) derived settling factors for helium $\log_{10}$SF(He) = 0.084 (a 22% increase from the present-day photospheric He/H ratio) and for all other heavy elements $\log_{10}$SF(Li-U) = 0.074 (18%). Lodders et al. (2009) and Palme et al. (2014) used 0.061 (15%) and 0.053 (13%), respectively. Asplund et al (2006, 2009) applied $\log_{10}$SF(He) = 0.05 (12%) and 0.04 (10%) for heavy elements from diffusion models by Turcotte & Wimmer-Schweingruber (2002); note that these are generally lower than used in several standard models (e.g., Table 8). Settling factors from standard solar models (independent of photospheric reference abundances used) for He are $\log_{10}$SF(He) = 0.070-0.075 (18-19%) and for the heavy elements $\log_{10}$SF(Li-U) = 0.064 (16-17%) by Bahcall et al. (2006) and Yang (2016, 2018). The standard solar model in Castro et al. (2007) correspond to $\log_{10}$SF(He) = 0.080 (20%) and 0.048 (12%) for the heavy elements. The settling corrections applied here are $\log_{10}$SF(He) = 0.07 (17%) and $\log_{10}$SF(Li-U) = 0.088 (23%) for the heavier elements. Note the lower settling factor for He than for heavier elements, opposite to what was previously assumed. The settling corrections are based on models by Yang (2019) who shows that the Sun's properties inferred from helioseismology observations (the sound-speeds, density profiles, and the depth of the solar convection zone for the present Sun) are reproduced with increased settling efficiencies by 50% (compared to previous models) and rotational mixing for solar models with lower metallicities

Increased settling of He from the solar convection zone below its base is almost completely counteracted by rotational forces. The settling of the heavier elements is counteracted less strongly by rotation, which is why the settling factors are now reversed from earlier estimates. The net effect of heavy element settling is an increase of the opacity in the interior below the convection zone. Yang's (2019) models agree well with observed sound-speeds and density profiles; even better than solar models using the older Grevesse & Sauval (1998) solar composition which had been a long-time favorite before more detailed rotational mixing was introduced into the solar models. The best-fit models by Yang (2019) are models with initial mass fraction of heavy elements to hydrogen ratios as obtained earlier by Lodders et al. (2009) and Caffau et al (2011, supplemented with data from Lodders 2009; see Table 6). More comprehensive models for element diffusion and settling from the solar convection zone and considering rotation will improve our understanding of element fractions that may become discernable with advanced spectroscopic methods and better meteoritic analyses.

## Recommended Solar System Abundances

The recommended elemental solar system abundances are summarized in Table 8. They are given for the present-day and for the time of solar system formation 4.567 Ga ago ("proto-solar") on the cosmochemical and the astronomical abundance scales. For comparison, the values for CI-chondrites and the sun (mainly photospheric values) are listed. Since the abundances in the solar convection zone and thus in the overlying photosphere have changed over the sun's lifetime, the present-day observed photospheric abundances for Helium and all other heavier elements needed to be readjusted with the settling factors to bring them back to proto-solar values. In addition, the abundances for elements with long-lived radioactive isotope were back-calculated to the 4.567 Ga ago.

Figure 4 shows the elemental abundances as a function of atomic number. Note the logarithmic scale which spans 13-orders of magnitude. The prominence of H and He is obvious, also the steep drop to Li, Be, and B, which are made of very fragile nuclei. The meteoritic and solar abundances of Li, Be, and B are exceptionally low when compared with their neighbors of lower and higher atomic numbers. Like the nuclides of lithium, those of beryllium and boron have low binding energies of their nuclei and are fragile in stellar interiors, and their major nucleosynthetic formation processes are relatively inefficient. Further, the Li abundance in the sun is 170-times less than in meteorites because Li is destroyed in the sun. The abundances of C, N, and O are high, then with increasing atomic numbers, elemental abundances drop with a minimum at Sc, to then steeply rise again to a maximum at Fe. After the “Fe-peak”, abundances drop steadily with smaller peaks at certain atomic numbers (see, e.g., Lodders & Fegley 2011 for detailed descriptions and references listed in Table 6.).

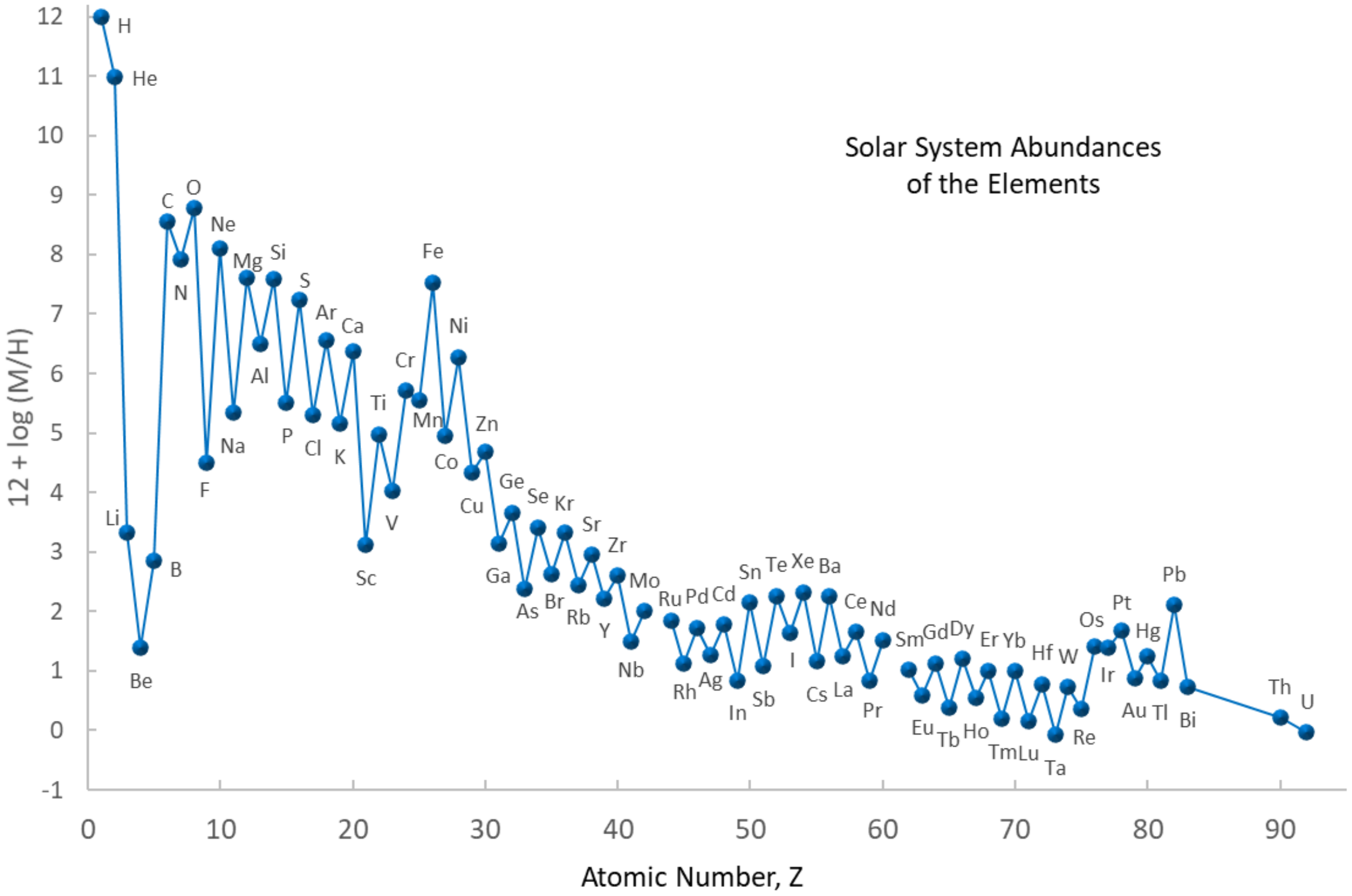

**Figure 4**. Solar system abundances of the elements as a function of atomic number.

**Table 8. Recommended Atomic Solar System Abundances: Present and Proto-Solar 4.567 Ga Ago**

| Z | E | CI-Chondrites | | Sun Outer Convection Zone (mainly photosphere) | | Solar System * Present | | Solar System* Proto-Solar †,‡ | | | Solar System * Present | | Solar System* Proto-Solar †,‡ | |
|---|---|---|---|---|---|---|---|---|---|---|---|---|---|---|
| | . | Si = $10^6$ | ±σ | Si = $10^6$ | ±σ | Si = $10^6$ | ±σ | Si = $10^6$ | ±σ | . | log (X/H)+12 | ±σ | log (X/H)+12 | ±σ |
| 1 | H | 4.83E+6 | 4.5E+5 | 3.09E+10 | | 3.09E+10 | | 2.52E+10 | | s | 12 | | 12 | |
| 2 | He | 0.599 | | 2.59E+9 | 1.2E+8 | 2.59E+9 | 1.2E+8 | 2.51E+9 | 1.2E+8 | s | 10.924 | 0.02 | 10.994 | 0.02 |
| 3 | Li | 56.9 | 3.4 | 0.339 | 0.088 | (0.339, sun) | 0.088 | 56.9 | 3.4 | | 3.27 | 0.03 | 3.35 | 0.03 |
| 4 | Be | 0.637 | 0.044 | 0.741 | 0.171 | 0.637 | 0.044 | 0.637 | 0.044 | | 1.31 | 0.04 | 1.40 | 0.04 |
| 5 | B | 18 | 1.3 | 15.5 | 9.1 | 18 | 1.3 | 18 | 1.3 | | 2.77 | 0.03 | 2.85 | 0.03 |
| 6 | C | 9.44E+5 | 1.79E+5 | 9.12E+6 | 1.35E+6 | 9.12E+6 | 1.8E+5 | 9.12E+6 | 1.8E+5 | s | 8.47 | 0.06 | 8.56 | 0.06 |
| 7 | N | 44830 | 9700 | 2.19E+6 | 7.0E+5 | 2.19E+6 | 9700 | 2.19E+6 | 9700 | s | 7.85 | 0.12 | 7.94 | 0.12 |
| 8 | O | 7.38E+6 | 1.3E+5 | 1.66E+7 | 2.9E+6 | 1.66E+7 | 1.3E+5 | 1.66E+7 | 1.3E+5 | s | 8.73 | 0.07 | 8.82 | 0.07 |
| 9 | F | 1270 | 275 | 780 | 600 | 1270 | 275 | 1270 | 275 | | 4.61 | 0.09 | 4.79 | 0.09 |
| 10 | Ne | 0.00234 | | 4.37E+6 | 1.13E+6 | 4.37E+6 | 1.13E+6 | 4.37E+6 | 1.13E+6 | s,o | 8.15 | 0.10 | 8.24 | 0.10 |
| 11 | Na | 57800 | 4700 | 50120 | 4840 | 57800 | 4700 | 57800 | 4700 | | 6.27 | 0.03 | 6.36 | 0.03 |
| 12 | Mg | 1.03E+6 | 4.4E+4 | 1.12E+6 | 1.4E+5 | 1.03E+6 | 4.4E+4 | 1.03E+6 | 4.4E+4 | | 7.52 | 0.02 | 7.61 | 0.02 |
| 13 | Al | 81820 | 6110 | 83180 | 8030 | 81820 | 6110 | 81820 | 6110 | | 6.42 | 0.03 | 6.51 | 0.03 |
| 14 | Si | 1.00E+6 | 3.4E+4 | 1.00E+6 | 7E+4 | 1.00E+6 | 3.4E+4 | 1.00E+6 | 3.4E+4 | | 7.51 | 0.01 | 7.60 | 0.01 |
| 15 | P | 8260 | 530 | 8430 | 810 | 8260 | 530 | 8260 | 530 | | 5.43 | 0.03 | 5.52 | 0.03 |
| 16 | S | 4.37E+5 | 2.6E+4 | 4.37E+5 | 5.3E+4 | 4.37E+5 | 2.6E+4 | 4.37E+5 | 2.6E+4 | | 7.15 | 0.03 | 7.24 | 0.03 |
| 17 | Cl | 5290 | 810 | 5495 | 1750 | 5290 | 810 | 5290 | 810 | | 5.23 | 0.06 | 5.32 | 0.06 |
| 18 | Ar | 0.00959 | | … | | 97700 | 25300 | 97700 | 25300 | s,o | 6.50 | 0.10 | 6.59 | 0.10 |
| 19 | K | 3610 | 190 | 3685 | 850 | 3606 | 190 | 3611 | 190 | | 5.07 | 0.02 | 5.16 | 0.02 |
| 20 | Ca | 57239 | 4500 | 64565 | 4620 | 57239 | 4500 | 57234 | 4500 | | 6.27 | 0.03 | 6.36 | 0.03 |
| 21 | Sc | 33.7 | 2.2 | 44.7 | 4.3 | 33.7 | 2.2 | 33.7 | 2.2 | | 3.04 | 0.03 | 3.13 | 0.03 |
| 22 | Ti | 2460 | 150 | 2630 | 254 | 2459 | 150 | 2459 | 150 | | 4.90 | 0.03 | 4.99 | 0.03 |
| 23 | V | 275 | 18 | 240 | 49 | 275 | 18 | 275 | 18 | | 3.95 | 0.03 | 4.04 | 0.03 |
| 24 | Cr | 13130 | 460 | 12880 | 1240 | 13130 | 460 | 13130 | 460 | | 5.63 | 0.02 | 5.72 | 0.02 |
| 25 | Mn | 9090 | 620 | 10230 | 730 | 9090 | 620 | 9090 | 620 | | 5.47 | 0.03 | 5.56 | 0.03 |
| 26 | Fe | 8.72E+5 | 3.8E+4 | 9.33E+5 | 9.0E+4 | 8.72E+5 | 3.8E+4 | 8.72E+5 | 3.8E+4 | | 7.45 | 0.02 | 7.54 | 0.02 |
| 27 | Co | 2260 | 100 | 2630 | 320 | 2260 | 100 | 2260 | 100 | | 4.86 | 0.02 | 4.95 | 0.02 |
| 28 | Ni | 48670 | 2940 | 48980 | 4730 | 48670 | 2940 | 48670 | 2940 | | 6.20 | 0.03 | 6.29 | 0.03 |
| 29 | Cu | 535 | 50 | 468 | 57 | 535 | 50 | 535 | 50 | | 4.24 | 0.04 | 4.33 | 0.04 |
| 30 | Zn | 1260 | 180 | 1120 | 140 | 1260 | 180 | 1260 | 180 | | 4.61 | 0.60 | 4.70 | 0.60 |
| 31 | Ga | 36.2 | 1.8 | 32.4 | 3.9 | 36.2 | 1.8 | 36.2 | 1.8 | | 3.07 | 0.02 | 3.16 | 0.02 |
| 32 | Ge | 120 | 7 | 132 | 23 | 120 | 7 | 120 | 7 | | 3.59 | 0.03 | 3.68 | 0.03 |
| 33 | As | 6.07 | 0.50 | … | | 6.07 | 0.50 | 6.07 | 0.50 | | 2.29 | 0.03 | 2.38 | 0.03 |
| 34 | Se | 67.6 | 5.0 | … | | 67.6 | 5.0 | 67.6 | 5.0 | | 3.34 | 0.03 | 3.43 | 0.03 |
| 35 | Br | 12.3 | 2.9 | … | | 12.3 | 2.9 | 12.3 | 2.9 | | 2.60 | 0.09 | 2.69 | 0.09 |
| 36 | Kr | 1.63E-04 | | … | | 51.3 | 10.4 | 51.3 | 10.4 | o | 3.22 | 0.08 | 3.31 | 0.08 |
| 37 | Rb | 7.04 | 0.46 | 9.12 | 1.60 | 7.04 | 0.46 | 7.17 | 0.47 | | 2.36 | 0.03 | 2.45 | 0.03 |
| 38 | Sr | 23.4 | 1.3 | 20.9 | 3.1 | 23.4 | 1.3 | 23.3 | 1.3 | | 2.88 | 0.02 | 2.97 | 0.02 |
| 39 | Y | 4.35 | 0.24 | 5.01 | 0.61 | 4.35 | 0.24 | 4.35 | 0.24 | | 2.15 | 0.02 | 2.24 | 0.02 |
| 40 | Zr | 10.9 | 1.0 | 12.5 | 1.84 | 10.9 | 1.0 | 10.9 | 1.0 | | 2.55 | 0.04 | 2.64 | 0.04 |
| 41 | Nb | 0.780 | 0.070 | 0.912 | 0.135 | 0.78 | 0.070 | 0.78 | 0.070 | | 1.4 | 0.04 | 1.49 | 0.04 |
| 42 | Mo | 2.6 | 0.26 | 2.34 | 0.54 | 2.6 | 0.26 | 2.6 | 0.26 | | 1.92 | 0.04 | 2.01 | 0.04 |
| 43 | Tc | | | | | | | | | | | | | |
| 44 | Ru | 1.81 | 0.02 | 1.74 | 0.35 | 1.81 | 0.02 | 1.81 | 0.02 | | 1.77 | 0.01 | 1.86 | 0.01 |
| 45 | Rh | 0.338 | 0.015 | 0.240 | 0.049 | 0.338 | 0.015 | 0.338 | 0.015 | | 1.04 | 0.02 | 1.13 | 0.02 |
| 46 | Pd | 1.38 | 0.07 | 1.10 | 0.16 | 1.38 | 0.07 | 1.38 | 0.07 | | 1.65 | 0.02 | 1.74 | 0.02 |
| 47 | Ag | 0.497 | 0.022 | 0.282 | 0.073 | 0.497 | 0.022 | 0.497 | 0.022 | | 1.21 | 0.02 | 1.29 | 0.02 |
| 48 | Cd | 1.58 | 0.06 | 1.82 | 0.75 | 1.58 | 0.06 | 1.58 | 0.06 | | 1.71 | 0.02 | 1.80 | 0.02 |
| 49 | In | 0.179 | 0.008 | 0.195 | 0.114 | 0.179 | 0.008 | 0.179 | 0.008 | | 0.76 | 0.02 | 0.85 | 0.02 |
| 50 | Sn | 3.59 | 0.22 | 3.24 | 0.84 | 3.59 | 0.22 | 3.59 | 0.22 | | 2.07 | 0.03 | 2.15 | 0.03 |
| 51 | Sb | 0.359 | 0.045 | … | | 0.359 | 0.045 | 0.359 | 0.045 | | 1.06 | 0.05 | 1.15 | 0.05 |
| 52 | Te | 4.72 | 0.23 | … | | 4.72 | 0.23 | 4.72 | 0.23 | | 2.18 | 0.02 | 2.27 | 0.02 |
| 53 | I | 1.59 | 0.64 | … | | 1.59 | 0.64 | 1.59 | 0.64 | | 1.71 | 0.15 | 1.80 | 0.15 |
| 54 | Xe | 3.47E-04 | | … | | 5.50 | 1.11 | 5.50 | 1.11 | o | 2.25 | 0.08 | 2.34 | 0.08 |
| 55 | Cs | 0.368 | 0.043 | … | | 0.368 | 0.043 | 0.368 | 0.043 | | 1.08 | 0.02 | 1.16 | 0.02 |
| 56 | Ba | 4.55 | 0.27 | 5.5 | 0.92 | 4.55 | 0.27 | 4.55 | 0.27 | | 2.17 | 0.02 | 2.26 | 0.02 |
| 57 | La | 0.459 | 0.024 | 0.398 | 0.038 | 0.459 | 0.024 | 0.459 | 0.024 | | 1.17 | 0.02 | 1.26 | 0.02 |
| 58 | Ce | 1.16 | 0.05 | 1.17 | 0.11 | 1.16 | 0.05 | 1.16 | 0.05 | | 1.58 | 0.02 | 1.66 | 0.02 |
| 59 | Pr | 0.175 | 0.012 | 0.162 | 0.016 | 0.175 | 0.012 | 0.175 | 0.012 | | 0.75 | 0.01 | 0.84 | 0.01 |
| 60 | Nd | 0.865 | 0.022 | 0.813 | 0.078 | 0.865 | 0.022 | 0.864 | 0.022 | | 1.45 | 0.01 | 1.53 | 0.01 |
| 61 | Pm | … | | … | | … | | … | | | | … | | … |
| 62 | Sm | 0.271 | 0.012 | 0.275 | 0.027 | 0.271 | 0.012 | 0.273 | 0.012 | | 0.94 | 0.02 | 1.03 | 0.02 |
| 63 | Eu | 0.100 | 0.005 | 0.102 | 0.010 | 0.1 | 0.005 | 0.1 | 0.005 | | 0.51 | 0.02 | 0.60 | 0.02 |
| 64 | Gd | 0.346 | 0.013 | 0.372 | 0.036 | 0.346 | 0.013 | 0.346 | 0.013 | | 1.05 | 0.02 | 1.14 | 0.02 |
| 65 | Tb | 0.0625 | 0.0025 | 0.0631 | 0.0163 | 0.0625 | 0.0025 | 0.0625 | 0.0025 | | 0.31 | 0.02 | 0.39 | 0.02 |
| 66 | Dy | 0.407 | 0.016 | 0.389 | 0.038 | 0.407 | 0.016 | 0.407 | 0.016 | | 1.12 | 0.02 | 1.21 | 0.02 |
| 67 | Ho | 0.0891 | 0.0035 | 0.0933 | 0.0269 | 0.0891 | 0.0035 | 0.0891 | 0.0035 | | 0.46 | 0.02 | 0.55 | 0.02 |
| 68 | Er | 0.256 | 0.009 | 0.263 | 0.032 | 0.256 | 0.009 | 0.256 | 0.009 | | 0.92 | 0.02 | 1.01 | 0.02 |
| 69 | Tm | 0.0403 | 0.0014 | 0.0398 | 0.0038 | 0.0403 | 0.0014 | 0.0403 | 0.0014 | | 0.11 | 0.02 | 0.20 | 0.02 |
| 70 | Yb | 0.252 | 0.009 | 0.219 | 0.063 | 0.252 | 0.009 | 0.252 | 0.009 | | 0.91 | 0.02 | 1.00 | 0.02 |
| 71 | Lu | 0.038 | 0.0018 | 0.0389 | 0.0090 | 0.038 | 0.0018 | 0.0381 | 0.0018 | | 0.09 | 0.02 | 0.18 | 0.02 |
| 72 | Hf | 0.155 | 0.011 | 0.219 | 0.027 | 0.155 | 0.011 | 0.155 | 0.011 | | 0.70 | 0.03 | 0.79 | 0.03 |
| 73 | Ta | 0.0215 | 0.0010 | … | | 0.0215 | 0.0010 | 0.0215 | 0.0010 | | -0.16 | 0.02 | -0.07 | 0.02 |
| 74 | W | 0.144 | 0.013 | 0.209 | 0.060 | 0.144 | 0.013 | 0.144 | 0.013 | | 0.67 | 0.04 | 0.76 | 0.04 |
| 75 | Re | 0.0521 | 0.0042 | … | | 0.0521 | 0.0042 | 0.0547 | 0.0042 | | 0.23 | 0.03 | 0.34 | 0.03 |
| 76 | Os | 0.655 | 0.040 | 0.776 | 0.095 | 0.655 | 0.040 | 0.652 | 0.040 | | 1.33 | 0.03 | 1.41 | 0.03 |
| 77 | Ir | 0.633 | 0.029 | 0.813 | 0.142 | 0.633 | 0.029 | 0.633 | 0.029 | | 1.31 | 0.02 | 1.40 | 0.02 |
| 78 | Pt | 1.24 | 0.10 | … | | 1.24 | 0.10 | 1.24 | 0.10 | | 1.60 | 0.03 | 1.69 | 0.03 |
| 79 | Au | 0.195 | 0.016 | 0.251 | 0.051 | 0.195 | 0.016 | 0.195 | 0.016 | | 0.80 | 0.03 | 0.89 | 0.03 |
| 80 | Hg | 0.376 | 0.156 | … | | 0.376 | 0.156 | 0.376 | 0.156 | | 1.08 | 0.15 | 1.17 | 0.15 |
| 81 | Tl | 0.179 | 0.015 | 0.275 | 0.161 | 0.179 | 0.015 | 0.179 | 0.015 | | 0.76 | 0.04 | 0.85 | 0.04 |
| 82 | Pb | 3.33 | 0.20 | 2.57 | 0.52 | 3.33 | 0.20 | 3.31 | 0.20 | | 2.03 | 0.03 | 2.12 | 0.03 |
| 83 | Bi | 0.141 | 0.010 | … | | 0.141 | 0.010 | 0.141 | 0.010 | | 0.66 | 0.03 | 0.75 | 0.03 |
| 90 | Th | 0.0336 | 0.0017 | ≤ 0.033 | | 0.0336 | 0.0017 | 0.0421 | 0.0021 | | 0.04 | 0.02 | 0.22 | 0.02 |
| 92 | U | 0.00897 | 0.00064 | … | | 0.00897 | 0.00064 | 0.02389 | 0.00170 | | -0.54 | 0.03 | -0.02 | 0.03 |

**Footnote for Table 8**: * based on CI-chondrites except as indicated: s = from sun (Table 6, recommended 3D values), o = by other means. † corrected for radioactive decay. ‡ photospheric values corrected element settling from the outer convection zone.

## Solar System Abundances of the Isotopes

The elemental abundance features in Figure 4 are better discernable in a plot of the isotopic (or nuclide) abundances as a function of mass number, which is shown in Figure 5. This type of plot has been shown in many references indicated in Table 6, and was more influential in the development of nuclear theory and nucleosynthesis (e.g., Suess 1947a,b,  Burbidge et al. 1957, Cameron 1957) than the atomic elemental abundance plot in Figure 4 (which, however, instigated early theories, e.g., Harkins 1917). The corresponding nuclide plot reveals the control of abundances by nuclear stability and peaks at certain mass-numbers correspond to nuclei with "magic" proton and neutron numbers.

The The nuclide abundances listed in Table 9 and shown in Figure 5 were calculated from the elemental proto-solar abundances in Table 8 and the isotopic composition (atom-percent) of the elements largely taken from the review by Meija et al. (2016) and some references described below. Several updates and changes to Meija (2016) are included but not all listed here and will be described elsewhere. Compositions for elements with long-lived isotopes were computed back to an age of 4.567 Ga, and chondritic isotopic compositions were preferred, as described in Lodders (2003) and Lodders et al. (2009).

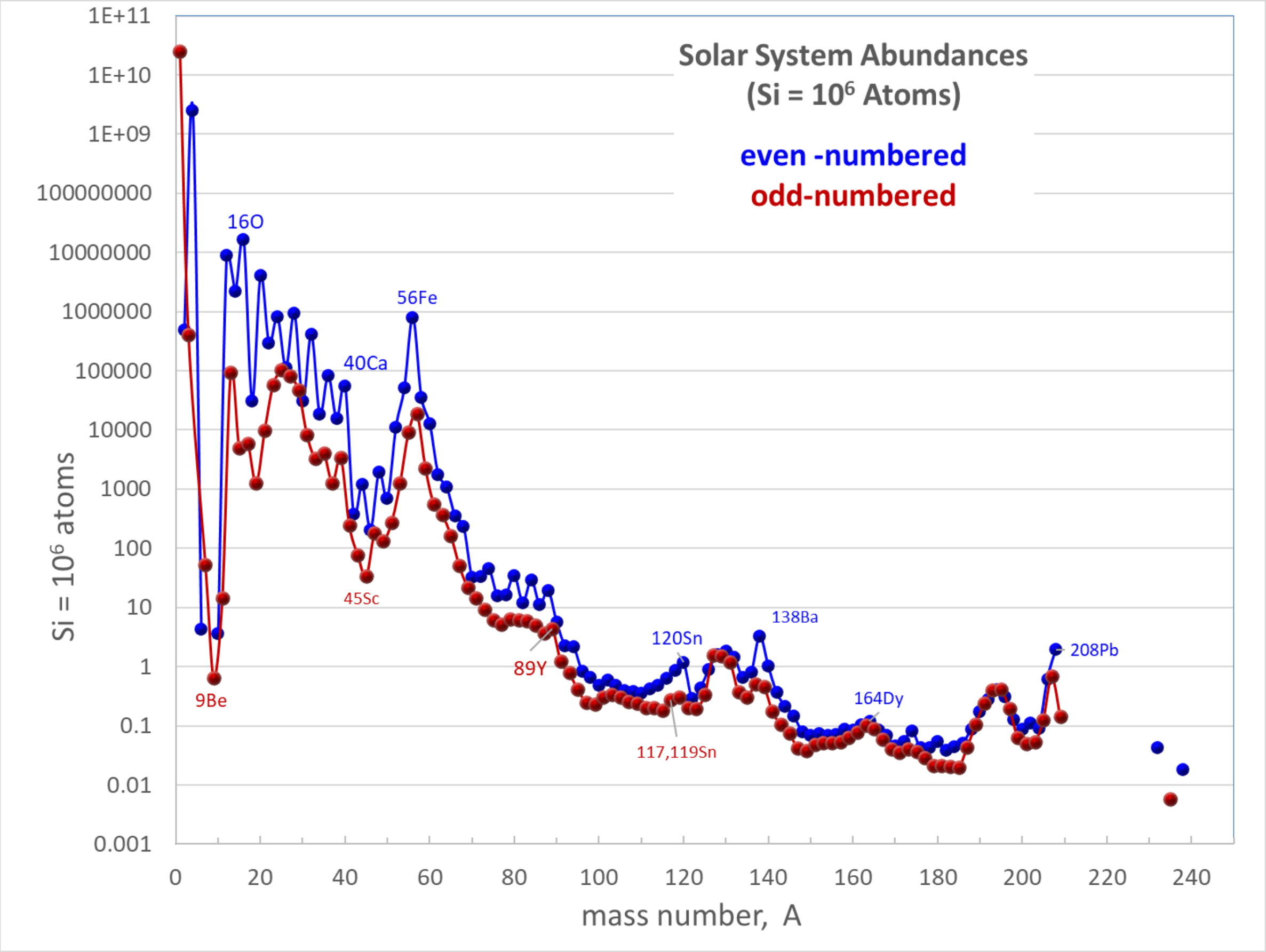

**Figure 5.** Abundances of the isotopes as a function of mass number. Data from Table 9. Even numbered nuclides are shown in blue, odd-numbered nuclides in red.

Except for H, the noble gases, C, N, and O, the isotopic composition of materials within the solar system are relatively homogeneous on a one-percent level. Isotopic variations in meteorites and planetary materials are often at the per-mil level for heavier elements, which are much smaller than the uncertainties (usually more than 5-10%) associated with the elemental abundances. In these cases variations in the relative nuclide abundance distribution (Figure 5) are much more sensitive to elemental variations than to isotopic variations of a given element.

Exceptions with large ranges in observed isotopic compositions are H and D, because deuterium burning in the pre-main sequence sun left the sun essentially D-free, and D enrichments in condensed materials in meteorites and comets often stem from ion-molecule reactions. The proto-solar D/H ratio ($1.97(\pm 0.35)\times 10^{-5}$;Geiss & Gloeckler 2003) is derived from comparing the helium isotopic measurements of Jupiter's atmosphere with that of the solar wind and mass-balance calculations accounting for $^{3}$He produced by deuterium burning. However, direct measurements of the D/H ratio from Jupiter may not be representative according to Geiss & Gloeckler (2003). There is a large range in D/H ratios in cometary and meteoritic materials, and an increase of D/H ratios in the giant planets' atmospheres with increasing distance from the sun and the "correct" D/H ratio for the solar system is debatable. Related to the D/H ratio is the helium isotopic composition. The value of $^{3}$He/$^{4}$He of $1.66(\pm 0.05)\times 10^{-4}$ of Jupiter's atmosphere is adopted as proto-solar value (Mahaffy et al. 1998, Geiss & Gloeckler 2003). The present solar $^{3}$He/$^{4}$He ratio is higher than the proto-solar value because $^{3}$He is a product of D-burning. The solar wind $^{3}$He/$^{4}$He from in-situ measurements shows that the various solar wind regimes are isotopically fractionated. Further very $^{3}$He-enriched SEPs from solar flares have been observed (e.g. Mason et al. 2004). This is also seen with much higher precision in Genesis regime targets (Heber et al. 2012) and the fractionation is large for He, and smaller for the heavier Ne and Ar. Note that the derivation of the proto-solar D/H ratio depends on the $^{3}$He/$^{4}$He ratio of the present-day sun, hence differences in reported proto-solar D/H values exist (see Geiss & Gloeckler 2003, Wieler 2016, Trieloff 2018, this encyclopedia, for alternate choices).

The C, N, and O isotopic compositions in solar system materials are more variable and can reach 10-100 per-mil levels. Measurements of the photospheric C and O isotopic compositions have large uncertainties. The recent solar wind measurements give a light isotopic composition of the solar wind (i.e., the lower-mass isotope of an element is more abundant). Relative to Earth and most meteorites, the solar wind nitrogen is poor in $^{15}$N, and has a similar $^{14}$N/$^{15}$N = 440 as found for Jupiter (for more discussion see Marty et al 2010, Füri & Marty 2015), which is adopted for the isotope Table here. The oxygen isotopic composition of the solar wind is depleted in $^{17}$O, $^{18}$O by about 7% relative to Earth and most meteoritic materials (McKeegan et al. 2011, Laming et al. 2017). The C-isotopic composition is still relatively close to the terrestrial and meteoritic values. In Table 9, the noble gases (except helium) data are in Crowther & Gilmour 2013, Heber et al. (2009, 2012), Pepin et al. (2012), Vogel et al. (2011), and Meshik et al. (2014). Terrestrial atmospheric noble gas isotopic compositions are probably a worse proxy to the unknown isotopic noble gas solar compositions than are the solar wind measurements. However, the solar wind isotopic compositions may not be representative for the solar system as a whole because the solar wind is fractionated in elemental composition relative to the photosphere and meteorites, and corresponding mass-dependent isotopic fractionations are observed in the solar wind. However, theoretical corrections to mass-dependent fractionations of the adopted solar-wind isotopic compositions were not applied here because the they are not yet well-understood and vary with different authors (e.g., Heber et al. 2012, Ott 2015). On the other hand, if the solar wind is isotopically light (as seen in C,N,O the noble gases and probably other elements) and were representative for the solar system as a whole, the reasons for the differences between the sun and the planets, moons, asteroids, and comets await more explanations. More needs to be explored for the elemental and isotopic abundances of the solar system, and while great progress has been achieved in the past century, more discovery awaits.

**Table 9. Abundances of the Isotopes in the Solar System Scaled to Si=$10^6$ Atoms.**

| Z | E | A | atom% | N |
|---|---|---|---|---|
| 1 | H | 1 | 99.9980 | 2.52E+10 |
| 1 | H | 2 | 0.00197 | 4.97E+5 |
| | | | 100 | 2.52E+10 |
| 2 | He | 3 | 0.0166 | 4.13E+5 |
| 2 | He | 4 | 99.9834 | 2.49E+9 |
| | | | 100 | 2.49E+9 |
| 3 | Li | 6 | 7.589 | 4.3 |
| 3 | Li | 7 | 92.411 | 52.6 |
| | | | 100 | 56.9 |
| 4 | Be | 9 | 100 | 0.637 |
| 5 | B | 10 | 19.827 | 3.6 |
| 5 | B | 11 | 80.173 | 14.4 |
| | | | 100 | 18.0 |
| 6 | C | 12 | 98.965 | 9.03E+6 |
| 6 | C | 13 | 1.035 | 9.44E+4 |
| | | | 100 | 9.12E+6 |
| 7 | N | 14 | 99.774 | 2.18E+6 |
| 7 | N | 15 | 0.226 | 4.94E+3 |
| | | | 100 | 2.19E+6 |
| 8 | O | 16 | 99.777 | 1.66E+7 |
| 8 | O | 17 | 0.035 | 5.80E+3 |
| 8 | O | 18 | 0.188 | 3.13E+4 |
| | | | 100 | 1.66E+7 |
| 9 | F | 19 | 100 | 1267 |
| 10 | Ne | 20 | 93.1251 | 4.07E+6 |
| 10 | Ne | 21 | 0.2236 | 9.76E+3 |
| 10 | Ne | 22 | 6.6513 | 2.90E+5 |
| | | | 100 | 4.37E+6 |
| 11 | Na | 23 | 100 | 57800 |
| 12 | Mg | 24 | 78.992 | 8.10E+5 |
| 12 | Mg | 25 | 10.003 | 1.03E+5 |
| 12 | Mg | 26 | 11.005 | 1.13E+5 |
| | | | 100 | 1.03E+6 |
| 13 | Al | 27 | 100 | 8.18E+4 |
| 14 | Si | 28 | 92.230 | 9.22E+5 |
| 14 | Si | 29 | 4.683 | 4.68E+4 |
| 14 | Si | 30 | 3.087 | 3.09E+4 |
| | | | 100 | 1.00E+6 |
| 15 | P | 31 | 100 | 8260 |
| 16 | S | 32 | 95.04074 | 415600 |
| 16 | S | 33 | 0.74869 | 3270 |
| 16 | S | 34 | 4.19599 | 18300 |
| 16 | S | 36 | 0.01458 | 64 |
| | | | 100 | 437300 |
| 17 | Cl | 35 | 75.7647 | 4010 |
| 17 | Cl | 37 | 24.2353 | 1280 |
| | | | 100 | 5290 |
| 18 | Ar | 36 | 84.298 | 82400 |
| 18 | Ar | 38 | 15.698 | 15300 |
| 18 | Ar | 40 | 0.004 | 4 |
| | | | 100 | 97700 |
| 19 | K | 39 | 93.132 | 3360 |
| 19 | K* | 40 | 0.147 | 5 |
| 19 | K | 41 | 6.721 | 243 |
| | | | 100 | 3610 |

| Z | E | A | atom% | N |
|---|---|---|---|---|
| 20 | Ca | 40 | 96.941 | 55500 |
| 20 | Ca | 42 | 0.647 | 370 |
| 20 | Ca | 43 | 0.135 | 77 |
| 20 | Ca | 44 | 2.086 | 1194 |
| 20 | Ca | 46 | 0.004 | 2 |
| 20 | Ca | 48 | 0.187 | 107 |
| | | | 100 | 57200 |
| 21 | Sc | 45 | 100 | 33.7 |
| 22 | Ti | 46 | 8.249 | 203 |
| 22 | Ti | 47 | 7.437 | 183 |
| 22 | Ti | 48 | 73.72 | 1810 |
| 22 | Ti | 49 | 5.409 | 133 |
| 22 | Ti | 50 | 5.185 | 128 |
| | | | 100 | 2460 |
| 23 | V | 50 | 0.2497 | 0.7 |
| 23 | V | 51 | 99.7503 | 274.5 |
| | | | 100 | 275.2 |
| 24 | Cr | 50 | 4.3452 | 571 |
| 24 | Cr | 52 | 83.7895 | 11000 |
| 24 | Cr | 53 | 9.5006 | 1250 |
| 24 | Cr | 54 | 2.3647 | 310 |
| | | | 100 | 13100 |
| 25 | Mn | 55 | 100 | 9090 |
| 26 | Fe | 54 | 5.845 | 51000 |
| 26 | Fe | 56 | 91.754 | 8.01E+5 |
| 26 | Fe | 57 | 2.1191 | 18500 |
| 26 | Fe | 58 | 0.2819 | 2460 |
| | | | 100 | 8.73E+5 |
| 27 | Co | 59 | 100 | 2260 |
| 28 | Ni | 58 | 68.0769 | 33100 |
| 28 | Ni | 60 | 26.2231 | 12800 |
| 28 | Ni | 61 | 1.1399 | 555 |
| 28 | Ni | 62 | 3.6345 | 1770 |
| 28 | Ni | 64 | 0.9256 | 450 |
| | | | 100 | 48700 |
| 29 | Cu | 63 | 69.174 | 370 |
| 29 | Cu | 65 | 30.826 | 165 |
| | | | 100 | 535 |
| 30 | Zn | 64 | 49.1704 | 620 |
| 30 | Zn | 66 | 27.7306 | 349 |
| 30 | Zn | 67 | 4.0401 | 51 |
| 30 | Zn | 68 | 18.4483 | 233 |
| 30 | Zn | 70 | 0.6106 | 8 |
| | | | 100 | 1260 |
| 31 | Ga | 69 | 60.108 | 21.7 |
| 31 | Ga | 71 | 39.892 | 14.4 |
| | | | 100 | 36.2 |
| 32 | Ge | 70 | 20.526 | 24.7 |
| 32 | Ge | 72 | 27.446 | 33.0 |
| 32 | Ge | 73 | 7.76 | 9.3 |
| 32 | Ge | 74 | 36.523 | 43.9 |
| 32 | Ge | 76 | 7.745 | 9.3 |
| | | | 100 | 120 |
| 33 | As | 75 | 100 | 6.07 |

| Z | E | A | atom% | N |
|---|---|---|---|---|
| 34 | Se | 74 | 0.86 | 0.58 |
| 34 | Se | 76 | 9.22 | 6.23 |
| 34 | Se | 77 | 7.59 | 5.13 |
| 34 | Se | 78 | 23.69 | 16.00 |
| 34 | Se | 80 | 49.81 | 33.66 |
| 34 | Se | 82 | 8.83 | 5.96 |
| | | | 100 | 67.6 |
| 35 | Br | 79 | 50.686 | 6.26 |
| 35 | Br | 81 | 49.314 | 6.09 |
| | | | 100 | 12.3 |
| 36 | Kr | 78 | 0.362 | 0.19 |
| 36 | Kr | 80 | 2.326 | 1.20 |
| 36 | Kr | 82 | 11.655 | 5.99 |
| 36 | Kr | 83 | 11.546 | 5.94 |
| 36 | Kr | 84 | 56.903 | 29.18 |
| 36 | Kr | 86 | 17.208 | 8.79 |
| | | | 100 | 51.3 |
| 37 | Rb | 85 | 70.844 | 5.080 |
| 37 | Rb* | 87 | 29.156 | 2.091 |
| | | | 100 | 7.17 |
| 38 | Sr | 84 | 0.5584 | 0.13 |
| 38 | Sr | 86 | 9.8708 | 2.30 |
| 38 | Sr | 87 | 6.8982 | 1.61 |
| 38 | Sr | 88 | 82.6725 | 19.2 |
| | | | 100 | 23.3 |
| 39 | Y | 89 | 100 | 4.35 |
| 40 | Zr | 90 | 51.452 | 5.621 |
| 40 | Zr | 91 | 11.223 | 1.226 |
| 40 | Zr | 92 | 17.146 | 1.873 |
| 40 | Zr | 94 | 17.38 | 1.899 |
| 40 | Zr^ | 96 | 2.799 | 0.306 |
| | | | 100 | 10.92 |
| 41 | Nb | 93 | 100 | 0.780 |
| 42 | Mo | 92 | 14.525 | 0.380 |
| 42 | Mo | 94 | 9.151 | 0.238 |
| 42 | Mo | 95 | 15.838 | 0.412 |
| 42 | Mo | 96 | 16.672 | 0.433 |
| 42 | Mo | 97 | 9.599 | 0.249 |
| 42 | Mo | 98 | 24.391 | 0.630 |
| 42 | Mo^ | 100 | 9.824 | 0.253 |
| | | | 100 | 2.60 |
| 44 | Ru | 96 | 5.542 | 0.100 |
| 44 | Ru | 98 | 1.869 | 0.034 |
| 44 | Ru | 99 | 12.758 | 0.230 |
| 44 | Ru | 100 | 12.599 | 0.228 |
| 44 | Ru | 101 | 17.060 | 0.308 |
| 44 | Ru | 102 | 31.552 | 0.570 |
| 44 | Ru | 104 | 18.621 | 0.336 |
| | | | 100 | 1.81 |
| 45 | Rh | 103 | 100 | 0.338 |
| 46 | Pd | 102 | 1.02 | 0.0141 |
| 46 | Pd | 104 | 11.14 | 0.1536 |
| 46 | Pd | 105 | 22.33 | 0.3079 |
| 46 | Pd | 106 | 27.33 | 0.377 |
| 46 | Pd | 108 | 26.46 | 0.365 |
| 46 | Pd | 110 | 11.72 | 0.162 |
| | | | 100 | 1.38 |

| Z | E | A | atom% | N |
|---|---|---|---|---|
| 47 | Ag | 107 | 51.839 | 0.258 |
| 47 | Ag | 109 | 48.161 | 0.239 |
| | | | 100 | 0.497 |
| 48 | Cd | 106 | 1.249 | 0.020 |
| 48 | Cd | 108 | 0.89 | 0.014 |
| 48 | Cd | 110 | 12.485 | 0.197 |
| 48 | Cd | 111 | 12.804 | 0.202 |
| 48 | Cd | 112 | 24.117 | 0.381 |
| 48 | Cd^ | 113 | 12.225 | 0.193 |
| 48 | Cd | 114 | 28.729 | 0.454 |
| 48 | Cd^ | 116 | 7.501 | 0.119 |
| | | | 100 | 1.58 |
| 49 | In | 113 | 4.281 | 0.008 |
| 49 | In^ | 115 | 95.719 | 0.171 |
| | | | 100 | 0.179 |
| 50 | Sn | 112 | 0.971 | 0.035 |
| 50 | Sn | 114 | 0.659 | 0.024 |
| 50 | Sn | 115 | 0.339 | 0.012 |
| 50 | Sn | 116 | 14.536 | 0.522 |
| 50 | Sn | 117 | 7.676 | 0.276 |
| 50 | Sn | 118 | 24.223 | 0.870 |
| 50 | Sn | 119 | 8.585 | 0.308 |
| 50 | Sn | 120 | 32.593 | 1.171 |
| 50 | Sn | 122 | 4.629 | 0.166 |
| 50 | Sn | 124 | 5.789 | 0.208 |
| | | | 100 | 3.59 |
| 51 | Sb | 121 | 57.213 | 0.205 |
| 51 | Sb | 123 | 42.787 | 0.154 |
| | | | 100 | 0.359 |
| 52 | Te | 120 | 0.096 | 0.005 |
| 52 | Te | 122 | 2.603 | 0.123 |
| 52 | Te^ | 123 | 0.908 | 0.043 |
| 52 | Te | 124 | 4.816 | 0.227 |
| 52 | Te | 125 | 7.139 | 0.337 |
| 52 | Te | 126 | 18.952 | 0.894 |
| 52 | Te^ | 128 | 31.687 | 1.494 |
| 52 | Te^ | 130 | 33.799 | 1.594 |
| | | | 100 | 4.72 |
| 53 | I | 127 | 100 | 1.59 |
| 54 | Xe^ | 124 | 0.129 | 0.007 |
| 54 | Xe | 126 | 0.110 | 0.006 |
| 54 | Xe | 128 | 2.220 | 0.122 |
| 54 | Xe | 129 | 27.428 | 1.507 |
| 54 | Xe | 130 | 4.349 | 0.239 |
| 54 | Xe | 131 | 21.763 | 1.196 |
| 54 | Xe | 132 | 26.360 | 1.449 |
| 54 | Xe | 134 | 9.730 | 0.535 |
| 54 | Xe^ | 136 | 7.911 | 0.435 |
| | | | 100 | 5.495 |
| 55 | Cs | 133 | 100 | 0.368 |
| 56 | Ba^ | 130 | 0.106 | 0.005 |
| 56 | Ba | 132 | 0.101 | 0.005 |
| 56 | Ba | 134 | 2.417 | 0.110 |
| 56 | Ba | 135 | 6.592 | 0.300 |
| 56 | Ba | 136 | 7.853 | 0.357 |
| 56 | Ba | 137 | 11.232 | 0.511 |
| 56 | Ba | 138 | 71.699 | 3.264 |
| | | | 100 | 4.552 |
| 57 | La* | 138 | 0.092 | 0.000 |
| 57 | La | 139 | 99.908 | 0.459 |
| | | | 100 | 0.459 |

| Z | E | A | atom% | N |
|---|---|---|---|---|
| 58 | Ce | 136 | 0.186 | 0.002 |
| 58 | Ce^ | 138 | 0.250 | 0.003 |
| 58 | Ce | 140 | 88.450 | 1.030 |
| 58 | Ce^ | 142 | 11.114 | 0.129 |
| | | | 100 | 1.165 |
| 59 | Pr | 141 | 100 | 0.175 |
| 60 | Nd | 142 | 27.045 | 0.234 |
| 60 | Nd | 143 | 12.021 | 0.104 |
| 60 | Nd^ | 144 | 23.729 | 0.205 |
| 60 | Nd | 145 | 8.763 | 0.076 |
| 60 | Nd | 146 | 17.130 | 0.148 |
| 60 | Nd | 148 | 5.716 | 0.049 |
| 60 | Nd^ | 150 | 5.596 | 0.048 |
| | | | 100 | 0.864 |
| 62 | Sm | 144 | 3.083 | 0.008 |
| 62 | Sm* | 147 | 15.017 | 0.042 |
| 62 | Sm^ | 148 | 11.254 | 0.031 |
| 62 | Sm | 149 | 13.830 | 0.038 |
| 62 | Sm | 150 | 7.351 | 0.020 |
| 62 | Sm | 152 | 26.735 | 0.073 |
| 62 | Sm | 154 | 22.730 | 0.062 |
| | | | 100 | 0.273 |
| 63 | Eu^ | 151 | 47.81 | 0.0479 |
| 63 | Eu | 153 | 52.19 | 0.0523 |
| | | | 100 | 0.1002 |
| 64 | Gd | 154 | 2.181 | 0.0075 |
| 64 | Gd | 155 | 14.800 | 0.0512 |
| 64 | Gd | 156 | 20.466 | 0.0708 |
| 64 | Gd | 157 | 15.652 | 0.0542 |
| 64 | Gd | 158 | 24.835 | 0.0859 |
| 64 | Gd | 160 | 21.864 | 0.0756 |
| | | | 100 | 0.345 |
| 65 | Tb | 159 | 100 | 0.0625 |
| 66 | Dy | 158 | 0.095 | 0.0004 |
| 66 | Dy | 160 | 2.329 | 0.0095 |
| 66 | Dy | 161 | 18.889 | 0.0769 |
| 66 | Dy | 162 | 25.479 | 0.1038 |
| 66 | Dy | 163 | 24.895 | 0.1014 |
| 66 | Dy | 164 | 28.260 | 0.1151 |
| | | | 100 | 0.407 |
| 67 | Ho | 165 | 100 | 0.0891 |
| 68 | Er | 162 | 0.139 | 0.0004 |
| 68 | Er | 164 | 1.601 | 0.0041 |
| 68 | Er | 166 | 33.503 | 0.086 |
| 68 | Er | 167 | 22.869 | 0.059 |
| 68 | Er | 168 | 26.978 | 0.069 |
| 68 | Er | 170 | 14.910 | 0.038 |
| | | | 100 | 0.256 |
| 69 | Tm | 169 | 100 | 0.0403 |
| 70 | Yb | 168 | 0.12 | 0.0003 |
| 70 | Yb | 170 | 2.98 | 0.0075 |
| 70 | Yb | 171 | 14.09 | 0.0356 |
| 70 | Yb | 172 | 21.69 | 0.0547 |
| 70 | Yb | 173 | 16.10 | 0.0406 |
| 70 | Yb | 174 | 32.03 | 0.0808 |
| 70 | Yb | 176 | 13.00 | 0.0328 |
| | | | 100 | 0.252 |
| 71 | Lu | 175 | 97.1795 | 0.0370 |
| 71 | Lu* | 176 | 2.8205 | 0.0011 |
| | | | 100 | 0.0381 |

| Z | E | A | atom% | N |
|---|---|---|---|---|
| 72 | Hf^ | 174 | 0.161 | 0.0003 |
| 72 | Hf | 176 | 5.205 | 0.0081 |
| 72 | Hf | 177 | 18.604 | 0.0289 |
| 72 | Hf | 178 | 27.297 | 0.0424 |
| 72 | Hf | 179 | 13.627 | 0.0212 |
| 72 | Hf | 180 | 35.105 | 0.0545 |
| | | | 100 | 0.155 |
| 73 | Ta* | 180 | 0.0120 | 2.6E-06 |
| 73 | Ta | 181 | 99.9880 | 0.0215 |
| | | | 100 | 0.0215 |
| 74 | W^ | 180 | 0.120 | 0.0002 |
| 74 | W | 182 | 26.499 | 0.0381 |
| 74 | W | 183 | 14.314 | 0.0206 |
| 74 | W | 184 | 30.642 | 0.0440 |
| 74 | W | 186 | 28.426 | 0.0408 |
| | | | 100 | 0.144 |
| 75 | Re | 185 | 35.662 | 0.0195 |
| 75 | Re* | 187 | 64.338 | 0.0352 |
| | | | 100 | 0.0547 |
| 76 | Os^ | 184 | 0.020 | 0.0001 |
| 76 | Os^ | 186 | 1.597 | 0.0104 |
| 76 | Os | 187 | 1.298 | 0.0085 |
| 76 | Os | 188 | 13.325 | 0.0869 |
| 76 | Os | 189 | 16.252 | 0.106 |
| 76 | Os | 190 | 26.433 | 0.172 |
| 76 | Os | 192 | 41.076 | 0.268 |
| | | | 100 | 0.652 |
| 77 | Ir | 191 | 37.272 | 0.236 |
| 77 | Ir | 193 | 62.728 | 0.397 |
| | | | 100 | 0.633 |
| 78 | Pt* | 190 | 0.013 | 0.0002 |
| 78 | Pt | 192 | 0.794 | 0.010 |
| 78 | Pt | 194 | 32.808 | 0.407 |
| 78 | Pt | 195 | 33.787 | 0.419 |
| 78 | Pt | 196 | 25.290 | 0.314 |
| 78 | Pt | 198 | 7.308 | 0.091 |
| | | | 100 | 1.24 |
| 79 | Au | 197 | 100 | 0.195 |
| 80 | Hg | 196 | 0.16 | 0.001 |
| 80 | Hg | 198 | 10.04 | 0.038 |
| 80 | Hg | 199 | 16.94 | 0.064 |
| 80 | Hg | 200 | 23.14 | 0.087 |
| 80 | Hg | 201 | 13.17 | 0.049 |
| 80 | Hg | 202 | 29.74 | 0.112 |
| 80 | Hg | 204 | 6.82 | 0.026 |
| | | | 100 | 0.376 |
| 81 | Tl | 203 | 29.524 | 0.053 |
| 81 | Tl | 205 | 70.476 | 0.126 |
| | | | 100 | 0.179 |
| 82 | Pb^ | 204 | 1.997 | 0.066 |
| 82 | Pb | 206 | 18.582 | 0.615 |
| 82 | Pb | 207 | 20.563 | 0.680 |
| 82 | Pb | 208 | 58.858 | 1.947 |
| | | | 100 | 3.308 |
| 83 | Bi^ | 209 | 100 | 0.1414 |
| 90 | Th* | 232 | 100 | 0.0421 |
| 92 | U* | 234 | 0.004 | 9.9E-07 |
| 92 | U* | 235 | 24.302 | 0.0058 |
| 92 | U* | 238 | 75.694 | 0.0181 |
| | | | 100 | 0.0239 |

**Footnote for Table 9**: Entries marked with "* "involve long-lived radioactive nuclides with half-lives up to $10^{12}$ years and abundances are for 4.567 Ga ago. Isotopes with half-lives above $10^{12}$ years (marked with "^") can be regarded as stable compared to the age of the solar system and are of interest of studies of double-beta decay. Isotopic compositions mainly adopted from Meija et al. (2016); see text for H, He, C, N, O, Ne, Ar, Kr, and Xe isotopic compositions.

## Acknowledgements

I thank Bruce Fegley, Ryan Ogliore, and Rainer Wieler for thoughtful comments and suggestions for this paper. I thank Don Burnett for communications about the Genesis data. Work supported in part by NSF AST 1517541.